\documentclass[aps,prd,groupedaddress,preprint,eqsecnum,nofootinbib,tightenlines]{revtex4}
\usepackage{calligra}
\usepackage{graphicx,epsf,amssymb,amsbsy,amsfonts,amssymb,amsmath}
\usepackage{verbatim}
\usepackage{pstricks}
\usepackage{color}
\usepackage[font=normalsize]{subfig}

\newif\ifdraft
\drafttrue
\newif\ifpreprint
\preprinttrue

\newbox\charbox
\newbox\slabox
\def\s#1{{      % Feynman slash
        \setbox\charbox=\hbox{$#1$}
        \setbox\slabox=\hbox{$/$}
        \dimen\charbox=\ht\slabox
        \advance\dimen\charbox by -\dp\slabox
        \advance\dimen\charbox by -\ht\charbox
        \advance\dimen\charbox by \dp\charbox
        \divide\dimen\charbox by 2
        \raise-\dimen\charbox\hbox to \wd\charbox{\hss/\hss}
        \llap{$#1$} }}

\begin{document}
\hfuzz 20pt

\title{One-Loop QCD and Higgs to Partons Processes Using Six-Dimensional Helicity and Generalized Unitarity\\}

\author{Scott Davies}

\affiliation{
Department of Physics and Astronomy, UCLA, Los Angeles, CA
90095-1547, USA\\
}

\begin{abstract}

We combine the six-dimensional helicity formalism of Cheung and O'Connell with $D$-dimensional generalized unitarity to obtain a new formalism for computing one-loop amplitudes in dimensionally regularized QCD.  With this procedure, we simultaneously obtain the pieces that are constructible from four-dimensional unitarity cuts and the rational pieces that are missed by them, while retaining a helicity formalism.  We illustrate the procedure using four- and five-point one-loop amplitudes in QCD, including examples with external fermions.  We also demonstrate the technique's effectiveness in next-to-leading order QCD corrections to Higgs processes by computing the next-to-leading order correction to the Higgs plus three positive-helicity gluons amplitude in the large top-quark mass limit.

\end{abstract}

\maketitle

\section{Introduction}

With the LHC recently surpassing the Tevatron for the world record in hadron collider beam intensity and energy, we stand ready to test and expand our current understanding of particle physics.  The LHC is well on its way in hunting for the Higgs boson and for physics beyond the Standard Model.  To engage in these ambitious searches, however, a detailed picture of Standard Model backgrounds is necessary.  Higher order QCD is an important theoretical tool in both Higgs production and the Standard Model background.

The QCD corrections are large and much effort has been dedicated to calculating important cross sections \cite{NLOMultileg}.  While some processes will require next-to-next-to-leading order precision and resummation of large logarithms, for many processes knowledge through next-to-leading order (NLO) in the perturbative expansion will be sufficient.  This need for NLO calculations has been discussed and codified in an experimenter's wishlist \cite{NLOMultileg}.  One-loop calculations in QCD are therefore a crucial part of theoretical studies of backgrounds and signals at the LHC.

As the number of external partons increases, Feynman-diagrammatic techniques encounter difficulties, especially at loop level.  This has motivated many attempts to seek improved techniques.  In recent years there has been rapid progress significantly reducing the computational intensity behind NLO calculations.  The unitarity method, pioneered by Bern, Dixon, Dunbar and Kosower in the mid-nineties \cite{BDDK1, BDDK2}, offers an alternative using the branch-cut structure of loop integrals to find the coefficients of the integrals in terms of products of on-shell tree amplitudes.  Use of generalized unitarity is natural in this framework, giving further enhancements to the method~\cite{Zqqgg}.  The tree amplitudes are sewn together across various generalized channels to reconstruct the loop amplitude.  In this way only physical degrees of freedom are used.  One-loop amplitudes for various supersymmetric theories can be constructed in their entirety through the use of four-dimensional cuts, though non-supersymmetric theories, including QCD, have cut-free rational terms that must be handled through other means~\cite{BernMorgan,Zqqgg}.  By using complex momenta, Britto, Cachazo and Feng showed that generalized quadruple cuts determine all scalar box integral coefficients as products of four on-shell tree amplitudes \cite{genUnit}, motivating new reduction procedures on the integrand.  Ossola, Papadopoulos and Pittau (OPP) followed suit with a purely algebraic method to reduce tensor integrals to a basis of master scalar integrals \cite{OPP1} in a form directly usable in numerical programs.  The OPP method uses a particular parametrization of the loop momentum to find coefficients for all scalar integrals by solving a system of linear equations.  Forde followed with a method for simple extraction of triangle and bubble coefficients in four-dimensional cuts that exploits the analytic structure of the integrand \cite{Forde}.  Coefficients for all scalar integrals can be found through the limiting behavior of the products of tree amplitudes by using the momentum parametrization in ref.~\cite{OPP1}, first given in ref.~\cite{AguilaPittau}.  A numerical variant of Forde's method was then developed and applied in ref.~\cite{BlackHatI}.  There have also been important improvements in Feynman diagrammatic techniques~\cite{RecentFeynman}.

Another important technique that greatly simplifies amplitude calculations is helicity methods~\cite{SpinorHelicity}.  At tree level these lead to enormous simplifications, especially for massless amplitudes.  The massless spinor-helicity formalism in four dimensions has been an invaluable tool for simplifying calculations and exposing the structure of amplitudes starting with the discovery of the Parke-Taylor formula for maximally helicity-violating tree amplitudes~\cite{ParkeTaylor}.  At loop level helicity methods have also played an important role in pushing back the calculational frontiers.  Recently the first NLO calculation of a hadron-collider process with five final-state objects was presented: $W$+4-jet production~\cite{Wp4Jets}.  This calculation relied on using spinor variables in the tree amplitudes composing the unitarity cuts.

At loop level the application of four-dimensional helicity methods is much trickier than at tree level because of the use of dimensional regularization, which requires that the loop momenta be outside of four dimensions.  This can drop rational pieces not captured by four-dimensional unitarity cuts.  A purely four-dimensional method relies instead on on-shell recursion relations to reconstruct the rational pieces~\cite{recursionRational, BernLastOfTheFinite, BlackHatI}.  Another approach is to abandon four-dimensional helicity in the cuts and use $D$-dimensional generalized cuts, which simultaneously give cut-constructible and rational terms~\cite{BernMorgan, BDDK3, OtherDDimUnitarity}.  Rational terms arise through the interference of the divergences in loop integrals with the order $\epsilon$ pieces of $(4-2\epsilon)$-dimensional loop momenta.  A means for tracking these within the OPP framework in terms of a set of Feynman-like rules has been given in ref.~\cite{RationFeynmanRules}.  Work by Ellis, Giele, Kunszt and Melnikov combines $D$-dimensional generalized unitarity with the OPP method for numerical calculations~\cite{KunsztNumerical, FullOneLoopGiele, KunsztFermionsDUnitarity}.  A related approach due to Badger \cite{Badger} associates the $(D-4)$-dimensional pieces to masses to give the rational terms, and a numerical version of this has been developed and applied in refs.~\cite{Wp3Jets,Wp4Jets}.

Given the success of helicity methods when combined with generalized unitarity at pushing back the frontiers~\cite{Wp4Jets}, it seems reasonable to try to find improved means of combining the two techniques.  In this paper we bring $D$-dimensional unitarity and spinor helicity together: we work in $D=6$ using a six-dimensional helicity formalism due to Cheung and O'Connell \cite{COC6D} to calculate QCD amplitudes, including NLO corrections to Higgs processes, at one loop.  We capture both the cut-constructible and rational pieces while enjoying the conveniences of a spinor-helicity formalism, and are in fact able to write the answer in terms of massless four-dimensional spinors through a decomposition of their six-dimensional counterparts.  At the end of the calculation, we analytically continue the loop momenta to $D=4-2\epsilon$ dimensions and perform a state-sum reduction to reduce the spin states of six dimensions to match our desired regularization scheme.  In this paper we focus on the four-dimensional helicity (FDH) scheme \cite{BernFDH, BernMorgan, BDDK3}.  Loop computations combining six-dimensional helicity and unitarity to obtain supersymmetric amplitudes have been carried out in refs.~\cite{Bern6D, Tristan, Koschade}, though for the cases treated there, there was no need to apply a state-sum reduction procedure as we do here.

We illustrate the technique using various four-point QCD amplitudes as examples, including ones with external quarks.  We also present a simple five-point example.  As a more sophisticated example, we evaluate an amplitude involving the Higgs boson and three gluons.  This process is mediated by a heavy-quark loop \cite{GeorgiGlashow}, so NLO involves a two-loop process.  However, in the large top-quark mass limit, we can replace the top loop with an effective operator \cite{Djouadi, Dawson, Wilczek, Shifman}, thereby bringing our NLO computation down to one loop.  In all cases we reproduce known results confirming the validity of the formalism.

This paper is organized as follows: In section \ref{SixD} we review the six-dimensional spinor-helicity formalism of Cheung and O'Connell \cite{COC6D}.  We cover the embedding of four-dimensional spinors into six-dimensional spinors, a crucial step to obtaining amplitudes in terms of four-dimensional objects.  In section \ref{IntBasis} we outline our choice of integral basis in $4-2\epsilon$ dimensions for the FDH scheme and identify coefficients that must be found to obtain the full loop amplitude.  Then in section \ref{IntExtract} we review integral coefficient extraction for $(4-2\epsilon)$-dimensional generalized unitarity following Badger \cite{Badger}.  We give the appropriate loop momentum parametrization and coefficient formulas, leaving the details behind the procedure to an appendix.
Section \ref{4pointglue} brings us to sample calculations of the one-loop four-gluon amplitude $A_4^{(1)}(1^-,2^+,3^+,4^+)$ for various internal-particle states.  We find that a state-sum reduction is necessary to eliminate the extra-dimensional polarization states of six dimensions compared to four dimensions.  We also present our results for a maximally helicity-violating configuration for a gluon loop, as well as for the five-point amplitude $A_4^{(1)}(1^+,2^+,3^+,4^+,5^+)$.  In section \ref{ExtFerm} we discuss the case of external massless quarks, also subject to a similar state-sum reduction procedure.  In section \ref{HiggsPartons} we apply our technique to a simple Higgs to partons NLO QCD correction process, a topic of great interest at the LHC.  All Higgs plus four partons processes have been calculated analytically using four-dimensional cuts and various other methods for the rational pieces \cite{Higgs4Parton}.  We show the utility of our method with the simple example $A^{(1)}(H,1^+,2^+,3^+)$.  In section \ref{Conclusion} we give our concluding remarks.

\section{SIX-DIMENSIONAL HELICITY} \label{SixD}

We begin with a summary of Cheung and O'Connell's six-dimensional helicity formalism \cite{COC6D}, which gives us a spinor-helicity formalism to use in unitarity cuts where internal particles must be kept in $D\neq 4$.  After a brief review of four-dimensional spinor helicity, we make the analogous construction in six dimensions using solutions to the massless six-dimensional Dirac equation in a similar ``Weyl'' basis.  We then show how the six-dimensional spinors can be decomposed into four-dimensional spinors.  In Section~\ref{4pointglue}, we will explain how to correct for the state sums being in six dimensions instead of the dimensionally regularized values.

\subsection{Spinor helicity}

The four-dimensional spinor-helicity formalism has been widely used in scattering amplitude calculations, and a useful review can be found in ref.~\cite{Dixon}.  We use massless chiral and anti-chiral spinors in a Weyl basis to represent a light-like momentum $p^{\mu}$ as a bi-spinor,
\begin{align}
p_{\alpha\dot{\alpha}}=p_{\mu}\sigma_{\alpha\dot{\alpha}}^{\mu}=\lambda_{\alpha}\tilde{\lambda}_{\dot{\alpha}}.
\end{align}
In addition to its momentum $p^{\mu}$, a massless particle in four dimensions is labeled by its helicity.  We define polarization vectors in terms of spinors $\lambda_p=|p^+\rangle$ and $\tilde{\lambda}_p=|p^-\rangle$, which can be done as
%
\begin{comment}
\begin{align}
\epsilon_{\mu}^+(p,q)=\frac{\langle q^-|\bar{\sigma}_{\mu}|p^-\rangle}{\sqrt{2}\langle q^-|p^+\rangle}, \hspace{2cm} \epsilon_{\mu}^-(p,q)=\frac{\langle q^+|\sigma_{\mu}|p^+\rangle}{\sqrt{2}\langle p^+|q^-\rangle}
\end{align}
\end{comment}
%
\begin{align}
\epsilon_{\mu}^{\pm}(p,q)=\pm\frac{\langle q^{\mp}|\gamma_{\mu}|p^{\mp}\rangle}{\sqrt{2}\langle q^{\mp}|p^{\pm}\rangle},
\end{align}
where $q^{\mu}$ is some arbitrary light-like reference momentum representing a gauge freedom.  Following this formalism, we are able to write compact expressions for amplitudes in terms of Lorentz-invariant spinor products,
\begin{align}
\epsilon^{\alpha\beta}\lambda_{i\beta}\lambda_{j\alpha}=\langle i\,j\rangle\,\mathrm{~and~}\,\epsilon_{\dot{\alpha}\dot{\beta}}\tilde{\lambda}_i^{\dot{\beta}}\tilde{\lambda}_j^{\dot{\alpha}}=[i\,j],
\end{align}
where $\epsilon^{12}=\epsilon_{21}=1$.  Here we are following the convention $\langle i\,j\rangle[j\,i]=2p_i\cdot p_j=s_{ij}$.

In six dimensions the Lorentz group is SO(5,1).  We can then decompose the Dirac equation into chiral and anti-chiral pieces using a ``Weyl'' basis,
\begin{align}
p_{\mu}\sigma_{AB}^{\mu}\lambda_p^{Ba}=0, \hspace{3cm} p_{\mu}\tilde{\sigma}^{\mu AB}\tilde{\lambda}_{pB\dot{a}}=0,
\end{align}
where $\{A, B,\cdots\}$ are fundamental representation indices of the covering group, SU*(4).  The sigma matrices $\sigma_{AB}$ and $\tilde{\sigma}^{AB}$ are 4x4 antisymmetric matrices playing a part analogous to the Pauli matrices of four dimensions.  Explicit forms and some useful relations can be found in Appendix A of ref.~\cite{COC6D}.  The chiral and anti-chiral Weyl spinors $\lambda^{Ba}$ and $\tilde{\lambda}_{B\dot{a}}$ each have two solutions labeled by the indices $a=1, 2$ and $\dot{a}=1, 2$.  These are indices of the little group SO(4), corresponding to SU(2)$\times$SU(2).  They can be raised and lowered using the matrices $\epsilon_{ab}$ and $\epsilon^{\dot{a}\dot{b}}$,
\begin{align}
\lambda_a=\epsilon_{ab}\lambda^b, \hspace{3cm} \tilde{\lambda}^{\dot{a}}=\epsilon^{\dot{a}\dot{b}}\tilde{\lambda}_{\dot{b}},
\end{align}
where once again $\epsilon^{12}=\epsilon_{21}=1$.

Using these spinors we can write a six-dimensional light-like momentum vector in a bi-spinor representation similar to that of four dimensions,
\begin{align}
p^{AB}&=p_{\mu}\tilde{\sigma}^{\mu AB}=\lambda^{Aa}\epsilon_{ab}\lambda^{Bb}=|p^a\rangle\epsilon_{ab}\langle p^b|, \notag \\
p_{AB}&=p_{\mu}\sigma^{\mu}_{AB}=\tilde{\lambda}_{A\dot{a}}\epsilon^{\dot{a}\dot{b}}\tilde{\lambda}_{B\dot{b}}=|p_{\dot{a}}]\epsilon^{\dot{a}\dot{b}}[p_{\dot{b}}|,
\end{align}
where we have adopted a bra-ket notation $\lambda^{Aa}=|p^a\rangle$, $\tilde{\lambda}_{A\dot{a}}=|p_{\dot{a}}]$.  We can also express the momentum vector directly as
\begin{align}
p^{\mu}=-\frac{1}{4}\langle p^a|\sigma^{\mu}|p^b\rangle\epsilon_{ab}=-\frac{1}{4}[p_{\dot{a}}|\tilde{\sigma}^{\mu}|p_{\dot{b}}]\epsilon^{\dot{a}\dot{b}},
\end{align}
where there is a contraction of SU*(4) indices between the sigma matrix and the spinors.  Lorentz-invariant spinor inner-products are also defined by contractions of SU*(4) indices,
\begin{align}
\langle i^a|j_{\dot{b}}]&=\lambda_i^{Aa}\tilde{\lambda}_{jA\dot{b}}=[j_{\dot{b}}|i^a\rangle, \notag \\
\langle i^a|i_{\dot{a}}]&=0.
\end{align}
Other important quantities showing up in amplitude calculations are the spinor contractions with the SU*(4)-invariant Levi-Civita tensor,
\begin{align}
\langle i^aj^bk^cl^d\rangle&\equiv\epsilon_{ABCD}\lambda_i^{Aa}\lambda_j^{Bb}\lambda_k^{Cc}\lambda_l^{Dd}, \notag \\
[i_{\dot{a}}j_{\dot{b}}k_{\dot{c}}l_{\dot{d}}]&\equiv\epsilon^{ABCD}\tilde{\lambda}_{iA\dot{a}}\tilde{\lambda}_{jB\dot{b}}\tilde{\lambda}_{kC\dot{c}}\tilde{\lambda}_{lD\dot{d}},
\end{align}
and spinor strings,
\begin{align}
\langle i^a|\s{p}_1\s{p}_2\cdots\s{p}_{2n+1}|j^b\rangle&=(\lambda_i)^{A_1a}(p_1)_{A_1A_2}(p_2)^{A_2A_3}\cdots(p_{2n+1})_{A_{2n+1}A_{2n+2}}(\lambda_j)^{A_{2n+2}b}, \notag \\
\langle i^a|\s{p}_1\s{p}_2\cdots\s{p}_{2n}|j_{\dot{b}}]&=(\lambda_i)^{A_1a}(p_1)_{A_1A_2}(p_2)^{A_2A_3}\cdots(p_{2n})^{A_{2n}A_{2n+1}}(\tilde{\lambda}_j)_{A_{2n+1}\dot{b}}.
\end{align}

Finally we look at polarization vectors, which, as in four dimensions, can be written in terms of spinors.  Following ref.~\cite{COC6D},
\begin{align}
\epsilon^{\mu}_{a\dot{a}}(p,q)=-\frac{1}{\sqrt{2}}\langle p_a|\sigma^{\mu}|q_b\rangle(\langle q_b|p^{\dot{a}}])^{-1}=\frac{1}{\sqrt{2}}(\langle p^a|q_{\dot{b}}])^{-1}[q_{\dot{b}}|\tilde{\sigma^{\mu}}|p_{\dot{a}}].
\end{align}
Unlike the four-dimensional case, we cannot label these as simply $+$ or $-$ because gluons in six dimensions have four polarization states.  The states are in fact labeled by SU(2)$\times$SU(2) little-group indices.  For the Weyl spinors $\lambda^{Aa}$ and $\tilde{\lambda}_{A\dot{a}}$ then, the indices $a$ and $\dot{a}$ label two helicity states respectively.  We notice that the six-dimensional polarization vectors have a gauge freedom through the null reference momentum $q$, as is the case in four dimensions.
\subsection{Decomposing six-dimensional spinors into four-dimensional ones}
We find it convenient to write six-dimensional spinors in terms of four-dimensional ones, allowing amplitudes to be expressed in terms of the more familiar four-dimensional spinors.  From a four-dimensional perspective, we view six-dimensional null vectors as being massive.  Making the associations,
\begin{align}
m\equiv p_5-ip_4, \hspace{3cm} \tilde{m}\equiv p_5+ip_4,
\label{eq:mmTilde}
\end{align}
our six-dimensional massless condition is
\begin{align}
p^2=\,\bar{p}^2-p_4^2-p_5^2\equiv\,\bar{p}^2-m\tilde{m}=0,
\end{align}
where $\bar{p}$ denotes a momentum vector with only the first four components.  This is the on-shell condition for a four-dimensional massive momentum.  We can write a bi-spinor representation of massive momenta in terms of two pairs of four-dimensional spinors: $\lambda$, $\tilde{\lambda}$ and $\mu$, $\tilde{\mu}$, as
\begin{align}
\bar{p}_{\alpha\dot{\alpha}}=\lambda_{\alpha}\tilde{\lambda}_{\dot{\alpha}}+\rho\,\mu_{\alpha}\tilde{\mu}_{\dot{\alpha}},
\end{align}
where
\begin{align}
\rho=\kappa\tilde{\kappa}=\kappa'\tilde{\kappa}',  \hspace{6mm} \kappa\equiv\frac{m}{\langle\lambda\,\mu\rangle},  \hspace{6mm} \tilde{\kappa}\equiv\frac{\tilde{m}}{[\mu\,\lambda]}, \hspace{6mm} \kappa'\equiv\frac{\tilde{m}}{\langle\lambda\,\mu\rangle}, \hspace{6mm} \tilde{\kappa}'\equiv\frac{m}{[\mu\,\lambda]},
\end{align}
which leads to a form for the six-dimensional spinors.  Treating them as 4$\times$2 matrices, we decompose them in terms of the above four-dimensional spinors as~\cite{Bern6D, Tristan, Boels}
\begin{align}
\lambda^A_{\,\,\,a}=\left(\begin{array}{cc}
-\kappa\mu_{\alpha}&\lambda_{\alpha} \\
\tilde{\lambda}^{\dot{\alpha}}&\tilde{\kappa}\tilde{\mu}^{\dot{\alpha}}\end{array}\right),
\hspace{3cm} \tilde{\lambda}_{A\dot{a}}=\left(\begin{array}{cc}
\kappa'\mu^{\alpha}&\lambda^{\alpha} \\
-\tilde{\lambda}_{\dot{\alpha}}&\tilde{\kappa}'\tilde{\mu}_{\dot{\alpha}}\end{array}\right).
\label{eq:6Dspinors}
\end{align}
The SU*(4) indices label the rows while the little group indices $a$ and $\dot{a}$ take on the values 1 and 2 to label the columns.  The embedding is specific to the form of the $\sigma^{\mu}_{AB}$ matrices as taken from ref.~\cite{COC6D}.

If we take the momenta to be in the four-dimensional subspace with $p^{4,5}=0$, or equivalently $m=\tilde{m}=0$, the spinors reduce to the four-dimensional forms,
\begin{align}
\lambda^A_{\,\,a}=\left(\begin{array}{cc}
0&\lambda_{\alpha} \\
\tilde{\lambda}^{\dot{\alpha}}&0\end{array}\right),
\hspace{3cm} \tilde{\lambda}_{A\dot{a}}=\left(\begin{array}{cc}
0&\lambda^{\alpha} \\
-\tilde{\lambda}_{\dot{\alpha}}&0\end{array}\right).
\label{eq:4D6Dspinors}
\end{align}
When we do unitarity calculations, we will keep the external particles in the four-dimensional subspace, so they will have this simpler form.  The internal particles, on the other hand, will have the more complicated form of eq.~\eqref{eq:6Dspinors}.

\subsection{Tree-level amplitude examples}

We now consider some tree-level amplitudes that we will need for unitarity cuts.  Tree-level amplitudes in six dimensions have a remarkable chiral-conjugate structure.  The color-ordered four-gluon tree amplitude is given by \cite{COC6D}
\begin{align}
A_4^{(0)}(1^g_{a\dot{a}},2^g_{b\dot{b}},3^g_{c\dot{c}},4^g_{d\dot{d}})=-\frac{i}{s_{12}s_{23}}\langle 1_a2_b3_c4_d\rangle[1_{\dot{a}}2_{\dot{b}}3_{\dot{c}}4_{\dot{d}}].
\label{eq:4gTree}
\end{align}
This was found in ref.~\cite{COC6D} by building up from three-point amplitudes using the Britto-Cachazo-Feng-Witten recursion relations \cite{BCFW}.  Notice that, unlike four-dimensional amplitudes, the helicity is not specified.  The little group SU(2)$\times$SU(2) connects all helicities together so this simple result holds for all helicity arrangements.  For the two-chiral-quark two-gluon tree amplitude, we have \cite{Bern6D}
\begin{align}
A_4^{(0)}(1^g_{a\dot{a}},2^g_{b\dot{b}},3^q_c,4^q_d)=-\frac{i}{2s_{12}s_{23}}\langle 1_a2_b3_c4_d\rangle[1_{\dot{a}}2_{\dot{b}}3_{\dot{e}}3^{\dot{e}}],
\label{eq:2g2fTree}
\end{align}
while for the two-real-scalar two-gluon tree amplitude,
\begin{align}
A_4^{(0)}(1^g_{a\dot{a}},2^g_{b\dot{b}},3^s,4^s)=-\frac{i}{4s_{12}s_{23}}\langle 1_a2_b3_e3^e\rangle[1_{\dot{a}}2_{\dot{b}}3_{\dot{e}}3^{\dot{e}}].
\label{eq:2g2sTree}
\end{align}
One can easily check that these reduce to the known four-dimensional results using the reduced spinors in eq.~\eqref{eq:4D6Dspinors}.

\section{$(4-2\epsilon)$-DIMENSIONAL INTEGRAL BASIS} \label{IntBasis}

A general one-loop amplitude can be written in terms of a basis of master integrals.  We give our choice of basis here and leave the details to Appendix \ref{IntBasisApp}, where relevant references can be found as well.  Following ref.~\cite{Badger}, for $n$ particles in four dimensions with all internal particles massless and in $D=4-2\epsilon$, a one-loop amplitude can be written as
\begin{align}
A_n^{(1)}=&\frac{\mu^{2\epsilon}}{(4\pi)^{2-\epsilon}}\left(\sum_{K_4}C_{4;K_4}^{[0]}I_{4;K_4}^{4-2\epsilon}+\sum_{K_4}C_{4;K_4}^{[4]}I_{4;K_4}^{4-2\epsilon}[\mu^4]+\sum_{K_3}C_{3;K_3}^{[0]}I_{3;K_3}^{4-2\epsilon}\right. \notag \\
&\left.+\sum_{K_3}C_{3;K_3}^{[2]}I_{3;K_3}^{4-2\epsilon}[\mu^2]+\sum_{K_2}C_{2;K_2}^{[0]}I_{2;K_2}^{4-2\epsilon}+\sum_{K_2}C_{2;K_2}^{[2]}I_{2;K_2}^{4-2\epsilon}[\mu^2]\right)+ O(\epsilon), \label{eq:intBasis}
\end{align}
where $K_r$ refers to the set of all ordered partitions of the external momenta into $r$ distinct groups.  The integrals are given by
\begin{align}
I_n^{4-2\epsilon}[f(\mu^2)]=i(-1)^{n+1}(4\pi)^{2-\epsilon}\int\frac{d^{4-2\epsilon}l}{(2\pi)^{4-2\epsilon}}\frac{f(\mu^2)}{l^2(l-K_1)^2(l-K_1-K_2)^2\cdots(l+K_n)^2}.
\label{Ints}
\end{align}
For simplicity in eq.~\eqref{eq:intBasis}, we have written $I_n^{4-2\epsilon}$ without the explicit $\mu^2$ dependence when $f(\mu^2)=1$.  The scalar integrals have been evaluated and can be found in refs.~\cite{PentInts, BDDK2}; the integrals with $f(\mu^2)\neq 1$ are straightforward to evaluate and are given in Appendix \ref{IntBasisApp}.  Because all the integrals in eq.~\eqref{eq:intBasis} are known, calculating one-loop amplitudes boils down to finding the coefficients in front of the integrals.

\section{INTEGRAL COEFFICIENT EXTRACTION} \label{IntExtract}

With up to box integrals in our basis, the largest number of cuts necessary for generalized unitarity in $4-2\epsilon$ dimensions is four.  Quadruple cuts give us box integral coefficients, while three- and two-particle cuts give us, respectively, triangle and bubble integral coefficients.  However, there is some entanglement depending on the cut.  Two-particle cuts, for example, are also contained in box and triangle integrals, and the bubble coefficient must be isolated through some extraction procedure.  An efficient method for coefficient extraction in $D=4$ based on the analytic properties of the integrand is given by Forde \cite{Forde}.  Badger \cite{Badger} generalizes this method to $D=4-2\epsilon$ with an emphasis on rational terms obtained by treating the extra-dimensional components of momenta effectively as masses, in a  form related to extraction of integral coefficients in the presence of masses~\cite{Kilgore}.  We follow the construction of these references, providing the momentum solutions to the cut conditions and the final forms of the coefficients.  Further details on the procedure can be found in Appendix \ref{ExtractAppend}.

\subsection{Box integral coefficients} \label{Boxes}

%%%%%%%%% FIGURE %%%%%%%%%%%%%%%
\begin{figure}[ht]
\centerline{\epsfxsize 2 truein \epsfbox{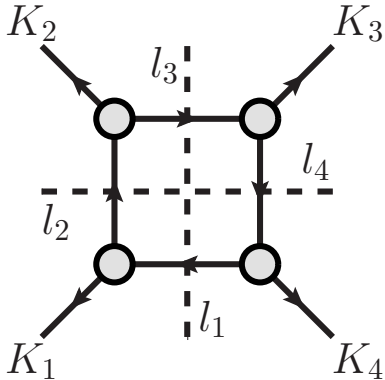}}
\caption[a]{\small A general quadruple cut.  Loop momenta flow clockwise.}
\label{Fig:BoxCut}
\end{figure}
%%%%%%%%%%%%%%%%%%%%%%%%%%%%%%%%

For the quadruple cut (Fig. \ref{Fig:BoxCut}) in $4-2\epsilon$ dimensions, the on-shell cut conditions are
\begin{align}
l_1^2=l_2^2=l_3^2=l_4^2=0,
\end{align}
or equivalently,
\begin{align}
\bar{l}_1^2=\bar{l}_2^2=\bar{l}_3^2=\bar{l}_4^2=\mu^2,
\end{align}
where $\bar{l}_i$ are the momenta truncated to four dimensions and $\mu$ represents the $(-2 \epsilon)$-dimensional components.  It is convenient to view $\mu^2$ as a mass term and the internal momenta as four-dimensionally massive. (This is the view taken in ref.~\cite{Badger} and as we saw a useful one for six-dimensional helicity as well.)  We parametrize the four-momentum $\bar{l}_1$ using a construction first given in refs.~\cite{OPP1, OPP2, AguilaPittau} and also applied in refs.~\cite{Forde, Badger, Kilgore},
\begin{align}
\bar{l}_1^{\mu}=\frac{1}{2}(a\langle K_4^{\flat -}|\gamma^{\mu}|K_4^{\flat -}\rangle+b\langle K_1^{\flat -}|\gamma^{\mu}|K_1^{\flat -}\rangle+c\langle K_4^{\flat -}|\gamma^{\mu}|K_1^{\flat -}\rangle+d\langle K_1^{\flat -}|\gamma^{\mu}|K_4^{\flat -}\rangle),
\end{align}
where $K_{1,4}^{\flat}$ is the massless projection of one of the external legs in the direction of the other masslessly projected leg,
\begin{align}
K_1^{\flat\mu}=\frac{\gamma_{14}(\gamma_{14}K_1^{\mu}-S_1K_4^{\mu})}{\gamma_{14}^2-S_1S_4},
\hspace{1cm} K_4^{\flat\mu}=\frac{\gamma_{14}(\gamma_{14}K_4^{\mu}-S_4K_1^{\mu})}{\gamma_{14}^2-S_1S_4}, \notag
\label{eq:masslessProjectedLegs}
\end{align}
\begin{align}\gamma_{14}=K_1\cdot K_4\pm\sqrt{(K_1\cdot K_4)^2-S_1S_4},\hspace{1cm} S_i=K_i^2.
\end{align}
Solving the on-shell conditions, we find that $\bar{l}_1$ can be expanded as
\begin{align}
\bar{l}_1^{\mu}&=aK_4^{\flat\mu}+bK_1^{\flat\mu}+\frac{c_\pm}{2}\langle K_4^{\flat -}|\gamma^{\mu}|K_1^{\flat -}\rangle+\frac{\gamma_{14}ab-\mu^2}{2c_\pm\gamma_{14}}\langle K_1^{\flat -}|\gamma^{\mu}|K_4^{\flat -}\rangle \notag \\
&=\bar{l}_1^{\flat\mu}-\frac{\mu^2}{2c_\pm\gamma_{14}}\langle K_1^{\flat -}|\gamma^{\mu}|K_4^{\flat -}\rangle,
\label{eq:BoxSol}
\end{align}
where
\begin{align}
a=\frac{S_1(S_4+\gamma_{14})}{\gamma_{14}^2-S_1S_4}, \hspace{6mm} b=-\frac{S_4(S_1+\gamma_{14})}{\gamma_{14}^2-S_1S_4},\hspace{6mm} c_{\pm}=\frac{-c_1\pm\sqrt{c_1^2-4c_0c_2}}{2c_2}, \notag
\label{MomentumCoeffsab}
\end{align}
\begin{align}
c_2&=\langle K_4^{\flat -}|\s{K}_2|K_1^{\flat -}\rangle, \notag \\
c_1&=a\langle K_4^{\flat -}|\s{K}_2|K_4^{\flat -}\rangle+b\langle K_1^{\flat -}|\s{K}_2|K_1^{\flat -}\rangle-S_2-2K_1\cdot K_2, \notag \\
c_0&=\left(ab-\frac{\mu^2}{\gamma_{14}}\right)\langle K_1^{\flat -}|\s{K}_2|K_4^{\flat -}\rangle.
\end{align}
In general there are two solutions to the on-shell conditions.  It would appear initially that there are four, two each for $\gamma_{14}$ and $c_{\pm}$, but it turns out that,
\begin{align}
\bar{l}_1(\gamma_{14}^+,c_+)&=\bar{l}_1(\gamma_{14}^-,c_-), \notag \\
\bar{l}_1(\gamma_{14}^+,c_-)&=\bar{l}_1(\gamma_{14}^-,c_+).
\end{align}
In the case $S_1=0$ or $S_4=0$, there is only one solution for $\gamma_{14}$ (but still two solutions to the on-shell conditions).  To determine the full box coefficient, we must average over these solutions.

Using these solutions to define spinors, we can calculate the four tree amplitudes associated with the quadruple cut.  As explained in Appendix \ref{ExtractAppend}, the coefficients associated with our integral basis choice \eqref{eq:intBasis} are
\begin{align}
C_4^{[0]}&=\frac{i}{2}\sum_{\sigma}\left.A_1A_2A_3A_4(\bar{l}_1^{\sigma})\right|_{\mu^2\rightarrow 0}, \notag \\
C_4^{[4]}&=\frac{i}{2}\sum_{\sigma}\left.[\mathrm{Inf}_{\mu^2}A_1A_2A_3A_4](\mu^2)\right|_{\mu^4}.
\label{eq:BoxExtract}
\end{align}
The sum is over the two solutions to the quadruple cut; the product $A_1A_2A_3A_4$ must be computed for each.  To find $C_4^{[0]}$, our cut-constructible piece, we set $\mu^2=0$.  The Inf term in $C_4^{[4]}$, first given in ref.~\cite{BernInf}, contains the information from the boundary of the $\mu$ contour integral,
\begin{align}
\lim_{\mu\rightarrow\infty}([\mathrm{Inf}_{\mu^2}A_1A_2A_3A_4](\mu^2)-A_1(\mu)A_2(\mu)A_3(\mu)A_4(\mu))=0.
\end{align}
We expand around infinity and write it as a polynomial in $\mu^2$,
\begin{align}
[\mathrm{Inf}_{\mu^2}A_1A_2A_3A_4](\mu^2)=\sum_{i=0}^2c_i\mu^{2i},
\end{align}
then restrict $C_4^{[4]}$ to be the coefficient of the $\mu^4$ term, explaining the notation of eq.~\eqref{eq:BoxExtract}.

\subsection{Triangle integral coefficients} \label{Triangles}

%%%%%%%%% FIGURE %%%%%%%%%%%%%%%
\begin{figure}[ht]
\centerline{\epsfxsize 2 truein \epsfbox{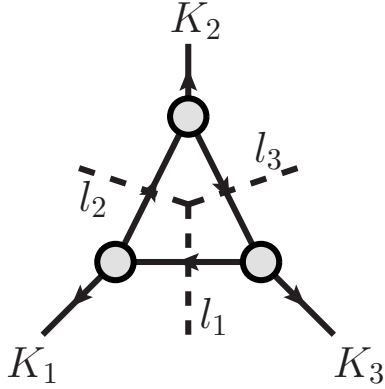}}
\caption[a]{\small A general triple cut.}
\label{Fig:TriCut}
\end{figure}
%%%%%%%%%%%%%%%%%%%%%%%%%%%%%%%%

To obtain the solution to the triple cut (Fig. \ref{Fig:TriCut}), we again parametrize our loop momentum in terms of adjacent projected external momenta,
\begin{align}
\bar{l}_1^{\mu}&=aK_3^{\flat\mu}+bK_1^{\flat\mu}+\frac{t}{2}\langle K_3^{\flat -}|\gamma^{\mu}|K_1^{\flat -}\rangle+\frac{\gamma_{13}ab-\mu^2}{2t\gamma_{13}}\langle K_1^{\flat -}|\gamma^{\mu}|K_3^{\flat -}\rangle,
\label{eq:TriCut}
\end{align}
where $K_1^{\flat}$ and $K_3^{\flat}$ are defined analogously to eq.~\eqref{eq:masslessProjectedLegs}.  The coefficients $a$ and $b$ are defined in eq.~\eqref{MomentumCoeffsab} and $\gamma_{13}$ is defined in eq.~\eqref{eq:masslessProjectedLegs} (with $K_3^{\flat}$ replacing $K_4^{\flat}$ in both).  We must average over the solutions for $\gamma_{13}$, though in the case that $S_1=0$ or $S_3=0$, there is only one solution.  For a fixed value of $\gamma_{13}$, we must also average over the coefficients given by the conjugate solution,
\begin{align}
\bar{l}_1^{*\mu}&=aK_3^{\flat\mu}+bK_1^{\flat\mu}+\frac{t}{2}\langle K_1^{\flat -}|\gamma^{\mu}|K_3^{\flat -}\rangle+\frac{\gamma_{13}ab-\mu^2}{2t\gamma_{13}}\langle K_3^{\flat -}|\gamma^{\mu}|K_1^{\flat -}\rangle.
\label{eq:ConjTriCut}
\end{align}
In both solutions the complex parameter $t$ is free.

Box integrals also contain triple cuts, so we must extract the triangle coefficients using the limiting behavior of the integrand.  The coefficients therefore contain an Inf term that is a polynomial expansion in $t$,
\begin{align}
C_3^{[0]}&=-\frac{1}{2n_{\gamma}}\sum_{\sigma}\left.[\mathrm{Inf}_tA_1A_2A_3(\bar{l}_1^{\sigma})](t)\right|_{\mu^2\rightarrow 0,t\rightarrow 0}, \notag \\
C_3^{[2]}&=-\frac{1}{2n_{\gamma}}\sum_{\sigma}\left.[\mathrm{Inf}_{\mu^2}[\mathrm{Inf}_tA_1A_2A_3(\bar{l}_1^{\sigma})](t)](\mu^2)\right|_{\mu^2,t\rightarrow 0},
\label{eq:TriCoeffs}
\end{align}
but only the order $t^0$ term is retained.  The sum is over the solutions, including the conjugate-momentum solution, to the cut conditions.  There may be either two or four solutions depending on the number of solutions $n_{\gamma}$ for $\gamma_{13}$.  In $C_3^{[0]}$, $\mu^2$ and $t$ are both set to zero, while the expansion in $C_3^{[2]}$ is restricted to the coefficients of the $\mu^2$ term (in addition to having $t=0$).

\subsection{Bubble integral coefficients} \label{Bubbles}

%%%%%%%%% FIGURE %%%%%%%%%%%%%%%
\begin{figure}[ht]
\centering
\subfloat[][]{\epsfxsize 2 truein \epsfbox{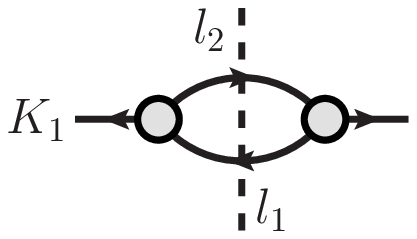}}
\hspace{1cm}
\subfloat[][]{\epsfxsize 2 truein \epsfbox{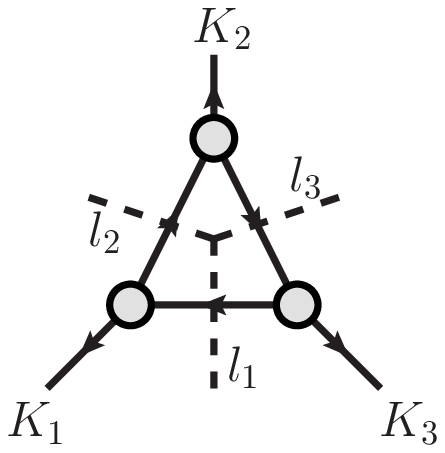}}
\caption[a]{Double and triple cuts contributing to bubble coefficients.}
\label{Fig:BubbCuts}
\end{figure}
%%%%%%%%%%%%%%%%%%%%%%%%%%%%%%%%

Two-particle cuts are contaminated by both boxes and triangles, so again an extraction procedure is necessary.  Furthermore, triple cuts that share two of their cuts with the double cut contribute to tensor triangle integrals that reduce to scalar bubbles, so we must take them into account for the full bubble coefficient (Fig. \ref{Fig:BubbCuts}).  The two-particle cut momentum solution has two free parameters $t$ and $y$,
\begin{align}
\bar{l}_1^{\mu}=yK_1^{\flat\mu}+\frac{S_1(1-y)}{\bar{\gamma}}\chi^{\mu}+\frac{t}{2}\langle K_1^{\flat -}|\gamma^{\mu}|\chi^-\rangle+\frac{y(1-y)S_1-\mu^2}{2t\bar{\gamma}}\langle\chi^-|\gamma^{\mu}|K_1^{\flat -}\rangle,
\label{eq:DoubCut}
\end{align}
where
\begin{align}
K_1^{\flat\mu}=K_1^{\mu}-\frac{S_1}{\bar{\gamma}}\chi^{\mu}, \hspace{1cm} \bar{\gamma}=2(K_1\cdot\chi),
\end{align}
and $\chi$ is some arbitrary massless vector.  We calculate the cut integrand $A_1A_2$ by taking a product of the two on-shell tree amplitudes, which will make up part of our coefficient after an extraction procedure.

For the triangle contribution to the bubble coefficient, we fix the parameter $y$ to put another propagator on shell,
\begin{align}
y_{\pm}=\frac{B_1\pm\sqrt{B_1^2+4B_0B_2}}{2B_2},
\label{eq:ySolns}
\end{align}
where
\begin{align}
B_2&=S_1\langle \chi^-|\s{K}_3|K_1^{\flat -}\rangle, \notag \\
B_1&=\bar\gamma t\langle K_1^{\flat -}|\s{K}_3|K_1^{\flat -}\rangle-S_1t\langle\chi^-|\s{K}_3|\chi^-\rangle+S_1\langle\chi^-|\s{K}_3|K_1^{\flat -}\rangle, \notag \\
B_0&=\bar\gamma t^2\langle K_1^{\flat -}|\s{K}_3|\chi^-\rangle-\mu^2\langle\chi^-|\s{K}_3|K_1^{\flat -}\rangle+\bar\gamma t S_3+t S_1\langle\chi^-|\s{K}_3|\chi^-\rangle.
\end{align}
We then calculate the triple-cut integrand $A_1A_2A_3$ for all triple cuts that share two cuts with the original double cut.  The bubble coefficients are then given by
\begin{align}
C_2^{[0]}=&-i[\mathrm{Inf}_t[\mathrm{Inf}_yA_1A_2](y)](t)|_{\mu^2\rightarrow 0,t\rightarrow 0, y^m\rightarrow Y_m}-\frac{1}{2}\sum_{C_{\mathrm{tri}}}\sum_{\sigma_y}[\mathrm{Inf}_tA_1A_2A_3](t)|_{\mu^2\rightarrow 0,t^j\rightarrow T_j}, \notag \\
C_2^{[2]}=&-i[\mathrm{Inf}_{\mu^2}[\mathrm{Inf}_t[\mathrm{Inf}_yA_1A_2](y)](t)](\mu^2)|_{\mu^2,t\rightarrow 0, y^m\rightarrow Y_m} \notag \\
&-\frac{1}{2}\sum_{C_{\mathrm{tri}}}\sum_{\sigma_y}[\mathrm{Inf}_{\mu^2}[\mathrm{Inf}_tA_1A_2A_3](t)](\mu^2)|_{\mu^2,t^j\rightarrow T_j}.
\label{eq:BubbCoeff}
\end{align}
In the series expansion $[\mathrm{Inf}_yA_1A_2](y)=\sum_{m=0}^kf_my^m$, we make the replacements $y^m\rightarrow Y_m$ where
\begin{align}
Y_0=1, \hspace{4mm} Y_1=\frac{1}{2}, \hspace{4mm} Y_2=\frac{1}{3}\left(1-\frac{\mu^2}{S_1}\right), \hspace{4mm} Y_3=\frac{1}{4}\left(1-2\frac{\mu^2}{S_1}\right), \hspace{4mm} Y_4=\frac{1}{5}\left(1-3\frac{\mu^2}{S_1}+\frac{\mu^4}{S_1^2}\right).
\end{align}
These double-cut terms are also expanded in $t$ and restricted to the $t^0$ term.  In $C_2^{[0]}$, $\mu^2$ is set to zero, while in $C_2^{[2]}$ there is an expansion in $\mu^2$, and $C_2^{[2]}$ is restricted to the order $\mu^2$ coefficient.  The triple-cut terms have a sum over the two solutions for each triple cut and a sum over all possible triple cuts that share two cuts with the double cut.  In the expansion in $t$, we replace $t^j\rightarrow T_j$ where
\begin{align}
T_0&=0, \notag \\
T_1&=-\frac{S_1\langle\chi^-|\s{K}_3|K_1^{\flat -}\rangle}{2\bar\gamma\Delta}, \notag \\
T_2&=-\frac{3S_1\langle\chi^-|\s{K}_3|K_1^{\flat -}\rangle^2}{8\bar{\gamma}^2\Delta^2}(S_1S_3+K_1\cdot K_3S_1), \notag \\
T_3&=-\frac{\langle\chi^-|\s{K}_3|K_1^{\flat -}\rangle^3}{48\bar{\gamma}^3\Delta^3}\left(15 S_1^3S_3^2+30 K_1\cdot K_3 S_1^3S_3+11(K_1\cdot K_3)^2S_1^3+4 S_1^4 S_3+16\mu^2 S_1^2\Delta\right), \label{eq:T}
\end{align}
and
\begin{align}
\Delta=(K_1\cdot K_3)^2-S_1S_3.
\end{align}
In a renormalizable gauge theory (such as QCD), double cuts can have terms up to order $y^4$ while triple cuts can have up to $t^3$, so all necessary integrals are evaluated above.\footnote{As it turns out, $y^3$ and $y^4$ terms come with factors of $1/t$ and $1/t^2$, respectively, so such terms drop out when the coefficient is restricted to the $t^0$ term, but we include the evaluations above so as not to break the procedural nature of the method.}  Again see Appendix \ref{ExtractAppend} for a more detailed explanation.

\section{COMPUTATION OF FOUR- AND FIVE-POINT GLUON AMPLITUDES} \label{4pointglue}

With the formalism established, we turn to sample calculations of one-loop amplitudes in QCD, starting with $A_4^{(1)}(1^-,2^+,3^+,4^+)$ for a scalar loop, gluon loop and fermion loop.  We use generalized unitarity to glue together tree amplitudes written in terms of six-dimensional spinors.  In keeping with Cheung and O'Connell \cite{COC6D}, these six-dimensional spinors are written in terms of four-dimensional ones based on the loop-momentum solutions of section \ref{IntExtract}.  The cuts then fall into a form that allows us to use our coefficient extraction techniques.  Our methodology for the cuts follows that of ref.~\cite{Bern6D}.

As mentioned earlier, internal particles are kept in six dimensions while external particles live in the four-dimensional subspace.  This leads to an additional complication: increasing the number of spacetime dimensions also increases the number of spin eigenstates.  We therefore do not expect the six-dimensional coefficients to exactly match the four-dimensional ones, and in fact we must perform a state-sum reduction to bring our result back to four dimensions.

We find that the scalar-loop diagram does not require adjustment, nor does the fermion loop.  For the gluon-loop diagram, however, we must subtract two factors of the scalar-loop diagram.   We motivate these procedures by tracking how the dimension makes its way into the integral coefficients and demonstrate their effectiveness in the one-loop amplitude $A_4^{(1)}(1^-,2^+,3^+,4^+)$.  We then include some additional results, but with fewer computational details.

\subsection{One-loop four-point cut conditions in $D=6$}

%%%%%%%%% FIGURE %%%%%%%%%%%%%%%
\begin{figure}[ht]
\centering
\subfloat[][]{\epsfxsize 1.7 truein \epsfbox{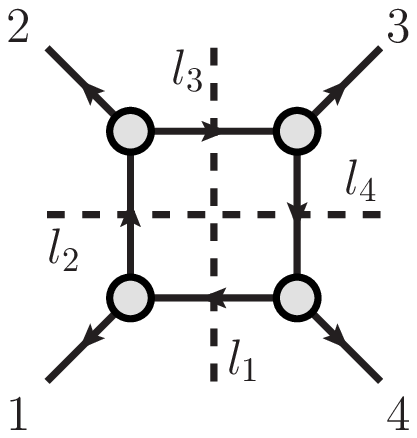}
\label{Fig:4pointCutsBox}}
\hspace{1.1cm}
\subfloat[][]{\epsfxsize 1.7 truein \epsfbox{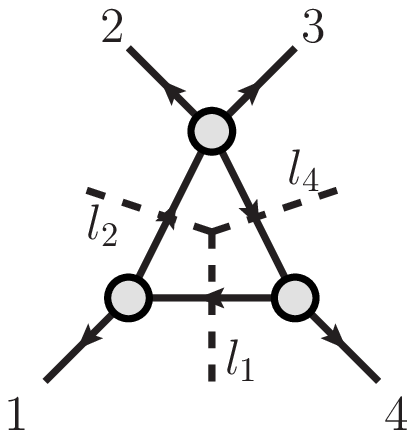}
\label{Fig:4pointCutsTri}}
\hspace{1.1cm}
\subfloat[][]{\epsfxsize 1.07 truein \epsfbox{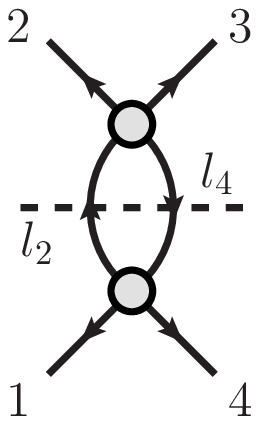}
\label{Fig:4pointCutsBubb}}
\caption[a]{Cuts for the four-point amplitude.  We show triple and double cuts in the $s_{23}$ channel; in general we must also evaluate cuts in the $s_{12}$ channel.}
\label{Fig:4pointCuts}
\end{figure}
%%%%%%%%%%%%%%%%%%%%%%%%%%%%%%%%

To compute the amplitudes using six-dimensional helicity, we use momentum solutions from section \ref{IntExtract} that fit with the coefficient extraction techniques.  There we interpreted the $(-2\epsilon)$-dimensional component $\mu$ as a mass and gave the four-dimensional components of the loop momenta the on-shell cut conditions, $\bar{l}_i^2=\mu^2$.  In section \ref{SixD} we also expressed the extra-dimensional components of six-dimensional momenta in terms of mass parameters $m$ and $\tilde{m}$ given in eq.~\eqref{eq:mmTilde},
\begin{align}
l_i^4=\frac{-m+\tilde{m}}{2i}, \hspace{3cm} l_i^5=\frac{m+\tilde{m}}{2},
\end{align}
giving the on-shell cut conditions, $\bar{l}_i^2=m\tilde{m}$.  This enabled us to embed four-dimensional spinors in the six-dimensional ones.  Making the association $\mu^2\leftrightarrow m\tilde{m}$, we see that we have a bi-spinor momentum representation that simultaneously allows us to use six-dimensional helicity and the coefficient extraction techniques.

Using eq.~\eqref{eq:BoxSol}, the momentum solution for the quadruple cut (Fig.~\ref{Fig:4pointCutsBox}) is
\begin{equation}
(\bar{l}^{\pm})_1^{\mu}=\frac{1}{2}\left(c_1^{\pm}\langle 4^-|\gamma^{\mu}|1^-\rangle-\frac{1}{c_1^{\pm}}\frac{m\tilde{m}}{s_{14}}\langle 1^-|\gamma^{\mu}|4^-\rangle\right),
\label{eq:4cutSol}
\end{equation}
where
\begin{equation}
c_1^{\pm}=\frac{\langle 1\,2\rangle}{2\langle 4\,2\rangle}\left(1\pm\sqrt{1+\frac{4m\tilde{m}s_{13}}{s_{12}s_{23}}}\right).
\end{equation}
The other loop momenta can be found through momentum conservation: $l_2=l_1-p_1$, $l_3=l_1-p_1-p_2$ and $l_4=l_1+p_4$, or by relabellings in eq.~\eqref{eq:4cutSol}.  It is easy to confirm that all are on shell.  There are in total two solutions to the quadruple cut, and we must average over them at the end of our calculation.

For the triple cut in Fig.~\ref{Fig:4pointCutsTri}, we use eq.~\eqref{eq:TriCut},
\begin{equation}
\bar{l}_1^{\mu}=\frac{1}{2}\left(t\langle 4^-|\gamma^{\mu}|1^-\rangle-\frac{1}{t}\frac{m\tilde{m}}{s_{14}}\langle 1^-|\gamma^{\mu}|4^-\rangle\right),
\end{equation}
which solves the on-shell conditions $\bar{l}_1^2=\bar{l}_2^2=\bar{l}_4^2=m\tilde{m}$.  As before $t$ parametrizes the remaining degree of freedom.  Notice that requiring the remaining internal propagator be on shell ($\bar{l}_3^2=m\tilde{m}$) recovers the box solution, i.e. $t=c_1^{\pm}$.  We will also need the conjugate solution for the cut in Fig.~\ref{Fig:4pointCutsTri},
\begin{equation}
\bar{l}_1^{*\mu}=\frac{1}{2}\left(t\langle 1^-|\gamma^{\mu}|4^-\rangle-\frac{1}{t}\frac{m\tilde{m}}{s_{14}}\langle 4^-|\gamma^{\mu}|1^-\rangle\right).
\end{equation}
Solutions for the other triple cuts can be constructed in the same manner.

Finally for the double cut in Fig.~\ref{Fig:4pointCutsBubb}, we have two free parameters for $\bar{l}_2^2=\bar{l}_4^2=m\tilde{m}$.  Choosing $K_1=p_1+p_4$ and $\chi=p_1$ in eq.~\eqref{eq:DoubCut}, we have $K_1^{\flat}=p_4$ and the momentum solution is
\begin{equation}
\bar{l}_4^{\mu}=\frac{1}{2}\left(y\langle 4^-|\gamma^{\mu}|4^-\rangle+(1-y)\langle 1^-|\gamma^{\mu}|1^-\rangle+t\langle 4^-|\gamma^{\mu}|1^-\rangle+\frac{y(1-y)-\frac{m\tilde{m}}{s_{14}}}{t}\langle 1^-|\gamma^{\mu}|4^-\rangle\right).
\end{equation}
Setting $y=1$ puts $l_1$ on shell and, in fact, gives us the momentum solution to the triple cut in Fig.~\ref{Fig:4pointCutsTri}.

\subsection{One-loop four-point solution for six-dimensional spinors}

With the momentum solutions above, it is straightforward to plug them into eq.~\eqref{eq:6Dspinors} and obtain six-dimensional spinors in terms of four-dimensional ones.  We work explicitly with the double cut; spinors for the triple and quadruple cuts can be found by setting $y=1$ and $t=c_1^{\pm}$ where appropriate.  We start by expressing the solution in two-component notation,
\begin{align}
\bar{l}_4&=y\,\lambda_4\tilde\lambda_4+(1-y)\,\lambda_1\tilde\lambda_1+t\,\lambda_4\tilde\lambda_1+\frac{y(1-y)-m\tilde{m}/s_{14}}{t}\,\lambda_1\tilde\lambda_4, \notag \\
\bar{l}_2&=-(1-y)\,\lambda_4\tilde\lambda_4-y\,\lambda_1\tilde\lambda_1+t\,\lambda_4\tilde\lambda_1+\frac{y(1-y)-m\tilde{m}/s_{14}}{t}\,\lambda_1\tilde\lambda_4\,.
\end{align}
Making the associations,
\begin{align}
\mu=\lambda_1, \hspace{8mm} &\tilde{\mu}=\frac{\tilde{\lambda}_4}{t}, \notag \\
 \lambda_{\bar{l}_4}=\frac{1-y}{t}\lambda_1+\lambda_4, \hspace{4mm}  \tilde{\lambda}_{\bar{l}_4}=t\tilde{\lambda}_1+y\tilde{\lambda}_4, \hspace{4mm} \lambda&{}_{\bar{l}_2}=-\frac{y}{t}\lambda_1+\lambda_4, \hspace{4mm}  \tilde{\lambda}_{\bar{l}_2}=t\tilde{\lambda}_1-(1-y)\tilde{\lambda}_4, \notag \\
\kappa_{14}=\frac{m}{\langle 4\,1\rangle}, \hspace{8mm} \tilde{\kappa}_{14}=\frac{\tilde{m}}{[4\,1]}, \hspace{8mm} &\kappa_{14}'=\frac{\tilde{m}}{\langle 4\,1\rangle}, \hspace{8mm} \tilde{\kappa}_{14}'=\frac{m}{[4\,1]},\hspace{2mm}
\end{align}
a comparison with eq.~\eqref{eq:6Dspinors} gives us the spinors,
\begin{align}
(\lambda_{l_4})^A_{\,\,\,a}&=\left(\begin{array}{cc}
-\kappa_{14}\lambda_1 & (1-y)\lambda_1/t+\lambda_4 \\
t\tilde{\lambda}_1+y\tilde{\lambda}_4 & \tilde{\kappa}_{14}\tilde{\lambda}_4/t \end{array} \right), \notag \\
(\tilde{\lambda}_{l_4})_{A\dot{a}}&=\left(\begin{array}{cc}
\kappa_{14}'\lambda_1 & (1-y)\lambda_1/t+\lambda_4 \\
-t\tilde{\lambda}_1-y\tilde{\lambda}_4 & \tilde{\kappa}_{14}'\tilde{\lambda}_4/t \end{array} \right), \notag \\
(\lambda_{l_2})^A_{\,\,\,a}&=\left(\begin{array}{cc}
-\kappa_{14}\lambda_1 & -y\lambda_1/t+\lambda_4 \\
t\tilde{\lambda}_1-(1-y)\tilde{\lambda}_4 & \tilde{\kappa}_{14}\tilde{\lambda}_4/t \end{array} \right), \notag \\
(\tilde{\lambda}_{l_2})_{A\dot{a}}&=\left(\begin{array}{cc}
\kappa_{14}'\lambda_1 & -y\lambda_1/t+\lambda_4 \\
-t\tilde{\lambda}_1+(1-y)\tilde{\lambda}_4 & \tilde{\kappa}_{14}'\tilde{\lambda}_4/t \end{array} \right).
\end{align}
With these it is simple to work out the spinor products required for a cut.  It should be noted that because we use an all outgoing convention, incoming momenta are labeled $-p$.  For purely gluonic amplitudes, it is sufficient to add a factor of $i$ to the negative momentum spinors to handle these cases.\footnote{Cut fermions do not have this factor of $i$.  We will generally define the spinor products with the factor of $i$ in the cases of negative momenta and manually remove it when we cut fermions (see ref.~\cite{BernLastOfTheFinite} for more detail).}  As an example, the spinor products relevant to the gluon-loop contribution to $A_4^{(1)}(1^-,2^+,3^+,4^+)$ are
\begin{align}
\langle(-l_2)_1,2_1,3_1,(l_4)_1\rangle&=\mathrm{det}\left(\begin{array}{cccc}
-i\kappa_{14}\lambda_1 & 0 & 0 & -\kappa_{14}\lambda_1 \\
it\tilde{\lambda}_1-i(1-y)\tilde{\lambda}_4 & \tilde{\lambda}_2 & \tilde{\lambda}_3 & t\tilde{\lambda}_1+y\tilde{\lambda}_4 \end{array}\right) \notag \\
&= 0, \notag \\
\langle(-l_2)_1,2_1,3_1,(l_4)_2\rangle&=\mathrm{det}\left(\begin{array}{cccc}
-i\kappa_{14}\lambda_1 & 0 & 0 & (1-y)\lambda_1/t+\lambda_4 \\
it\tilde{\lambda}_1-i(1-y)\tilde{\lambda}_4 & \tilde{\lambda}_2 & \tilde{\lambda}_3 & \tilde{\kappa}_{14}\tilde{\lambda}_4/t \end{array}\right) \notag \\
&= i\kappa_{14}\langle 1\,4\rangle[2\,3], \notag \\
\langle(-l_2)_2,2_1,3_1,(l_4)_1\rangle&=\mathrm{det}\left(\begin{array}{cccc}
-iy\lambda_1/t+i\lambda_4 & 0 & 0 & -\kappa_{14}\lambda_1 \\
i\tilde{\kappa}_{14}\tilde{\lambda}_4/t & \tilde{\lambda}_2 & \tilde{\lambda}_3 & t\tilde{\lambda}_1+y\tilde{\lambda}_4 \end{array}\right) \notag \\
&= -i\kappa_{14}\langle 1\,4\rangle[2\,3], \notag \\
\langle(-l_2)_2,2_1,3_1,(l_4)_2\rangle&=\mathrm{det}\left(\begin{array}{cccc}
-iy\lambda_1/t+i\lambda_4 & 0 & 0 & (1-y)\lambda_1/t+\lambda_4 \\
i\tilde{\kappa}_{14}\tilde{\lambda}_4/t & \tilde{\lambda}_2 & \tilde{\lambda}_3 & \tilde{\kappa}_{14}\tilde{\lambda}_4/t \end{array}\right) \notag \\
&= i\frac{1}{t}\langle 1\,4\rangle[2\,3], \notag
\end{align}
\begin{align}
\begin{array}{ll}
\langle(-l_4)_1,4_1,1_2,(l_2)_1\rangle= 0, &
\hspace{6mm}\langle(-l_4)_1,4_1,1_2,(l_2)_2\rangle= it \langle1\,4\rangle[4\,1], \\
\langle(-l_4)_2,4_1,1_2,(l_2)_1\rangle= -it \langle1\,4\rangle[4\,1], &
\hspace{6mm}\langle(-l_4)_2,4_1,1_2,(l_2)_2\rangle= 0.
\end{array}
\label{eq:spinProdsG1}
\end{align}
For the bracket spinor products, we make the replacement $\kappa_{14}\rightarrow\kappa'_{14}$ and change the sign where appropriate,
\begin{align}\begin{array}{ll}
\mathrm{[}(-l_2)_{\dot{1}},2_{\dot{1}},3_{\dot{1}},(l_4)_{\dot{1}}\mathrm{]}=0, &
\hspace{6mm}\mathrm{[}(-l_2)_{\dot{1}},2_{\dot{1}},3_{\dot{1}},(l_4)_{\dot{2}}\mathrm{]}=-i\kappa_{14}'\langle 1\,4\rangle[2\,3], \\
\mathrm{[}(-l_2)_{\dot{2}},2_{\dot{1}},3_{\dot{1}},(l_4)_{\dot{1}}\mathrm{]}=i\kappa_{14}'\langle 1\,4\rangle[2\,3], &
\hspace{6mm}\mathrm{[}(-l_2)_{\dot{2}},2_{\dot{1}},3_{\dot{1}},(l_4)_{\dot{2}}\mathrm{]}=i\frac{1}{t}\langle 1\,4\rangle[2\,3], \\
 \\
\mathrm{[}(-l_4)_{\dot{1}},4_{\dot{1}},1_{\dot{2}},(l_2)_{\dot{1}}\mathrm{]}=0, &
\hspace{6mm}\mathrm{[}(-l_4)_{\dot{1}},4_{\dot{1}},1_{\dot{2}},(l_2)_{\dot{2}}\mathrm{]}=it \langle1\,4\rangle[4\,1], \\
\mathrm{[}(-l_4)_{\dot{2}},4_{\dot{1}},1_{\dot{2}},(l_2)_{\dot{1}}\mathrm{]}=-it \langle1\,4\rangle[4\,1], &
\hspace{6mm}\mathrm{[}(-l_4)_{\dot{2}},4_{\dot{1}},1_{\dot{2}},(l_2)_{\dot{2}}\mathrm{]}=0.
\end{array}
\label{eq:spinProdsG2}
\end{align}

\subsection{Scalar loop} \label{ScalarCalc}

We are now ready to find the full amplitude for the scalar-loop contribution to $A_4^{(1)}(1^-,2^+,3^+,4^+)$.  
As mentioned before, increasing the dimension of spacetime for the virtual particles in the loop increases the number of spin eigenstates, potentially causing a difference between coefficients in six-dimensions and those in the FDH scheme.  However, scalar particles do not have spin, so we expect the coefficients here to match.  This is indeed what we find.

This fact can also be understood by examining the vertex factors in a one-loop Feynman diagram.  The two-scalar gluon vertex is
\begin{equation}
\frac{i}{\sqrt{2}}(p-q)_\mu,
\end{equation}
where $p$ and $q$ are the momenta of the scalars.  These momenta are contracted with the polarization vectors of the gluons, which, since they are external, are in four dimensions.  Any extra-dimensional components are annihilated in these contractions, so whether we are in six or $4-2\epsilon$ dimensions, we get the same contribution from a given vertex.  This argument can be extended to the two-scalar two-gluon vertex to handle all Feynman diagrams and show that there are no terms in the amplitude that are proportional to the dimension of the internal particles.

%%%%%%%%% FIGURE %%%%%%%%%%%%%%%
\begin{figure}[ht]
\centerline{\epsfxsize 1.7 truein \epsfbox{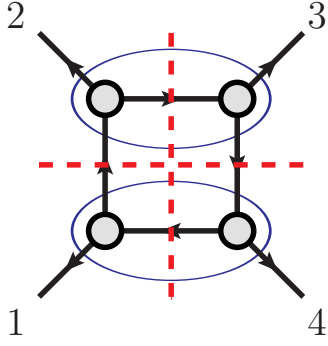}}
\caption[a]{The four-point quadruple cut.  Two pairs of three-point amplitudes are grouped together (indicated by blue ovals) to form two easier-to-use four-point tree amplitudes.  The cut propagators of the four-point tree amplitudes are canceled by multiplying by inverse propagators prior to imposing the cut conditions.}
\label{Fig:QuadCutTrees}
\end{figure}
%%%%%%%%%%%%%%%%%%%%%%%%%%%%%%%%

Moving on to the computation, we begin with the quadruple cut (Fig.~\ref{Fig:4pointCutsBox}).  The usual procedure is to sew four three-point amplitudes together, but working with three-point amplitudes in six dimensions can be complicated.  It is, in fact, easier to multiply simpler four-point tree amplitudes by inverse propagators making them equivalent to the sum of products of two three-point amplitudes (Fig.~\ref{Fig:QuadCutTrees}), given the cut conditions.  The quadruple cut is then
\begin{align}
C_{1234}&=\sum_{\mathrm{states}}A_3(-l_1^s,1^-,l_2^s)A_3(-l_2^s,2^+,l_3^s)A_3(-l_3^s,3^+,l_4^s)A_3(-l_4^s,4^+,l_1^s) \notag \\
&=\sum_{\mathrm{states}}(-i)^2(l_2-p_2)^2(l_2+p_1)^2A_4(-l_2^s,2^+,3^+,l_4^s)A_4(-l_4^s,4^+,1^-,l_2^s).
\end{align}
The two-real-scalar two-gluon tree amplitude was given in eq.~\eqref{eq:2g2sTree},
\begin{equation}
A_4^{(0)}(1_{a\dot{a}}^g,2_{b\dot{b}}^g,3^s,4^s)=-\frac{1}{4}\frac{i}{s_{12}s_{23}}\langle1_a2_b3_e3^e\rangle[1_{\dot{a}}2_{\dot{b}}3_{\dot{e}}3^{\dot{e}}],
\end{equation}
so the quadruple cut becomes
\begin{equation}
C_{1234}=\frac{\langle(-l_2)_a,2_1,3_1,(-l_2)^a\rangle[(-l_2)_{\dot{a}},2_{\dot{1}},3_{\dot{1}},(-l_2)^{\dot{a}}]\langle(-l_4)_b,4_1,1_2,(-l_4)^b\rangle[(-l_4)_{\dot{b}},4_{\dot{1}},1_{\dot{2}},(-l_4)^{\dot{b}}]}{16s_{23}^2}.
\end{equation}
The relevant spinor products are
\begin{align}
\langle(-l_2)_1,2_1,3_1,(-l_2)^1\rangle&=\langle(-l_2)_2,2_1,3_1,(-l_2)^2\rangle=-\kappa_{14}\langle 1\,4\rangle[2\,3], \notag \\
\langle(-l_4)_1,4_1,1_2,(-l_4)^1\rangle&=\langle(-l_4)_2,4_1,1_2,(-l_4)^2\rangle=-t\langle1\,4\rangle[4\,1], \notag \\
\mathrm{[}(-l_2)_{\dot{1}},2_{\dot{1}},3_{\dot{1}},(-l_2)^{\dot{1}}\mathrm{]}&=\mathrm{[}(-l_2)_{\dot{2}},2_{\dot{1}},3_{\dot{1}},(-l_2)^{\dot{2}}\mathrm{]}=\kappa_{14}'\langle 1\,4\rangle[2\,3], \notag \\
\mathrm{[}(-l_4)_{\dot{1}},4_{\dot{1}},1_{\dot{2}},(-l_4)^{\dot{1}}\mathrm{]}&=\mathrm{[}(-l_4)_{\dot{2}},4_{\dot{1}},1_{\dot{2}},(-l_4)^{\dot{2}}\mathrm{]}=-t\langle1\,4\rangle[4\,1].
\label{eq:mpppSpinProds}
\end{align}
Notice that we have not explicitly plugged in $t=c_1^{\pm}$.  We recommend that this be done only at the end of the calculation as the square-roots tend to fall out when we average over solutions.  Multiplying the products out, we arrive at
\begin{equation}
C_{1234}=-\frac{1}{s_{23}^2}\kappa_{14}\kappa_{14}'\langle 1\,4\rangle^4[4\,1]^2[2\,3]^2t^2.
\label{eq:mpppQuadCut}
\end{equation}
Now we take our sign convention into account using eq.~\eqref{eq:BoxExtract} and average over solutions that take $l_3$ on shell, i.e. $t=c_1^{\pm}$, arriving at
\begin{align}
C_4&=\frac{i}{2}\sum_{\sigma}C_{1234} \notag \\
&=i\frac{m\tilde{m}[2\,4]^2}{[1\,2]\langle 2\,3\rangle\langle 3\,4\rangle[4\,1]}\frac{s_{12}s_{23}}{s_{13}}\left(\frac{s_{12}s_{23}}{2s_{13}}+m\tilde{m}\right).
\end{align}
Associating $m\tilde{m}\rightarrow\mu^2$, we see that the cut is nicely in the form of a polynomial in $\mu^2$.  We identify the coefficients,
\begin{align}
C_4^{[0]}&=0, \notag \\
C_4^{[4]}&=i\frac{[2\,4]^2}{[1\,2]\langle 2\,3\rangle\langle 3\,4\rangle[4\,1]}\frac{s_{12}s_{23}}{s_{13}}.
\end{align}
These are in agreement with the previously known result \cite{BernMorgan}.

Next we examine the triple cut in Fig.~\ref{Fig:4pointCutsTri} to obtain triangle coefficients.  Conveniently we have most of the cut for one of our momentum solutions from the box computation.  We can use our quadruple cut in terms of the free parameter $t$ and multiply by the $l_3$ propagator.  This is the same as multiplying the three tree amplitudes in Fig.~\ref{Fig:4pointCutsTri} together,
\begin{align}
C_{124}&=A_3(-l_1^s,1^-,l_2^s)A_3(-l_2^s,2^+,3^+,l_4^s)A_3(-l_4^s,4^+,l_1^s) \notag \\
&=C_{1234}\frac{i}{l_3^2} \notag \\
&=-i\frac{1}{s_{23}^2}\kappa_{14}\kappa_{14}'\langle 1\,4\rangle^4[4\,1]^2[2\,3]^2t^2\left(t\langle 4\,2\rangle[1\,2]+s_{12}-\frac{m\tilde{m}}{ts_{14}}\langle 1\,2\rangle[4\,2]\right)^{-1}.
\end{align}
We also need the cut for the conjugate-momentum solution,
\begin{equation}
\bar{l}_1^{*\mu}=\frac{1}{2}\left(t\langle 1^-|\gamma^{\mu}|4^-\rangle-\frac{1}{t}\frac{m\tilde{m}}{s_{14}}\langle 4^-|\gamma^{\mu}|1^-\rangle\right).
\end{equation}
Instead of defining new spinors and spinor products, we recognize that the conjugate solution is obtained from the original solution through the substitution $t\rightarrow -\frac{m\tilde{m}}{ts_{14}}$.  We can then make this same substitution in the cut,
\begin{align}
C_{124}^*&=C_{124}|_{t\rightarrow -m\tilde{m}/ts_{14}} \notag \\
&=-i\frac{1}{s_{23}^2}\kappa_{14}\kappa_{14}'\langle 1\,4\rangle^4[4\,1]^2[2\,3]^2\frac{(m\tilde{m})^2}{t^2s_{14}^2}\left(t\langle 1\,2\rangle[4\,2]+s_{12}-\frac{m\tilde{m}}{ts_{14}}\langle 4\,2\rangle[1\,2]\right)^{-1}.
\end{align}
Finally we perform an expansion around $t\rightarrow\infty$ as lined out in section \ref{Triangles}, keeping only the order $t^0$ term,
\begin{align}
C_3&=-\frac{1}{2}\left\{[\mathrm{Inf}_t(C_{124})]|_{t\rightarrow0}+[\mathrm{Inf}_t(C_{124}^*)]|_{t\rightarrow0}\right\} \notag \\
&=-\frac{1}{2}\left\{-i\frac{m\tilde{m}[2\,4]^2}{[1\,2]\langle 2\,3\rangle\langle 3\,4\rangle[4\,1]}\frac{s_{12}s_{23}^2}{s_{13}^2}+0\right\},
\end{align}
which gives us the triangle coefficients,
\begin{align}
C_{3;23}^{[0]}&=0, \notag \\
C_{3;23}^{[2]}&=\frac{i}{2}\frac{[2\,4]^2}{[1\,2]\langle 2\,3\rangle\langle 3\,4\rangle[4\,1]}\frac{s_{12}s_{23}^2}{s_{13}^2},
\end{align}
where ``23'' indicates that legs 2 and 3 have been ``pinched'' to form the triangle.

We also need the triangle contributions from other triple-cut configurations.  To pinch legs 3 and 4, we use the momentum solution,
\begin{align}
\bar{l}_2^{\mu}=\frac{1}{2}\left(t\langle 1^-|\gamma^{\mu}|2^-\rangle-\frac{1}{t}\frac{m\tilde{m}}{s_{12}}\langle 2^-|\gamma^{\mu}|1^-\rangle\right),
\end{align}
along with its conjugate solution.  We define spinors and spinor products (many of which can be found through simple relabellings of our previous results) using this momentum solution and calculate cuts in a similar manner.  The remaining triangle coefficients are then
\begin{align}
C_{3;12}^{[0]}=C_{3;34}^{[0]}=C_{3;41}^{[0]}=0, \notag
\end{align}
\begin{align}
C_{3;12}^{[2]}&=\frac{i}{2}\frac{[2\,4]^2}{[1\,2]\langle 2\,3\rangle\langle 3\,4\rangle[4\,1]}\left(\frac{s_{23}^3}{s_{13}^2}+2\frac{s_{23}^2}{s_{13}}-s_{23}\right), \notag \\
C_{3;34}^{[2]}&=\frac{i}{2}\frac{[2\,4]^2}{[1\,2]\langle 2\,3\rangle\langle 3\,4\rangle[4\,1]}\frac{s_{12}^2s_{23}}{s_{13}^2}, \notag \\
C_{3;41}^{[2]}&=\frac{i}{2}\frac{[2\,4]^2}{[1\,2]\langle 2\,3\rangle\langle 3\,4\rangle[4\,1]}\left(\frac{s_{12}^3}{s_{13}^2}+2\frac{s_{12}^2}{s_{13}}-s_{12}\right).
\end{align}

Triangle integrals are dependent only on their kinematic invariant masses, so certain pairs of integrals corresponding to the above coefficients are in fact equal.  For the one-mass triangle with invariant mass square $s_{12}=(p_1+p_2)^2=s_{34}$,
\begin{align}
I_{3;s_{12}}^{4-2\epsilon}[f(\mu^2)]=I_{3;12}^{4-2\epsilon}[f(\mu^2)]=I_{3;34}^{4-2\epsilon}[f(\mu^2)].
\end{align}
It then makes sense to write only one coefficient $C_{3;s_{12}}=C_{3;12}+C_{3;34}$.  In this basis our results are
\begin{align}
C_{3;s_{12}}^{[2]}=i\frac{[2\,4]^2}{[1\,2]\langle 2\,3\rangle\langle 3\,4\rangle[4\,1]}\frac{s_{23}^2}{s_{13}^2}(s_{13}-s_{12}), \notag \\
C_{3;s_{23}}^{[2]}=i\frac{[2\,4]^2}{[1\,2]\langle 2\,3\rangle\langle 3\,4\rangle[4\,1]}\frac{s_{12}^2}{s_{13}^2}(s_{13}-s_{23}).
\end{align}

Finally we calculate the bubble integral coefficients using double and triple cuts.  For this special case, dependence on $y$ in the spinor products falls out and the products for these cuts are also given by eq.~\eqref{eq:mpppSpinProds} (this is of course not generally true).  We can then take the quadruple cut \eqref{eq:mpppQuadCut} and multiply by propagators for $l_1$ and $l_3$ to obtain the two-particle cut,
\begin{align}
C_{24}=&A_4(-l_2^s,2^+,3^+,l_4^s)A_4(-l_4^s,4^+,1^-,l_2^s) \notag \\
=&C_{1234}\frac{i}{l_1^2}\frac{i}{l_3^2} \notag \\
=&\frac{1}{s_{23}^2}\kappa_{14}\kappa_{14}'\langle 1\,4\rangle^4[4\,1]^2[2\,3]^2t^2(-(1-y)s_{14})^{-1} \notag \\
&\times\left(ys_{12}+(1-y)s_{13}+t\langle 4\,3\rangle[3\,1]+\left(\frac{y(1-y)}{t}-\frac{m\tilde{m}}{ts_{14}}\right)\langle 1\,3\rangle[3\,4]\right)^{-1}.
\end{align}
According to eq.~\eqref{eq:BubbCoeff}, the two-particle cut contribution to the bubble coefficient is
\begin{align}
-i[\mathrm{Inf}_t[\mathrm{Inf}_yA_1A_2](y)](t)|_{t\rightarrow 0,y^m\rightarrow Y_m}.
\end{align}
Expanding $C_{24}$ around infinity in $y$, we find
\begin{align}
\mathrm{Inf}_yC_{24}=0,
\end{align}
so the contribution in this case is $0$.

%%%%%%%%% FIGURE %%%%%%%%%%%%%%%
\begin{figure}[ht]
\centering
\subfloat[][]{\epsfxsize 1.7 truein \epsfbox{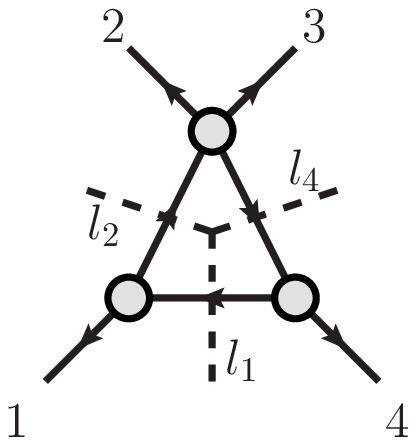}
\label{Fig:4pointBubbTriCutsBottom}}
\hspace{2cm}
\subfloat[][]{\epsfxsize 1.65 truein \epsfbox{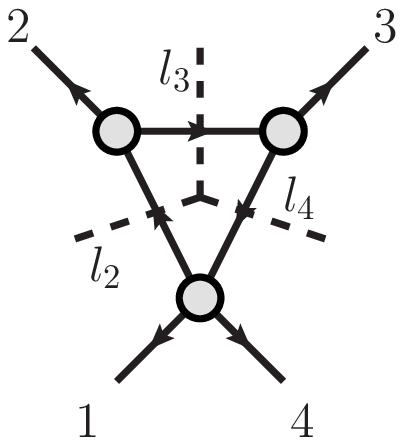}
\label{Fig:4pointBubbTriCutsTop}}
\caption[a]{Triple cuts contributing to the four-point amplitude bubble coefficient for the $s_{23}$-channel bubble integral.}
\label{Fig:4pointBubbTriCuts}
\end{figure}
%%%%%%%%%%%%%%%%%%%%%%%%%%%%%%%%

We also need to examine three-particle cuts that share two on-shell conditions with our double cut, as explained in section \ref{Bubbles}.  The two possible configurations are shown in Fig.~\ref{Fig:4pointBubbTriCuts}.  Fig.~\ref{Fig:4pointBubbTriCutsBottom} makes no contribution due to our convenient choice of momentum parametrization.  Any contribution must be proportional to $T_1$, $T_2$ and $T_3$ as given in eq.~\eqref{eq:T}.  Recall that for our momentum solution, $K_1^{\flat}=p_4$, and in Fig.~\ref{Fig:4pointBubbTriCutsBottom}, $K_3=p_4$.  We then find that $T_1=T_2=T_3=0$, and the contribution from this particular triple cut vanishes.

The only bubble contribution then comes from Fig.~\ref{Fig:4pointBubbTriCutsTop}, where $K_3=p_3$.  To make use of eq.~\eqref{eq:BubbCoeff}, we need to find the solutions to the parameter $y$ that take the third leg $l_3$ on shell.  These are given by eq.~\eqref{eq:ySolns},
\begin{align}
y_{\pm}=\frac{B_1\pm\sqrt{B_1^2+4B_0B_2}}{2B_2},
\end{align}
where
\begin{align}
B_2&=s_{23}\langle 1\,3\rangle[3\,4], \notag \\
B_1&=s_{23}t(s_{12}-s_{13})+s_{23}\langle 1\,3\rangle[3\,4], \notag \\
B_0&=s_{23}t^2\langle 4\,3\rangle[3\,1]-\mu^2\langle 1\,3\rangle[3\,4]+s_{13}s_{23}t.
\end{align}
Our cut is given by
\begin{align}
C_{234}&=A_4(-l_4^s,4^+,1^-,l_2^s)A_3(-l_2^s,2^+,l_3^s)A_3(-l_3^s,3^+,l_4^s) \notag \\
&=C_{1234}\frac{i}{l_1^2} \notag \\
&=-i\frac{1}{s_{14}^2}\kappa_{14}\kappa_{14}'\langle 1\,4\rangle^4[4\,1]^2[2\,3]^2t^2(-(1-y)s_{14})^{-1}.
\end{align}
Plugging in the solutions for $y$ and expanding in $t$, we find
\begin{align}
\frac{1}{2}\sum_{\sigma_y}[\mathrm{Inf}_tC_{234}](t)=\frac{i}{2}\frac{m\tilde{m}\langle 1\,3\rangle[2\,3][3\,4]}{\langle 2\,3\rangle}\frac{s_{13}-s_{12}}{s_{12}s_{13}}t+O(t^0).
\end{align}
Lastly we need to take $t\rightarrow T_1$, where $T_1$ is given by eq.~\eqref{eq:T},
\begin{align}
T_1=-2\frac{\langle 1\,3\rangle[3\,4]}{s_{23}^2}.
\end{align}
We then have the full coefficient expression,
\begin{align}
C_2&=-i[\mathrm{Inf}_t[\mathrm{Inf}_yA_1A_2](y)](t)|_{t\rightarrow 0, y^m\rightarrow Y_m}-\frac{1}{2}\sum_{C_{\mathrm{tri}}}\sum_{\sigma_y}[\mathrm{Inf}_tA_1A_2A_3](t)|_{t^j\rightarrow T_j} \notag \\
&=i\frac{m\tilde{m}[2\,4]^2}{[1\,2]\langle 2\,3\rangle\langle 3\,4\rangle[4\,1]}\frac{s_{12}}{s_{13}s_{23}}(s_{13}-s_{12}),
\end{align}
resulting in the coefficients,
\begin{align}
C_{2;23}^{[0]}&=0, \notag \\
C_{2;23}^{[2]}&=i\frac{[2\,4]^2}{[1\,2]\langle 2\,3\rangle\langle 3\,4\rangle[4\,1]}\frac{s_{12}}{s_{13}s_{23}}(s_{13}-s_{12}).
\end{align}
Through a similar computation, we find the coefficients for an invariant mass in the $s_{12}$ channel,
\begin{align}
C_{2;12}^{[0]}&=0, \notag \\
C_{2;12}^{[2]}&=i\frac{[2\,4]^2}{[1\,2]\langle 2\,3\rangle\langle 3\,4\rangle[4\,1]}\frac{s_{23}}{s_{12}s_{13}}(s_{13}-s_{23}).
\end{align}
All coefficients are in agreement with the previously known result \cite{BernMorgan}.

\subsection{Gluon loop} \label{GluonCalc}

Next we look at $A_4^{(1)}(1^-,2^+,3^+,4^+)$ with a gluon loop.  Unlike the scalar-loop amplitude, the gluon-loop amplitude requires a state-sum reduction to reduce from six dimensions to the FDH scheme.  Gluons in six dimensions have two additional polarization states compared to four dimensions.  The extra contribution of these states manifests itself in a term proportional to the dimension of spacetime, which is then the term that we must reduce.  We will find that we must subtract twice the contribution of the scalar-loop diagram to reduce from six dimensions to the FDH scheme.  The reasoning behind this can be found in ref.~\cite{FullOneLoopGiele}.  We summarize the argument before presenting the results of $A_4^{(1)}(1^-,2^+,3^+,4^+)$.

The dependence of a one-loop coefficient on the dimensionality of spacetime is at most linear.  Such terms arise from a closed loop of contracted metric tensors and/or gamma matrices from vertices and propagators making up the loop.  We can therefore separate coefficients into pieces dependent on $D$ and pieces independent of it,
\begin{align}
C^{(D)}=C_I+DC_D.
\label{eq:coeffProp}
\end{align}
As the authors of ref.~\cite{FullOneLoopGiele} point out, this can be used to find the coefficients in the FDH scheme by finding the coefficients in two dimensions $D>4$: $D_1$ and $D_2$ (doing the full calculation in $D=4$ of course loses the non-cut-constructible terms), then solving for $C_I$ and $C_D$,
\begin{align}
C_I&=\frac{D_2C^{(D_1)}-D_1C^{(D_2)}}{D_2-D_1}, \notag \\
C_D&=\frac{C^{(D_2)}-C^{(D_1)}}{D_2-D_1}.
\end{align}
The coefficient in the FDH scheme is then \cite{FullOneLoopGiele}
\begin{align}
C^{\mathrm{FDH}}=\frac{D_2C^{(D_2)}-D_1C^{(D_2)}}{D_2-D_1}+4\frac{C^{(D_2)}-C^{(D_1)}}{D_2-D_1}.
\end{align}

With this fact in mind, the authors of ref.~\cite{FullOneLoopGiele} study the gluon-loop amplitudes in five and six dimensions and show that they differ in a rather simple way.  Suppose we calculate in $D=5$ with five-dimensional loop momentum.  We will have three polarization states of the internal gluons, and we will obtain coefficients for five dimensions.  Now suppose we upgrade to $D=6$ but leave our loop momentum in five dimensions.  We view this as being the same loop momentum as in the $D=5$ case; the only difference is now we have added an additional polarization state of the gluon, which can be thought of as being along the sixth dimension of six-dimensional space.  

It should be noted that the restriction on the loop momentum does not prevent us from obtaining the correct $D=6$ coefficient; problems arise only when we take it down to four dimensions.  This can be understood through a fact mentioned in Appendix \ref{IntBasisApp}: the loop momentum components beyond four dimensions only enter coefficients in a  particular combination through contractions of the loop momentum with itself,
\begin{equation}
\tilde{l}^2=l^2-\bar{l}^2,
\end{equation}
where $\bar{l}$ and $\tilde{l}$ are the four- and $(D-4)$-dimensional pieces of the loop momentum respectively.  As long as we do not set this to $0$, we find the full coefficient (for a more in depth discussion see ref.~\cite{FullOneLoopGiele}).

So, the contribution to the $D=6$ coefficient due to the extra polarization state (with polarization along the sixth dimension) is exactly the difference between the $D=5$ and $D=6$ coefficients.  To see what this difference is, we go to the Feynman rules.

A relevant three-gluon vertex, with polarizations taken into account, is
\begin{equation}
\frac{i}{\sqrt{2}}(\eta^{\mu\nu}(k-l_2)^{\rho}+\eta^{\nu\rho}(l_2-l_1)^{\mu}+\eta^{\rho\mu}(l_1-k)^{\nu})\epsilon_{k\mu}\epsilon_{l_2\nu}\epsilon_{l_1\rho},
\label{eq:3gluon}
\end{equation}
where $k$ is an outgoing external momentum and $l_1$ and $l_2$ are outgoing internal loop momenta.  We wish to examine the contribution from the polarization in the sixth dimension, so we look at only those components of $\epsilon_{l_1}$ and $\epsilon_{l_2}$.  When these are contracted with $\epsilon_{k}$, $k$, $l_1$ or $l_2$, they vanish since none of those vectors contain a sixth-dimensional component.  The only non-vanishing contraction is that between the sixth-dimensional components of $\epsilon_{l_1}$ and $\epsilon_{l_2}$.  Our vertex then becomes
\begin{comment}
%
\begin{equation}
\frac{i}{\sqrt{2}}\,\epsilon_{l_1,6}\cdot\epsilon_{l_2,6}\,\,\epsilon_k\cdot(l_2-l_1)=\frac{i}{\sqrt{2}}\epsilon_k\cdot(l_1-l_2),\.
\label{eq:2scalarGluon}
\end{equation}
%
This is, up to a sign, the two-real-scalar gluon vertex (with polarization of the gluon included).  Therefore the contribution of the polarization in the sixth dimension is equal to that of a scalar particle.  After adjusting for the sign difference in scalar and gluon propagators, we see that the $D=5$ and $D=6$ coefficients differ by a coefficient found using cuts where the internal gluon loop has been replaced by a scalar loop,
\end{comment}
%
\begin{equation}
\frac{i}{\sqrt{2}}\,\epsilon_{l_1,6}\cdot\epsilon_{l_2,6}\,\,\epsilon_k\cdot(l_2-l_1)=\frac{i}{\sqrt{2}}\epsilon_k\cdot(l_2-l_1),
\label{eq:2scalarGluon}
\end{equation}
up to a sign depending on the normalization of the polarization vectors.  This is the two-real-scalar gluon vertex with polarization of the gluon included.  Therefore the contribution of the polarization in the sixth dimension is equal to that of a scalar particle, and the $D=5$ and $D=6$ coefficients differ by a coefficient found using cuts where the internal gluon loop has been replaced by a scalar loop,
\begin{equation}
C^{(6)}=C^{(5)}+C^{\mathrm{scalar}}.
\end{equation}
For our FDH coefficient then, we find
\begin{align}
C^{\mathrm{FDH}}=C^{(6)}-2C^{\mathrm{scalar}}.
\end{align}
Coefficients in the FDH scheme can be found from six-dimensional coefficients by subtracting twice the contribution of a scalar-loop diagram.  (Here we only studied the three-gluon vertex.  Similar arguments can be used for a four-gluon vertex; in fact, it is true that any $n$-gluon tree amplitude with two gluons polarized along the sixth dimension is equivalent to a tree amplitude with $n-2$ gluons and two scalars \cite{FullOneLoopGiele}.)  This leads us to the physical interpretation that the polarization states in six dimensions consist of the polarization states of four dimensions plus two scalars, but that they may be mixed in some nontrivial way.  We will not necessarily be able to identify the state $a=1$, $\dot{a}=\dot{1}$ in six dimensions, for example, with a state in four dimensions.

We now turn to the coefficients of $A_4^{(1)}(1^-,2^+,3^+,4^+)$ with a gluon loop and show that our state-sum reduction procedure works.  Using the four-gluon tree amplitude \eqref{eq:4gTree}, the quadruple cut is
\begin{align}
C_{1234}&=\sum_{\mathrm{states}}(-i)^2(l_2-p_2)^2(l_2+p_1)^2A_4(-l_2^g,2^+,3^+,l_4^g)A_4(-l_4^g,4^+,1^-,l_2^g) \notag \\
&=\frac{\langle(-l_2)_a,2_1,3_1,(l_4)_b\rangle[(-l_2)_{\dot{a}},2_{\dot{1}},3_{\dot{1}},(l_4)_{\dot{b}}]\langle(-l_4)^b,4_1,1_2,(l_2)^a\rangle[(-l_4)^{\dot{b}},4_{\dot{1}},1_{\dot{2}},(l_2)^{\dot{a}}]}{s_{23}^2}.
\end{align}
The spinor products required for this calculation are in eqs.~\eqref{eq:spinProdsG1} and \eqref{eq:spinProdsG2}.  In a manner similar to the scalar-loop calculation, we find all of the integral coefficients for the gluon-loop amplitude.  All scalar integral coefficients vanish, and we have
\begin{align}
C_4^{[4],6D}&=4i\frac{[2\,4]^2}{[1\,2]\langle 2\,3\rangle\langle 3\,4\rangle[4\,1]}\frac{s_{12}s_{23}}{s_{13}}, \notag \\
C_{3;s_{12}}^{[2],6D}&=4i\frac{[2\,4]^2}{[1\,2]\langle 2\,3\rangle\langle 3\,4\rangle[4\,1]}\frac{s_{23}^2}{s_{13}^2}(s_{13}-s_{12}), \notag \\
C_{3;s_{23}}^{[2],6D}&=4i\frac{[2\,4]^2}{[1\,2]\langle 2\,3\rangle\langle 3\,4\rangle[4\,1]}\frac{s_{12}^2}{s_{13}^2}(s_{13}-s_{23}), \notag \\
C_{2;12}^{[2],6D}&=4i\frac{[2\,4]^2}{[1\,2]\langle 2\,3\rangle\langle 3\,4\rangle[4\,1]}\frac{s_{23}}{s_{12}s_{13}}(s_{13}-s_{23}), \notag \\
C_{2;23}^{[2],6D}&=4i\frac{[2\,4]^2}{[1\,2]\langle 2\,3\rangle\langle 3\,4\rangle[4\,1]}\frac{s_{12}}{s_{13}s_{23}}(s_{13}-s_{12}). \label{eq:mppp6D}
\end{align}
Rational terms for a gluon-loop amplitude are equal to twice the rational terms for a scalar-loop amplitude \cite{Badger}, so it is not surprising that the six-dimensional coefficients for the gluon loop are 4 times those for the scalar loop (this is not true for cut-constructible pieces).  Subtracting twice the scalar-loop coefficients in section \ref{ScalarCalc} from eq.~\eqref{eq:mppp6D} according to our prescription, we arrive at our coefficients in the FDH scheme,
\begin{align}
C_4^{[4]}&=2i\frac{[2\,4]^2}{[1\,2]\langle 2\,3\rangle\langle 3\,4\rangle[4\,1]}\frac{s_{12}s_{23}}{s_{13}}, \notag \\
C_{3;s_{12}}^{[2]}&=2i\frac{[2\,4]^2}{[1\,2]\langle 2\,3\rangle\langle 3\,4\rangle[4\,1]}\frac{s_{23}^2}{s_{13}^2}(s_{13}-s_{12}), \notag \\
C_{3;s_{23}}^{[2]}&=2i\frac{[2\,4]^2}{[1\,2]\langle 2\,3\rangle\langle 3\,4\rangle[4\,1]}\frac{s_{12}^2}{s_{13}^2}(s_{13}-s_{23}), \notag \\
C_{2;12}^{[2]}&=2i\frac{[2\,4]^2}{[1\,2]\langle 2\,3\rangle\langle 3\,4\rangle[4\,1]}\frac{s_{23}}{s_{12}s_{13}}(s_{13}-s_{23}), \notag \\
C_{2;23}^{[2]}&=2i\frac{[2\,4]^2}{[1\,2]\langle 2\,3\rangle\langle 3\,4\rangle[4\,1]}\frac{s_{12}}{s_{13}s_{23}}(s_{13}-s_{12}).
\end{align}

\subsection{Fermion loop} \label{fermionLoop}

The final particle that we can have in the loop is a fermion.  Fermions in six dimensions also have additional spin states compared to their four-dimensional counterparts, but we compute using six-dimensional chiral quarks with only two states, eliminating the need for a state-sum reduction.  If non-chiral quarks were used, the loop amplitude in six dimensions would differ from that in four dimensions by a factor of two.  This can also be understood through the Feynman rules.  A quark loop gives rise to a trace over gamma matrices.  Gamma matrices in four dimensions are 4x4 while in six dimensions they are 8x8.  Where in four dimensions we have a factor of 4 then, we will have a factor of 8 in six dimensions.  However, when we calculate using six-dimensional chiral quarks, gamma matrices are replaced by Cheung and O'Connell's sigma matrices, which are 4x4 and therefore contribute the same factor as in four dimensions.

The two-chiral-quark two-gluon tree amplitude was given in eq.~\eqref{eq:2g2fTree}.  Using it for the quadruple cut, we have
\begin{align}
C_{1234}&=\sum_{\mathrm{states}}(-i)^2(l_2-p_2)^2(l_2+p_1)^2A_4(-l_2^q,2^+,3^+,l_4^q)A_4(-l_4^q,4^+,1^-,l_2^q) \notag \\
&=\frac{\langle(-l_2)_a,2_1,3_1,(l_4)_b\rangle[(-l_2)_{\dot{c}},2_{\dot{1}},3_{\dot{1}},(-l_2)^{\dot{c}}]\langle(-l_4)^b,4_1,1_2,(l_2)^a\rangle[(-l_4)_{\dot{d}},4_{\dot{1}},1_{\dot{2}},(-l_4)^{\dot{d}}]}{4s_{23}^2}.
\end{align}
Following our earlier calculation methods\footnote{As mentioned in the previous footnote, we must manually remove a factor of $i$ for each cut fermion.  With two cut fermions here, we manually insert a negative sign in the computation.}, we arrive at
\begin{align}
C_4^{[4]}&=-2i\frac{[2\,4]^2}{[1\,2]\langle 2\,3\rangle\langle 3\,4\rangle[4\,1]}\frac{s_{12}s_{23}}{s_{13}}, \notag \\
C_{3;s_{12}}^{[2]}&=-2i\frac{[2\,4]^2}{[1\,2]\langle 2\,3\rangle\langle 3\,4\rangle[4\,1]}\frac{s_{23}^2}{s_{13}^2}(s_{13}-s_{12}), \notag \\
C_{3;s_{23}}^{[2]}&=-2i\frac{[2\,4]^2}{[1\,2]\langle 2\,3\rangle\langle 3\,4\rangle[4\,1]}\frac{s_{12}^2}{s_{13}^2}(s_{13}-s_{23}), \notag \\
C_{2;12}^{[2]}&=-2i\frac{[2\,4]^2}{[1\,2]\langle 2\,3\rangle\langle 3\,4\rangle[4\,1]}\frac{s_{23}}{s_{12}s_{13}}(s_{13}-s_{23}), \notag \\
C_{2;23}^{[2]}&=-2i\frac{[2\,4]^2}{[1\,2]\langle 2\,3\rangle\langle 3\,4\rangle[4\,1]}\frac{s_{12}}{s_{13}s_{23}}(s_{13}-s_{12}),
\end{align}
where all other coefficients vanish.

\subsection{$A_4^{(1)}(1^-,2^-,3^+,4^+)$ with gluon loop}

Because there are no cut-constructible pieces, $A_4^{(1)}(1^-,2^+,3^+,4^+)$ with a gluon loop is just twice that of the amplitude with a scalar loop.  Here we demonstrate the effectiveness of six-dimensional helicity in computing an amplitude with both cut-constructible and rational pieces.  Specifically we look at $A_4^{(1)}(1^-,2^-,3^+,4^+)$.

The six-dimensional coefficients for the gluon-loop amplitude are
\begin{align}\begin{array}{ll}
C_4^{[0],6D}=-A_4^{(0)}s_{12}s_{23}, \hspace{1cm}&
C_4^{[4],6D}=-4A_4^{(0)}\frac{s_{23}}{s_{12}}, \\
C_{3;s_{12}}^{[0],6D}=0, &
C_{3;s_{23}}^{[0],6D}=0, \\
C_{3;s_{12}}^{[2],6D}=0, &
C_{3;s_{23}}^{[2],6D}=0, \\
C_{2;12}^{[0],6D}=0, &
C_{2;23}^{[0],6D}=-\frac{10}{3}A_4^{(0)}, \\
C_{2;12}^{[2],6D}=0, &
C_{2;23}^{[2],6D}=4A_4^{(0)}\left(-\frac{2}{3}\frac{1}{s_{23}}+\frac{1}{s_{12}}\right),
\end{array}
\end{align}
where
\begin{align}
A_4^{(0)}=i\frac{\langle 1\,2\rangle^4}{\langle 1\,2\rangle\langle 2\,3\rangle\langle 3\,4\rangle\langle 4\,1\rangle}.
\end{align}
For the scalar-loop amplitude, we have
\begin{align}\begin{array}{ll}
C_4^{[4]}= -A_4^{(0)}\frac{s_{23}}{s_{12}}, &
C_{2;23}^{[0]}=\frac{1}{6}A_4^{(0)},  \\
C_{2;23}^{[2]}=A_4^{(0)}\left(-\frac{2}{3}\frac{1}{s_{23}}+\frac{1}{s_{12}}\right), \hspace{1cm}&
\end{array}
\end{align}
where all other coefficients vanish.  Subtracting twice the scalar-loop coefficients from the gluon-loop coefficients, we arrive at our coefficients in the FDH scheme,
\begin{align}\begin{array}{ll}
C_4^{[0]}=-A_4^{(0)}s_{12}s_{23}, \hspace{1cm} &
C_4^{[4]}=-2A_4^{(0)}\frac{s_{23}}{s_{12}}, \\
C_{2;23}^{[0]}=-\frac{11}{3}A_4^{(0)}, &
C_{2;23}^{[2]}=2A_4^{(0)}\left(-\frac{2}{3}\frac{1}{s_{23}}+\frac{1}{s_{12}}\right),
\end{array}
\end{align}
in agreement with the previously known result \cite{BernMorgan}.

\subsection{$A_4^{(1)}(1^+,2^+,3^+,4^+,5^+)$ with gluon loop}

%%%%%%%%% FIGURE %%%%%%%%%%%%%%%
\begin{figure}[ht]
\centerline{\epsfxsize 1.7 truein \epsfbox{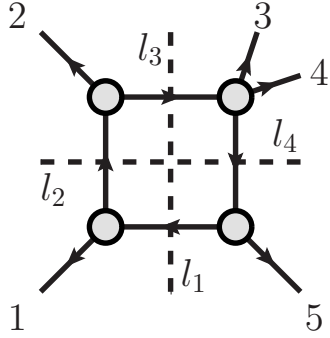}}
\caption[a]{A possible quadruple cut for a five-point amplitude.}
\label{Fig:5PointQuadCut}
\end{figure}
%%%%%%%%%%%%%%%%%%%%%%%%%%%%%%%%

As a five-point example we look at $A_4^{(1)}(1^+,2^+,3^+,4^+,5^+)$ with a gluon loop.  For one-loop four-point amplitudes, we multiplied two four-point tree amplitudes together along with inverse propagators to obtain the quadruple cut.  In a similar fashion, we multiply a four-point tree and a five-point tree together along with inverse propagators to obtain the quadruple cut here.  A simple form for the five-point tree amplitude is given in section VII of ref.~\cite{COC6D}.  It was found through Britto-Cachazo-Feng-Witten recursion and repeated use of the five-point Schouten identity.  However, we find it convenient to use an intermediate form before the use of the Schouten identity,
\begin{align}
A_5^{(0)}&(1_{a\dot{a}}^g,2_{b\dot{b}}^g,3_{c\dot{c}}^g,4_{d\dot{d}}^g,5_{e\dot{e}}^g) \notag \\
=&-\frac{i}{s_{12}s_{23}^2s_{34}s_{45}s_{15}} \notag \\
&\times\left\{\langle 1_a2_b3_c4_d\rangle[1_{\dot{a}}2_{\dot{b}}3_{\dot{c}}4_{\dot{d}}](s_{14}\langle 5_e|\s{p}_1\s{p}_3\s{p}_2\s{p}_4|5_{\dot{e}}]-s_{13}\langle 5_e|\s{p}_1\s{p}_4\s{p}_2\s{p}_4|5_{\dot{e}}]+s_{12}\langle 5_e|\s{p}_1\s{p}_4\s{p}_3\s{p}_4|5_{\dot{e}}])\right. \notag \\
&-\langle 1_a2_b3_c4_d\rangle[1_{\dot{a}}2_{\dot{b}}3_{\dot{c}}5_{\dot{e}}](s_{14}\langle 5_e|\s{p}_1\s{p}_3\s{p}_2\s{p}_5|4_{\dot{d}}]-s_{13}\langle 5_e|\s{p}_1\s{p}_4\s{p}_2\s{p}_5|4_{\dot{d}}]+s_{12}\langle 5_e|\s{p}_1\s{p}_4\s{p}_3\s{p}_5|4_{\dot{d}}]) \notag \\
&+\langle 1_a2_b3_c5_e\rangle[1_{\dot{a}}2_{\dot{b}}3_{\dot{c}}4_{\dot{d}}](s_{14}\langle 4_d|\s{p}_5\s{p}_1\s{p}_2\s{p}_3|5_{\dot{e}}]-s_{13}\langle 4_d|\s{p}_5\s{p}_1\s{p}_2\s{p}_4|5_{\dot{e}}]+s_{12}\langle 4_d|\s{p}_5\s{p}_1\s{p}_3\s{p}_4|5_{\dot{e}}]) \notag \\
&+s_{15}\langle 1_a2_b3_c5_e\rangle[1_{\dot{a}}2_{\dot{b}}3_{\dot{c}}5_{\dot{e}}]\langle 4_d|\s{p}_5\s{p}_1\s{p}_2\s{p}_3|4_{\dot{d}}]-s_{45}\langle 2_b3_c4_d5_e\rangle[2_{\dot{b}}3_{\dot{c}}4_{\dot{d}}5_{\dot{e}}]\langle 1_a|\s{p}_5\s{p}_4\s{p}_3\s{p}_2|1_{\dot{a}}] \notag \\
&+s_{45}\langle 2_b3_c4_d5_e\rangle[1_{\dot{a}}2_{\dot{b}}3_{\dot{c}}4_{\dot{d}}]\langle 1_a|\s{p}_2\s{p}_3\s{p}_4\s{p}_1|5_{\dot{e}}]+s_{45}\langle 2_b3_c4_d5_e\rangle[1_{\dot{a}}2_{\dot{b}}3_{\dot{c}}5_{\dot{e}}]\langle 1_a|\s{p}_5\s{p}_1\s{p}_2\s{p}_3|4_{\dot{d}}] \notag \\
&-s_{45}\langle 1_a2_b3_c4_d\rangle[2_{\dot{b}}3_{\dot{c}}4_{\dot{d}}5_{\dot{e}}]\langle 5_e|\s{p}_1\s{p}_4\s{p}_3\s{p}_2|1_{\dot{a}}]\left.-s_{45}\langle 1_a2_b3_c5_e\rangle[2_{\dot{b}}3_{\dot{c}}4_{\dot{d}}5_{\dot{e}}]\langle 4_d|\s{p}_3\s{p}_2\s{p}_1\s{p}_5|1_{\dot{a}}]\right\}.
\label{eq:5PointSpecial}
\end{align}
The five-point tree amplitude used to find the quadruple cut will have three adjacent external gluons of positive helicity.  Choosing these to be particles $2$, $3$ and $4$, the form in eq.~\eqref{eq:5PointSpecial} leaves us with only one non-vanishing term,
\begin{align}
A_5^{(0)}(1_{a\dot{a}}^g,2_{1\dot{1}}^{g,4D},3_{1\dot{1}}^{g,4D},4_{1\dot{1}}^{g,4D},5_{e\dot{e}}^g)=&-\frac{i}{s_{12}s_{23}^2s_{34}s_{45}}\langle 1_a2_13_15_e\rangle[1_{\dot{a}}2_{\dot{1}}3_{\dot{1}}5_{\dot{e}}]\langle 4_1|\s{p}_5\s{p}_1\s{p}_2\s{p}_3|4_{\dot{1}}],
\end{align}
where we have indicated that the external particles reside in the four-dimensional subspace.

There are five possible quadruple cuts for a five-point amplitude.  We will calculate the cut in Fig.~\ref{Fig:5PointQuadCut} explicitly and obtain the rest through cyclic symmetry.  The quadruple cut momentum solution is
\begin{align}
\bar{l}_1^{\pm}=c_1^{\pm}\lambda_5\tilde{\lambda}_1-\frac{1}{c_1^{\pm}}\frac{m\tilde{m}}{s_{15}}\lambda_1\tilde{\lambda}_5,
\end{align}
where
\begin{align}
c_1^{\pm}=\frac{\langle 1\,2\rangle}{2\langle 5\,2\rangle}\left(1\pm\sqrt{1+\frac{4m\tilde{m}s_{25}}{s_{12}s_{23}}}\right).
\end{align}
The cut is given by
\begin{align}
C_{1235}=&\sum_{\mathrm{states}}(-i)^2l_1^2l_3^2A_5(-l_2^g,2^+,3^+,4^+,l_4^g)A_4(-l_4^g,5^+,1^+,l_2^g) \notag \\
=&\frac{1}{s_{23}^2s_{34}s_{15}(l_4+p_4)^2}\langle(-l_2)_a,2_1,3_1,(l_4)_b\rangle\mathrm{[}(-l_2)_{\dot{a}},2_{\dot{1}},3_{\dot{1}},(l_4)_{\dot{b}}\mathrm{]}\langle 4_1|\s{l}_4(-\s{l}_2)\s{p}_2\s{p}_3|4_{\dot{1}}] \notag \\
&\times\langle(-l_4)^b,5_1,1_1,(l_2)^a\rangle\mathrm{[}(-l_4)^{\dot{b}},5_{\dot{1}},1_{\dot{1}},(l_2)^{\dot{a}}\mathrm{]}.
\label{eq:fivePointBoxCut}
\end{align}
This does not fall conveniently into a polynomial in $\mu^2$, so we must use the extraction formulas in eq.~\eqref{eq:BoxExtract}, which give
\begin{align}
C_{4;34}^{[0],6D}&=0, \notag \\
C_{4;34}^{[4],6D}&=\frac{4i}{\langle 1\,2\rangle\langle 2\,3\rangle\langle 3\,4\rangle\langle 4\,5\rangle\langle 5\,1\rangle}\frac{\mathrm{tr}_+[3452]}{\mathrm{tr}_5[1234]}s_{12}s_{15},
\end{align}
where $\mathrm{tr}_5[A]=\mathrm{tr}[\gamma_5A]$ and $\mathrm{tr}_+[A]=\frac{1}{2}\mathrm{tr}[(1+\gamma_5)A]$.  After subtracting twice the scalar contribution, we arrive at
\begin{align}
C_{4;34}^{[0]}&=0, \notag \\
C_{4;34}^{[4]}&=\frac{2i}{\langle 1\,2\rangle\langle 2\,3\rangle\langle 3\,4\rangle\langle 4\,5\rangle\langle 5\,1\rangle}\frac{\mathrm{tr}_+[3452]}{\mathrm{tr}_5[1234]}s_{12}s_{15},
\end{align}
in agreement with the previously known results given in refs.~\cite{BernMorgan, Bern5Point, Badger}.  All triangle and bubble coefficients vanish, and the remaining box coefficients can be found through cyclic permutations.

\section{EXTERNAL FERMION AMPLITUDES} \label{ExtFerm}

%%%%%%%%% FIGURE %%%%%%%%%%%%%%%
\begin{figure}[ht]
\centering
\subfloat[][]{\epsfxsize 1.7 truein \epsfbox{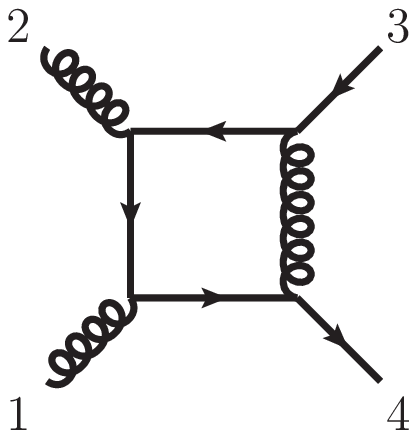}
\label{Fig:FermionRight}}
\hspace{2cm}
\subfloat[][]{\epsfxsize 1.7 truein \epsfbox{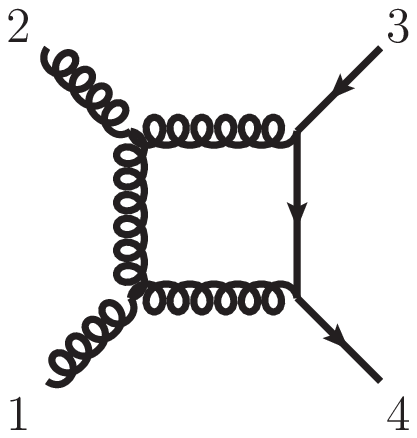}
\label{Fig:FermionLeft}}
\caption[a]{External fermion box diagrams for (a) the right-turning fermion and (b) the left-turning fermion.}
\label{Fig:Fermion}
\end{figure}
%%%%%%%%%%%%%%%%%%%%%%%%%%%%%%%%

%%%%%%%%% FIGURE %%%%%%%%%%%%%%%
\begin{figure}[ht]
\centering
\subfloat[][]{\epsfxsize 1.7 truein \epsfbox{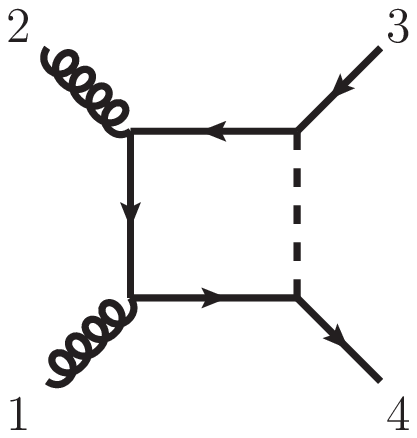}
\label{Fig:FermionScalarRight}}
\hspace{2cm}
\subfloat[][]{\epsfxsize 1.7 truein \epsfbox{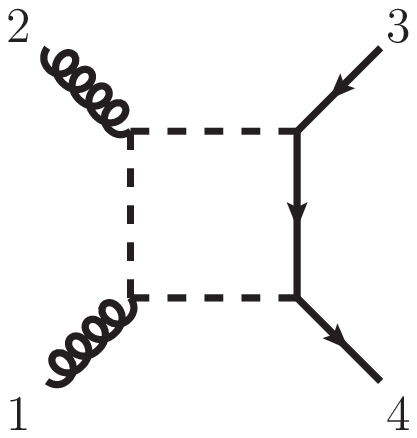}
\label{Fig:FermionScalarLeft}}
\caption[a]{External fermion box diagrams with internal gluons replaced by scalars.}
\label{Fig:FermionScalar}
\end{figure}
%%%%%%%%%%%%%%%%%%%%%%%%%%%%%%%%

Next we turn to the case of external fermions.  The leading-color contribution to a one-loop amplitude with two external fermions can be decomposed as~\cite{Bern2g3f}
\begin{align}
A_n^f=A_n^L-\frac{1}{N_c^2}A_n^R+\frac{N_f}{N_c}A_n^{L,[1/2]}+\frac{N_s}{N_c}A_n^{L,[0]}.
\end{align}
Only amplitudes with adjacent external fermions contribute at leading color.  $A_n^L$ is the ``left-turning'' fermion piece where the fermion follows the shortest path through the loop (see Fig.~\ref{Fig:FermionLeft}), while for $A_n^R$ the fermion follows the longest path (Fig.~\ref{Fig:FermionRight}).  $A_n^{L,[1/2]}$ and $A_n^{L,[0]}$ have, respectively, a fermion and scalar loop and are therefore proportional to the number of each.  We focus on the left- and right-turning pieces and calculate the cuts in Fig.~\ref{Fig:Fermion}.  (For our sample helicity arrangement to follow, $A_n^{L,[1/2]}$ and $A_n^{L,[0]}$ happen to vanish.)

Once again the extra spin states of six dimensions must be treated with a state-sum reduction procedure.  Fortunately our findings in section \ref{4pointglue} give us good physical intuition hinting at the appropriate procedure.  Fermions have additional states, but since we calculate with a six-dimensional chiral quark with only two states, no reduction procedure is necessary.  For gluons, we interpret the extra-dimensional polarization states as scalars and subtract twice the contribution of an amplitude with a scalar loop replacing the gluon loop.  The natural generalization then is to replace any internal gluons by scalars, even if they do not comprise the entire loop, and subtract twice the contribution of this diagram from a diagram with internal gluons.  For example, cuts from the diagrams in Fig.~\ref{Fig:FermionScalar} should be subtracted from their corresponding diagrams in Fig.~\ref{Fig:Fermion} to reduce to the FDH scheme.

For a more rigorous motivation, we again appeal to the Feynman rules.  The dimensional dependence for Feynman loop diagrams involving external fermions arises through a contraction of gamma matrices.  If we isolate the terms with dimensional dependence, we find them to differ only by a factor of $D$ from equivalent diagrams with all internal gluons replaced by scalars.  We can therefore summarize our procedure as
\begin{align}
C^{\mathrm{FDH}}=C^{(6)}-2C^{\mathrm{scalar}}.
\end{align}

\subsection{Right-turning $A_4^{(1)}(1^+,2^-,3_{\bar{q}}^-,4_q^+)$}

As an example we calculate $A_4^{(1)}(1^+,2^-,3_{\bar{q}}^-,4_q^+)$ for both the right-turning and left-turning configurations.  For the right-turning, we require the two-chiral-quark two-gluon tree amplitude \eqref{eq:2g2fTree} to sew four-point amplitudes across the $s_{23}$ channel.  We also need the four-chiral-quark tree amplitude with a gluon propagator\footnote{If we wanted to compute a four-dimensional tree amplitude using this, we would need to reduce the state-sum by subtracting two times the diagrams with a scalar propagator replacing the gluon propagator. The six-dimensional chirality of the external quarks is not important in four dimensions since each chirality separately carries both helicity states of the four-dimensional quark.},
\begin{align}
A_4^{(0);g}(1_a^q,2_b^q,3_c^q,4_d^q)=-i\frac{s_{13}}{s_{12}s_{23}}\langle 1_a2_b3_c4_d\rangle,
\end{align}
for sewing across the $s_{12}$ channel.  Using these in our normal procedure, we find the six-dimensional coefficients,
\begin{align}\begin{array}{ll}
C_4^{[0],6D}=\frac{1}{2}A_4^{(0)}\frac{s_{12}^2s_{23}}{s_{13}}, &
C_4^{[4],6D}=0, \\
C_{3;12}^{[0],6D}=\frac{1}{2}A_4^{(0)}\left(s_{12}-\frac{s_{12}s_{23}}{s_{13}}\right), &
C_{3;12}^{[2],6D}=A_4^{(0)}\left(2+\frac{s_{13}}{s_{23}}-\frac{s_{23}}{s_{13}}\right), \\
C_{3;23}^{[0],6D}=\frac{1}{2}A_4^{(0)}\frac{s_{12}s_{23}}{s_{13}}, &
C_{3;23}^{[2],6D}=-A_4^{(0)}\frac{s_{12}}{s_{13}}, \\
C_{3;34}^{[0],6D}=-\frac{1}{2}A_4^{(0)}\left(s_{12}+\frac{s_{12}s_{23}}{s_{13}}\right), \hspace{1cm}&
C_{3;34}^{[2],6D}=-A_4^{(0)}\left(2+\frac{s_{13}}{s_{23}}+\frac{s_{23}}{s_{13}}\right), \\
C_{3;41}^{[0],6D}=\frac{1}{2}A_4^{(0)}\frac{s_{12}s_{23}}{s_{13}}, &
C_{3;41}^{[2],6D}=-A_4^{(0)}\frac{s_{12}}{s_{13}}, \\
C_{2;12}^{[0],6D}=-A_4^{(0)}, &
C_{2;12}^{[2],6D}=0, \\
C_{2;23}^{[0],6D}=0, &
C_{2;23}^{[2],6D}=0,
\end{array}
\end{align}
where
\begin{align}
A_4^{(0)}&=i\frac{\langle 2\,3\rangle^3\langle 2\,4\rangle}{\langle 1\,2\rangle\langle 2\,3\rangle\langle 3\,4\rangle\langle 4\,1\rangle}.
\end{align}
To perform the state-sum reduction, our required tree amplitudes are the two-chiral-quark two-anti-chiral-quark amplitude with a scalar propagator (with adjacent chiral quarks) and the chiral-quark anti-chiral-quark scalar gluon amplitude,
\begin{align}
A_4^{(0);s}(1_a^q,2_b^q,3_{\dot{c}}^q,4_{\dot{d}}^q)&=-\frac{i}{2s_{23}}\langle 1_a|4_{\dot{d}}]\langle 2_b|3_{\dot{c}}], \notag \\
A_4^{(0)}(1_{a\dot{a}}^g,2_b^q, 3^s,4_{\dot{d}}^q)&=-\frac{i}{4\sqrt{2}s_{12}s_{23}}\langle 1_a2_b3_c3^c\rangle[1_{\dot{a}}3^{\dot{c}}3_{\dot{c}}4_{\dot{d}}],
\end{align}
as computed from Feynman diagrams.  The coefficients for diagrams with internal gluons replaced by scalars are
\begin{align}\begin{array}{ll}
C_4^{[0]}=0, &
C_4^{[4]}=0, \\
C_{3;12}^{[0]}=0, &
C_{3;12}^{[2]}=\frac{1}{4}A_4^{(0)}\left(2+\frac{s_{13}}{s_{23}}-\frac{s_{23}}{s_{13}}\right), \\
C_{3;23}^{[0]}=0, &
C_{3;23}^{[2]}=-\frac{1}{4}A_4^{(0)}\frac{s_{12}}{s_{13}}, \\
C_{3;34}^{[0]}=0, &
C_{3;34}^{[2]}=-\frac{1}{4}A_4^{(0)}\left(2+\frac{s_{13}}{s_{23}}+\frac{s_{23}}{s_{13}}\right), \\
C_{3;41}^{[0]}=0, &
C_{3;41}^{[2]}=-\frac{1}{4}A_4^{(0)}\frac{s_{12}}{s_{13}}, \\
C_{2;12}^{[0]}=\frac{1}{4}A_4^{(0)}, \hspace{1cm} &
C_{2;12}^{[2]}=0, \\
C_{2;23}^{[0]}=0, &
C_{2;23}^{[2]}=0.
\end{array}
\end{align}
After utilizing our state-sum reduction procedure and combining terms with invariants in the same channel, we finally have,
\begin{align}\begin{array}{ll}
C_4^{[0]}=\frac{1}{2}A_4^{(0)}\frac{s_{12}^2s_{23}}{s_{13}}, &
C_4^{[4]}=0, \\
C_{3;s_{12}}^{[0]}=-A_4^{(0)}\frac{s_{12}s_{23}}{s_{13}}, \hspace{1cm} &
C_{3;s_{12}}^{[2]}=-A_4^{(0)}\frac{s_{23}}{s_{13}}, \\
C_{3;s_{23}}^{[0]}=A_4^{(0)}\frac{s_{12}s_{23}}{s_{13}}, &
C_{3;s_{23}}^{[2]}=-A_4^{(0)}\frac{s_{12}}{s_{13}}, \\
C_{2;12}^{[0]}=-\frac{3}{2}A_4^{(0)}, &
C_{2;12}^{[2]}=0, \\
C_{2;23}^{[0]}=0, &
C_{2;23}^{[2]}=0,
\end{array}
\end{align}
in agreement with the previously known result \cite{KunsztResults}.

\subsection{Left-turning $A_4^{(1)}(1^+,2^-,3_{\bar{q}}^-,4_q^+)$}

For the left-turning diagram (Fig.~\ref{Fig:FermionLeft}), the only new tree amplitude that we require is the two-chiral-quark two-scalar amplitude,
\begin{align}
A_4^{(0)}(1^s,2^s,3_c^q,4_d^q)=\frac{i(s_{13}-s_{23})}{4s_{12}s_{23}}\langle 1_a1^a3_c4_d\rangle.
\end{align}
Proceeding as usual, we find for our coefficients,
\begin{align}\begin{array}{ll}
C_4^{[0]}=-A_4^{(0)}\left(s_{12}s_{23}+\frac{1}{2}\frac{s_{12}^2s_{23}}{s_{13}}\right), \hspace{1cm} &
C_4^{[4]}=0, \\ 
C_{3;s_{12}}^{[0]}=A_4^{(0)}\frac{s_{12}s_{23}}{s_{13}}, &
C_{3;s_{12}}^{[2]}=A_4^{(0)}\frac{s_{23}}{s_{13}}, \\
C_{3;s_{23}}^{[0]}=-A_4^{(0)}\frac{s_{12}s_{23}}{s_{13}}, &
C_{3;s_{23}}^{[2]}=A_4^{(0)}\frac{s_{12}}{s_{13}}, \\
C_{2;12}^{[0]}=-\frac{3}{2}A_4^{(0)}, &
C_{2;12}^{[2]}=0, \\
C_{2;23}^{[0]}=0, &
C_{2;23}^{[2]}=0,
\end{array}
\end{align}
once again in agreement with ref.~\cite{KunsztResults}.

\section{HIGGS TO PARTONS PROCESSES: $A^{(1)}_3(H,1^+,2^+,3^+)$} \label{HiggsPartons}

An important process for the Higgs boson search is production via the strong interaction.  In the Standard Model, the Higgs couples to two gluons via a heavy-quark loop~\cite{GeorgiGlashow}.  The dominant contribution is from the top quark; contributions from other quarks are suppressed by at least a factor of $O(m_b^2/m_t^2)$ where $m_b$ is the mass of the bottom quark and $m_t$ is the mass of the top.

Computations that keep explicit dependence on $m_t$ can be difficult, especially at NLO since they involve two-loop amplitudes.  Fortunately in the large-$m_t$ limit, where the mass of the Higgs $m_H$ is smaller than the threshold for top-quark pair production, $m_H<2m_t$, computations are greatly simplified by letting $m_t\rightarrow\infty$ \cite{Wilczek, Djouadi, Dawson}.  Accurate approximations (to within $10\%$ of the Higgs-mass range at the LHC) of NLO QCD corrections can be found by finding the NLO QCD correction factor, $K^{NLO}\equiv\sigma^{NLO}/\sigma^{LO}$ in the limit $m_t\rightarrow\infty$, then multiplying by the exact leading-order cross section \cite{Kramer}.  The approximation is valid for multi-jet processes provided the transverse momentum of each jet is less than $m_t$ \cite{DelDucaKilgore}.

Taking the limit $m_t\rightarrow\infty$ allows us to use an effective Lagrangian to express the coupling of the Higgs to gluons \cite{Wilczek},
\begin{align}
\mathcal{L}^{\mathrm{int}}_H=\frac{C}{2}H\,\mathrm{tr}\,G_{\mu\nu}G^{\mu\nu}.
\end{align}
This reduces two-loop calculations at NLO to one-loop calculations.  The Higgs two-gluon color-ordered vertex is
\begin{align}
-2i(\eta^{\mu_1\mu_2}k_1\cdot k_2-k_1^{\mu_2}k_2^{\mu_1}),
\label{eq:Hgg}
\end{align}
which gives the six-dimensional helicity tree amplitude,
\begin{align}
A_2^{(0)}(H,1_{a\dot{a}},2_{b\dot{b}})=i\langle 1_a|2_{\dot{b}}]\langle 2_b|1_{\dot{a}}].
\end{align}
The Higgs three-gluon vertex factor is
\begin{align}
-i\sqrt{2}(\eta^{\mu_1\mu_2}(k_1-k_2)^{\mu_3}+\eta^{\mu_2\mu_3}(k_2-k_3)^{\mu_1}+\eta^{\mu_3\mu_1}(k_3-k_1)^{\mu_2}),
\label{eq:Hggg}
\end{align}
which, when combined with other relevant Feynman diagrams, gives the tree amplitude,
\begin{align}
A_3^{(0)}(H,1_{a\dot{a}},2_{b\dot{b}},3_{c\dot{c}})=\frac{-i}{s_{12}s_{23}^2s_{13}^2}&\left\{(s_{12}m_H^2+s_{13}s_{23})\langle 1_a|\s{k}_3|2_b\rangle[1_{\dot{a}}|\s{k}_3|2_{\dot{b}}]\langle 3_c|\s{k}_2\s{k}_1|3_{\dot{c}}]\right. \notag \\
&-s_{23}(s_{13}m_H^2+s_{12}s_{23})\langle 1_a|3_{\dot{c}}]\langle 3_c|1_{\dot{a}}]\langle 2_b|\s{k}_1\s{k}_3|2_{\dot{b}}] \notag \\
&+s_{13}(s_{23}m_H^2+s_{12}s_{13})\langle 2_b|3_{\dot{c}}]\langle 3_c|2_{\dot{b}}]\langle 1_a|\s{k}_2\s{k}_3|1_{\dot{a}}] \notag \\
&\left.-s_{12}\langle 1_a|\s{k}_2\s{k}_3|1_{\dot{a}}]\langle 2_b|\s{k}_1\s{k}_3|2_{\dot{b}}]\langle 3_c|\s{k}_2\s{k}_1|3_{\dot{c}}]\right\}.
\end{align}
We use these to calculate $A^{(1)}_3(H,1^+,2^+,3^+)$, first given in ref.~\cite{Schmidt}.

\subsection{Momentum solutions}

%%%%%%%%% FIGURE %%%%%%%%%%%%%%%
\begin{figure}[ht]
\centerline{\epsfxsize 1.7 truein \epsfbox{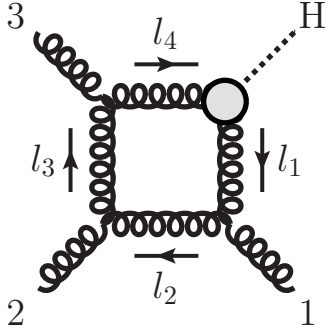}}
\caption[a]{A possible box diagram for $A^{(1)}_3(H,1^+,2^+,3^+)$ using the effective coupling of the Higgs to two gluons.  We denote this cut with ordering ``123H''.}
\label{Fig:HiggsBox}
\end{figure}
%%%%%%%%%%%%%%%%%%%%%%%%%%%%%%%%

Momentum solutions can be a bit more complicated with a massive leg.  We give explicit forms in line with those in section \ref{IntExtract}.  For the box in Fig.~\ref{Fig:HiggsBox}, we use
\begin{align}
\bar{l}_2=c_{\pm}\lambda_1\tilde{\lambda}_2-\frac{1}{c_{\pm}}\frac{m\tilde{m}}{s_{12}}\lambda_2\tilde{\lambda}_1,
\end{align}
where
\begin{align}
c_{\pm}=\frac{\langle 2\,3\rangle}{2\langle 1\,3\rangle}\left(1\pm\sqrt{1+\frac{4m\tilde{m}s_{13}}{s_{12}s_{23}}}\right).
\end{align}
The triple-cut solution for cutting propagators $l_1$, $l_2$ and $l_3$ can be found by letting $c_{\pm}\rightarrow t$.  To cut $l_1$, $l_3$ and $l_4$, we use
\begin{align}
\bar{l}_4=\frac{m_H^2}{\gamma_{3H}}\lambda_3\tilde{\lambda}_3+t\lambda_3\tilde{\lambda}_H^{\flat}-\frac{m\tilde{m}}{t\gamma_{3H}}\lambda_H^{\flat}\tilde{\lambda}_3,
\end{align}
where
\begin{align}
K_H^{\flat}&=k_H-\frac{m_H^2}{\gamma_{3H}}k_3, \notag \\
\gamma_{3H}&=2k_H\cdot k_3 \notag \\
&=-s_{13}-s_{23}.
\end{align}
This allows us to set
\begin{align}
&\mu=\lambda_H^{\flat}, \hspace{8mm} \tilde{\mu}=\frac{\tilde{\lambda}_3}{t}, \hspace{8mm} \lambda=\lambda_3, \hspace{8mm} \tilde{\lambda}=\frac{m_H^2}{\gamma_{3H}}\tilde{\lambda}_3+t\tilde{\lambda}_H^{\flat}, \notag \\
\kappa_{3H}=&\frac{m}{\langle 3\lambda_H^{\flat}\rangle}, \hspace{8mm} \tilde{\kappa}_{3H}=\frac{\tilde{m}}{[3\lambda_H^{\flat}]}, \hspace{8mm} \kappa_{3H}'=\frac{m}{[3\lambda_H^{\flat}]}, \hspace{8mm} \tilde{\kappa}_{3H}'=\frac{\tilde{m}}{\langle 3\lambda_H^{\flat}\rangle}.
\end{align}
Making similar associations for $l_1$ and $l_3$ through the use of momentum conservation allows us to find the necessary spinor products.  We find the cut for the conjugate momentum solution by taking $t\rightarrow-\frac{m\tilde{m}}{t\gamma_{3H}}$.

To cut $l_1$ and $l_3$, we use
\begin{align}
\bar{l}_1=y\lambda_1\tilde{\lambda}_1+(1-y)\lambda_2\tilde{\lambda}_2+t\lambda_1\tilde{\lambda}_2+\frac{y(1-y)-m\tilde{m}/s_{12}}{t}\lambda_2\tilde{\lambda}_1,
\end{align}
whereas to cut $l_1$ and $l_4$ we use
\begin{align}
\bar{l}_4=y\lambda_H^{\flat}\tilde{\lambda}_H^{\flat}+\frac{m_H^2}{\bar{\gamma}}(1-y)\lambda_3\tilde{\lambda}_3+t\lambda_H^{\flat}\tilde{\lambda}_3+\frac{y(1-y)m_H^2-m\tilde{m}}{t\bar{\gamma}}\lambda_3\tilde{\lambda}_H^{\flat},
\end{align}
where
\begin{align}
K_H^{\flat}=k_H-\frac{m_H^2}{\bar{\gamma}}k_3, \hspace{2cm} \bar{\gamma}=-s_{13}-s_{23}.
\end{align}
All other cuts can be obtained from permutations of the ones given.  It should be noted that the Higgs can attach itself to any gluon line, so there are quite a few more cuts than there are for a four-gluon amplitude.  Fortunately many coefficients for our particular helicity arrangement are related through cyclic permutations of the external gluon momenta.

\subsection{Six-dimensional coefficients}

We explicitly calculate the box coefficients for Fig.~\ref{Fig:HiggsBox}.  The quadruple cut is given by,
\begin{align}
C_{1234}=&\sum_{\mathrm{states}}(-i)^2l_2^2l_4^2A_4^{(0)}(-l_1^g,1^+,2^+,l_3^g)A_4^{(0)}(-l_3^g,3^+,H,l_1^g) \notag \\
=&\frac{1}{(l_1\!+\!k_3)^4s_{12}^3}\langle(-l_1)^a,1_1,2_1,(l_3)^b\rangle\mathrm{[}(-l_1)^{\dot{a}},1_{\dot{1}},2_{\dot{1}},(l_3)^{\dot{b}}\mathrm{]} \notag \\
&\times\left\{(l_4^2m_H^2+s_{12}(l_1\!+\!k_3)^2)\right.\langle (-l_3)_b|\s{l}_1|3_1\rangle\mathrm{[}(-l_3)_{\dot{b}}|\s{l}_1|3_{\dot{1}}\mathrm{]}\langle (l_1)_a|\s{k}_3(-\s{l}_3)|(l_1)_{\dot{a}}] \notag \\
&-(l_1\!+\!k_3)^2(s_{12}m_H^2+l_4^2(l_1\!+\!k_3)^2)\langle (-l_3)_b|(l_1)_{\dot{a}}]\langle (l_1)_a|(-l_3)_{\dot{b}}]\langle 3_1|(-\s{l}_3)\s{l}_1|3_{\dot{1}}] \notag \\
&+s_{12}((l_1\!+\!k_3)^2m_H^2+l_4^2s_{12})\langle 3_1|(l_1)_{\dot{a}}]\langle (l_1)_a|3_{\dot{1}}]\langle (-l_3)_b|\s{k}_3\s{l}_1|(-l_3)_{\dot{b}}] \notag \\
&\left.-l_4^2\langle (-l_3)_b|\s{k}_3\s{l}_1|(-l_3)_{\dot{b}}]\langle 3_1|(-\s{l}_3)\s{l}_1|3_{\dot{1}}]\langle (l_1)_a|\s{k}_3(-\s{l}_3)|(l_1)_{\dot{a}}]\right\}.
\end{align}
After averaging over solutions, this becomes
\begin{align}
C_{1234}&=\frac{1}{\langle 1\,2\rangle\langle 2\,3\rangle\langle 3\,1\rangle}\left[m\tilde{m}s_{12}s_{23}m_H^2-\frac{1}{2}s_{12}s_{23}m_H^4\right].
\end{align}
We can then identify our scalar box coefficient (we neglect $C_4^{[2]}$ since its integral vanishes) for this cut and use permutations to find it for the other quadruple cuts,
\begin{align}
C_{4;123H}^{[0],6D}&=-\frac{i}{2}\frac{m_H^4}{\langle 1\,2\rangle\langle 2\,3\rangle\langle 3\,1\rangle}s_{12}s_{23}, \notag \\
C_{4;1H23}^{[0],6D}&=-\frac{i}{2}\frac{m_H^4}{\langle 1\,2\rangle\langle 2\,3\rangle\langle 3\,1\rangle}s_{23}s_{13}, \notag \\
C_{4;12H3}^{[0],6D}&=-\frac{i}{2}\frac{m_H^4}{\langle 1\,2\rangle\langle 2\,3\rangle\langle 3\,1\rangle}s_{12}s_{13},
\end{align}
where the subscript indicates the cyclic ordering of the external particles.  Our remaining coefficients are
\begin{align}\begin{array}{ll}
C_{3;123H;12}^{[0],6D}= \frac{1}{2}A_3^{(0)}(s_{13}+s_{23}), \hspace{1cm} & C_{3;123H;12}^{[2]}= -A_3^{(0)}\frac{s_{13}+s_{23}}{m_H^2}, \\
C_{3;123H;23}^{[0],6D}= \frac{1}{2}A_3^{(0)}(s_{12}+s_{13}), & C_{3;123H;23}^{[2]}= -A_3^{(0)}\frac{s_{12}+s_{13}}{m_H^2},  \\
C_{3;12H3;13}^{[0],6D}= \frac{1}{2}A_3^{(0)}(s_{12}+s_{23}), & C_{3;12H3;13}^{[2]}= -A_3^{(0)}\frac{s_{12}+s_{23}}{m_H^2}, \\
C_{3;12H3;12}^{[0],6D}= \frac{1}{2}A_3^{(0)}(s_{13}+s_{23}), & C_{3;12H3;12}^{[2]}= -A_3^{(0)}\frac{s_{13}+s_{23}}{m_H^2}, \\
C_{3;1H23;23}^{[0],6D}= \frac{1}{2}A_3^{(0)}(s_{12}+s_{13}), & C_{3;1H23;23}^{[2]}= -A_3^{(0)}\frac{s_{12}+s_{13}}{m_H^2}, \\
C_{3;1H23;13}^{[0],6D}= \frac{1}{2}A_3^{(0)}(s_{12}+s_{23}), & C_{3;1H23;13}^{[2]}= -A_3^{(0)}\frac{s_{12}+s_{23}}{m_H^2}, \\
C_{3;123H;3H}^{[2],6D}= -4A_3^{(0)}\frac{s_{13}s_{23}}{m_H^4}, & C_{3;12H3;2H}^{[2]}= -4A_3^{(0)}\frac{s_{12}s_{23}}{m_H^4}, \\
C_{3;1H23;1H}^{[2],6D}= -4A_3^{(0)}\frac{s_{12}s_{13}}{m_H^4}, & C_{2;123H;12}^{[2]}= 8A_3^{(0)}\frac{s_{13}s_{23}}{s_{12}m_H^4}, \\
C_{2;12H3;13}^{[2],6D}=8A_3^{(0)}\frac{s_{12}s_{23}}{s_{13}m_H^4}, & C_{2;1H23;23}^{[2],6D}=8A_3^{(0)}\frac{s_{12}s_{13}}{s_{23}m_H^4},
\end{array}
\end{align}
where all other coefficients, including $C_{2;123H;123}$, vanish, and
\begin{align}
A_3^{(0)}=i\frac{m_H^4}{\langle 1\,2\rangle\langle 2\,3\rangle\langle 3\,1\rangle}.
\label{eq:HgggTree}
\end{align}

\subsection{State-sum reduction}

%%%%%%%%% FIGURE %%%%%%%%%%%%%%%
\begin{figure}[ht]
\centerline{\epsfxsize 1.7 truein \epsfbox{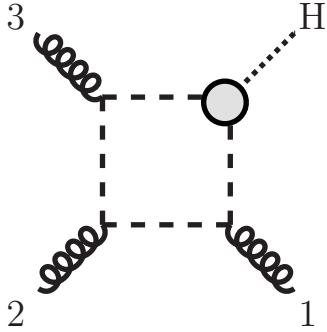}}
\caption[a]{The box diagram for the cut ordering ``123H'' with all internal gluons replaced by scalars.}
\label{Fig:HiggsScalarBox}
\end{figure}
%%%%%%%%%%%%%%%%%%%%%%%%%%%%%%%%

To reduce our coefficients to those of the FDH scheme, we again subtract twice the contribution of cuts where all internal gluons have been replaced by scalars (see Fig.~\ref{Fig:HiggsScalarBox}).  However, we need an effective Higgs two-scalar vertex.  Using the same argument as in section \ref{GluonCalc}, we can determine this vertex from the Higgs two-gluon factor by isolating the contribution due to internal gluons polarized along the sixth dimension while keeping the loop momenta in five dimensions.  This effectively picks out the term with a metric tensor containing Lorentz indices of the internal gluons.  This makes sense since it is the contractions of these metric tensors that gives rise to the dimensional dependence in a gluon loop when using the Feynman rules.  The Higgs two-scalar vertex factor, as determined from eq.~\eqref{eq:Hgg}, is then
%
\begin{comment}
Recall that replacing internal gluons by scalars is done to find the coefficients proportional to the dimension $D$ in eq.~\eqref{eq:coeffProp}.  This term arises through a closed loop of contracted metric tensors and/or gamma matrices.  For a gluon loop, we have only metric tensors.  The terms in the vertex factors multiplying the contracted metric tensors will be part of amplitude terms with explicit $D$ dependence.  We can therefore find the coefficients that are proportional to $D$ by replacing vertex factors containing internal gluons in the following way: we keep only the terms with a metric tensor containing Lorentz indices of the internal gluons, then we drop the metric tensor.  The more rigorous argument in section \ref{GluonCalc} effectively does this in going from eq.~\eqref{eq:3gluon} to eq.~\eqref{eq:2scalarGluon}.  This argument, or the more rigorous one of section \ref{GluonCalc}, gives us our Higgs two-scalar vertex factor from eq.~\eqref{eq:Hgg},
\end{comment}
%
\begin{align}
-2ik_1\cdot k_2=-is_{12}.
\label{eq:Hss}
\end{align}
We use eq.~\eqref{eq:Hggg} for the Higgs gluon two-scalar vertex factor,
\begin{align}
-i\sqrt{2}(k_1-k_2)^{\mu},
\end{align}
where $k_1$ and $k_2$ are the momenta of the scalars.  It should be noted that these are not the same factors that arise when taking the limit $m_t\rightarrow\infty$ in a loop diagram for $H\rightarrow ss$ mediated by a top-quark loop.

$A_2^{(0)}(H,1^s,2^s)$ is then simply given by eq.~\eqref{eq:Hss}, while for the Higgs gluon two-scalar tree amplitude we have
\begin{align}
A_3^{(0)}(H,1^g_{a\dot{a}},2^s,3^s)=-i\left(\frac{1}{s_{23}}+\frac{m_H^2}{s_{12}s_{13}}\right)\langle 1_a|\s{k}_3\s{k}_2|1_{\dot{a}}].
\end{align}
Using these, we can find our coefficients for $A_3^{(1)}(H,1^+,2^+,3^+)$ with a scalar loop,
\begin{align}\begin{array}{ll}
C_{3;123H;12}^{[2]}=-\frac{1}{2}A_3^{(0)}\frac{s_{13}+s_{23}}{m_H^2}, \hspace{1cm} & C_{3;123H;23}^{[2]}=-\frac{1}{2}A_3^{(0)}\frac{s_{12}+s_{13}}{m_H^2}, \\
C_{3;12H3;13}^{[2]}=-\frac{1}{2}A_3^{(0)}\frac{s_{12}+s_{23}}{m_H^2}, & C_{3;12H3;12}^{[2]}=-\frac{1}{2}A_3^{(0)}\frac{s_{13}+s_{23}}{m_H^2}, \\
C_{3;1H23;23}^{[2]}=-\frac{1}{2}A_3^{(0)}\frac{s_{12}+s_{13}}{m_H^2}, & C_{3;1H23;13}^{[2]}=-\frac{1}{2}A_3^{(0)}\frac{s_{12}+s_{23}}{m_H^2}, \\
C_{3;123H;3H}^{[2]}=-A_3^{(0)}\frac{s_{13}s_{23}}{m_H^4}, & C_{3;12H3;2H}^{[2]}=-A_3^{(0)}\frac{s_{12}s_{23}}{m_H^4}, \\
C_{3;1H23;1H}^{[2]}=-A_3^{(0)}\frac{s_{12}s_{13}}{m_H^4}, & C_{2;123H;12}^{[2]}=2A_3^{(0)}\frac{s_{13}s_{23}}{s_{12}m_H^4}, \\
C_{2;12H3;13}^{[2]}=2A_3^{(0)}\frac{s_{12}s_{23}}{s_{13}m_H^4}, & C_{2;1H23;23}^{[2]}=2A_3^{(0)}\frac{s_{12}s_{13}}{s_{23}m_H^4},
\end{array}
\end{align}
where $A_3^{(0)}$ is given by eq.~\eqref{eq:HgggTree}.  Subtracting 2 times these coefficients from our six-dimensional coefficients and performing the integrals, we arrive at our result in agreement with ref.~\cite{Schmidt},
\begin{align}
A_3^{(1)}(H,1^+,2^+,3^+)&=\frac{A_3^{(0)}}{(4\pi)^{2-\epsilon}}\left[U_3+\frac{1}{3}\frac{s_{12}s_{23}+s_{12}s_{13}+s_{13}s_{23}}{m_H^4}\right],
\end{align}
where
\begin{align}
U_3\equiv&-\frac{1}{\epsilon^2}\left[\left(\frac{\mu^2}{-s_{12}}\right)^{\epsilon}+\left(\frac{\mu^2}{-s_{23}}\right)^{\epsilon}+\left(\frac{\mu^2}{-s_{13}}\right)^{\epsilon}\right]+\frac{\pi^2}{2} \notag \\
&-\ln\left(\frac{-s_{12}}{-m_H^2}\right)\ln\left(\frac{-s_{23}}{-m_H^2}\right)-\ln\left(\frac{-s_{12}}{-m_H^2}\right)\ln\left(\frac{-s_{13}}{-m_H^2}\right)-\ln\left(\frac{-s_{23}}{-m_H^2}\right)\ln\left(\frac{-s_{13}}{-m_H^2}\right) \notag \\
&-2\,\mathrm{Li}_2\left(1-\frac{s_{12}}{m_H^2}\right)-2\,\mathrm{Li}_2\left(1-\frac{s_{23}}{m_H^2}\right)-2\,\mathrm{Li}_2\left(1-\frac{s_{13}}{m_H^2}\right).
\end{align}

\section{Conclusion} \label{Conclusion}

Four-dimensional spinor helicity has proven to be a powerful tool over the years, both for analytic and numerical computations.  It also exposes surprising structures in scattering amplitudes.  This motivates combining the six-dimensional spinor-helicity scheme of Cheung and O'Connell with $D$-dimensional generalized unitarity.  In this paper we presented such a formalism as a means for simultaneously calculating cut-constructible and rational pieces of one-loop amplitudes in QCD.  We illustrated this by computing various sample one-loop amplitudes in QCD and Higgs physics.

We obtain the amplitudes by finding the coefficients of a set of basis integrals by sewing together on-shell six-dimensional tree amplitudes and writing our answers in terms four-dimensional spinors through a decomposition of their six-dimensional counterparts.  We extract coefficients from their limiting behavior, taking advantage of the analytic properties of the loop integrals, following by now standard procedures.  Finally, to readjust the state sums to their proper values, we perform a necessary reduction procedure for internal particles with extra spin states in six dimensions.  This is accomplished by subtracting twice the contribution of cuts with all internal gluons replaced by real scalars.  To verify our formalism, we confirmed our sample calculations against known results.

Numerical implementation should also be possible and would be interesting to explore and compare to other available methods.  It should also be possible to apply these $D$-dimensional techniques to amplitudes with masses in the loops.  Our interpretation of the extra-dimensional pieces as mass terms makes such computations natural in our formalism, though a modification of the integral basis to include the dropped pentagon and box integrals, as well as tadpoles, would be necessary.

In summary, although many new tools are available, we believe six-dimensional helicity coupled with the unitarity method to be an additional powerful method for obtaining phenomenologically important loop amplitudes.

\section*{Acknowledgements}

I would like to thank Josh Samani for collaboration in the early stages of this work.  I would also like to thank Tristan Dennen, Lance Dixon, Yu-tin Huang, Harald Ita and Kemal Ozeren for many interesting and useful discussions.  I am especially grateful to Zvi Bern for countless ideas and discussions throughout the entire course of the work.

\appendix

\section{$(4-2\epsilon)$-DIMENSIONAL INTEGRAL BASIS} \label{IntBasisApp}

In this appendix we provide details behind the integral basis given in section \ref{IntBasis}.  A quick review of the Feynman rules shows that a general one-loop color-ordered amplitude for $n$ particles in $D$ dimensions, where all internal particles are massless, can be written as
\begin{equation}
A_n^{(1)}=\int\frac{d^Dl}{(2\pi)^D}\frac{N(\{p_i\},l)}{l^2(l-K_1)^2\ldots(l+K_n)^2},
\label{eq:genAmpl}
\end{equation}
where the numerator function $N$ contains all information from external momenta and polarization states, as well as tensor structure from the loop momenta.  We restrict external momenta to be in four dimensions while allowing internal momenta to be in arbitrary $D$.  Using $D$-dimensional Passarino-Veltman reduction techniques, we can then reduce eq.~\eqref{eq:genAmpl} to a scalar integral basis with $D$-dimensional rational coefficients with at most a pentagon integral \cite{Badger, FullOneLoopGiele, PassarinoVeltman},
\begin{align}
A_n^{(1)}=\frac{1}{(4\pi)^{D/2}}\left(\sum_{K_5}\right.&C_{5;K_5}(D)I_{5;K_5}^D+\sum_{K_4}C_{4;K_4}(D)I_{4;K_4}^D \notag \\
&\left.+\sum_{K_3}C_{3;K_3}(D)I_{3;K_3}^D+\sum_{K_2}C_{2;K_2}(D)I_{2;K_2}^D\right),
\end{align}
where $K_r$ refers to the set of all ordered partitions of the external momenta into $r$ distinct groups.  Passarino-Veltman reduction techniques were first used in ref.~\cite{PassarinoVeltman} and a useful outline can be found in a recent review \cite{BrittoReview}.

We are focused on internal loop momenta in $D=4-2\epsilon$.  With all external momenta and polarization vectors in four dimensions, any contractions that they have with loop momenta result in four-dimensional objects.  The $D\neq 4$ dependence of the numerators can then only arise through contractions of the loop momentum with itself,
\begin{equation}
l^2=\bar{l}^2+\tilde{l}^2=\bar{l}^2-\mu^2,
\label{eq:MoSq}
\end{equation}
where $\bar{l}$ contains the four-dimensional components of the loop momentum and $\tilde{l}$ refers to the $(-2\epsilon)$-dimensional components.  We see then that any dimensional dependence of the numerators arises only through dependence on $\mu^2$.  In a renormalizable gauge theory (such as QCD), the maximum number of powers of loop momentum appearing in the numerator of an $n$-point tensor integral is $n$, so the boxes can have at most a $\mu^4$ while the triangles and bubbles can have up to a $\mu^2$.  Tensor pentagon integrals are lost in the reduction, so we are left only with the scalar pentagon.  Our basis for $D=4-2\epsilon$ then becomes \cite{Badger}
\begin{align}
A_n^{(1)}=&\frac{\mu^{2\epsilon}}{(4\pi)^{2-\epsilon}}\left(\sum_{K_5}C_{5;K_5}^{[0]}I_{5;K_5}^{4-2\epsilon}+\sum_{K_4}C_{4;K_4}^{[0]}I_{4;K_4}^{4-2\epsilon}+\sum_{K_4}C_{4;K_4}^{[2]}I_{4;K_4}^{4-2\epsilon}[\mu^2]+\sum_{K_4}C_{4;K_4}^{[4]}I_{4;K_4}^{4-2\epsilon}[\mu^4]\right. \notag \\
&\left.+\sum_{K_3}C_{3;K_3}^{[0]}I_{3;K_3}^{4-2\epsilon}+\sum_{K_3}C_{3;K_3}^{[2]}I_{3;K_3}^{4-2\epsilon}[\mu^2]+\sum_{K_2}C_{2;K_2}^{[0]}I_{2;K_2}^{4-2\epsilon}+\sum_{K_2}C_{2;K_2}^{[2]}I_{2;K_2}^{4-2\epsilon}[\mu^2]\right).
\end{align}
Further decomposition of our basis is possible by writing the scalar pentagon integral in terms of scalar box integrals \cite{DimRegInts, PentInts},
\begin{align}
I_5^{4-2\epsilon}&=-\sum_{i=1}^5c_iI_5^{4-2\epsilon}[\mu^2]+\frac{1}{2}\sum_{i=1}^5c_iI_{4}^{(i),4-2\epsilon} \notag \\
&=\frac{1}{2}\sum_{i=1}^5c_iI_{4}^{(i),4-2\epsilon}+O(\epsilon), \label{eq:pentDecomp}
\end{align}
where the superscript $(i)$ is used to refer to the box obtained from the pentagon by removing the propagator between legs $i$ and $i-1$.  The form of the factors $c_i$ is unimportant because they are absorbed in the box integral coefficients in our basis, which becomes
\begin{align}
A_n^{(1)}=&\frac{\mu^{2\epsilon}}{(4\pi)^{2-\epsilon}}\left(\sum_{K_4}C_{4;K_4}^{[0]}I_{4;K_4}^{4-2\epsilon}+\sum_{K_4}C_{4;K_4}^{[4]}I_{4;K_4}^{4-2\epsilon}[\mu^4]+\sum_{K_3}C_{3;K_3}^{[0]}I_{3;K_3}^{4-2\epsilon}\right. \notag \\
&\left.+\sum_{K_3}C_{3;K_3}^{[2]}I_{3;K_3}^{4-2\epsilon}[\mu^2]+\sum_{K_2}C_{2;K_2}^{[0]}I_{2;K_2}^{4-2\epsilon}+\sum_{K_2}C_{2;K_2}^{[2]}I_{2;K_2}^{4-2\epsilon}[\mu^2]\right)+O(\epsilon),
\end{align}
where $\sum_{K_4}C_{4;K_4}^{[0]}I_{4;K_4}^{4-2\epsilon}$ has absorbed terms due to the pentagon integral via eq.~\eqref{eq:pentDecomp}.\footnote{It should be noted that eq.~\eqref{eq:pentDecomp} is only valid for $D=4-2\epsilon$.  In higher dimensions, the pentagon is independent and its coefficient must be found separately.  An additional extraction procedure is then required to remove its contribution to the box.  For the integral basis presented here, however, no such extraction should be performed as the box is meant to include the pentagon contribution, even for amplitudes with many legs.}  In the same way as $I_5^{4-2\epsilon}[\mu^2]$, $I_4^{4-2\epsilon}[\mu^2]$ is of order $\epsilon$ and has been pushed into the $O(\epsilon)$ term.

This recovers the basis in eq.~\eqref{eq:intBasis}.  As mentioned in section \ref{IntBasis}, evaluations of the scalar integrals can be found in ref.~\cite{BDDK2, PentInts}.  The integrals with $f(\mu^2)\neq 1$ evaluate to
\begin{align}
I_4^{4-2\epsilon}[\mu^4]&\overset{\epsilon\rightarrow 0}{\rightarrow}-\frac{1}{6}, \notag \\
I_3^{4-2\epsilon}[\mu^2]&\overset{\epsilon\rightarrow 0}{\rightarrow}-\frac{1}{2}, \notag \\
I_{2;s}^{4-2\epsilon}[\mu^2]&\overset{\epsilon\rightarrow 0}{\rightarrow}-\frac{1}{6}s,
\end{align}
where $I_{2;s}^{4-2\epsilon}[\mu^2]$ has an invariant mass square $s$ flowing through its external legs (when the loop momenta are massive, $I_{2;s}^{4-2\epsilon}[\mu^2]$ contains additional terms).  These can then be used to express the amplitude in terms of another common basis involving scalar integrals and rational terms \cite{ScalarBasis1, ScalarBasis2, FullOneLoopGiele},
\begin{align}
A_n^{(1)}=&\frac{\mu^{2\epsilon}}{(4\pi)^{2-\epsilon}}\left(\sum_{K_4}C_{4;K_4}^{[0]}I_{4;K_4}^{4-2\epsilon}+\sum_{K_3}C_{3;K_3}^{[0]}I_{3;K_3}^{4-2\epsilon}+\sum_{K_2}C_{2;K_2}^{[0]}I_{2;K_2}^{4-2\epsilon}\right)+\frac{1}{(4\pi)^2}R_n+ O(\epsilon), \label{eq:intBasisRat}
\end{align}
where
\begin{align}
R_n=-\frac{1}{6}\sum_{K_4}C_{4;K_4}^{[4]}-\frac{1}{2}\sum_{K_3}C_{3;K_3}^{[2]}-\frac{1}{6}\sum_{K_2}K_2^2C_{2;K_2}^{[2]}.
\end{align}
In either basis finding the one-loop amplitude comes down to finding the rational coefficients of the non-vanishing integrals.

\section{INTEGRAL COEFFICIENT EXTRACTION DETAILS} \label{ExtractAppend}

In this appendix, we expand upon our summary of coefficient integral extraction.  The loop-momentum parametrizations are given in section \ref{IntExtract}, but here we provide more details regarding the formulas for the coefficients.  We follow Badger \cite{Badger} directly for coefficients of integrals contributing to the rational piece of the amplitude, while simply taking $\mu^2\rightarrow 0$ as appropriate to find the coefficients of the scalar integrals.

\subsection{Box integral coefficients} \label{BoxesApp}

When we perform a unitarity cut, we make the replacements,
\begin{align}
\frac{i}{(l-K_i)^2}\rightarrow(2\pi)\delta((l-K_i)^2),
\end{align}
in the integral basis.  Only certain integrals contain cuts depending on how many we perform.  Unitarity tells us that the cut integrands are proportional to products of on-shell tree amplitudes, so we must decompose these products to match their pieces with cut integrals, thereby determining the integral coefficients.

The quadruple cut is given by
\begin{align}
(4\pi)^{2-\epsilon}(-2\pi i)^4&\int\frac{d^{4-2\epsilon}l_1}{(2\pi)^{4-2\epsilon}}\prod_{i=1}^4\delta(l_i^2)A_1A_2A_3A_4 \notag \\
&=(4\pi)^{2-\epsilon}(-2\pi i)^4\int\frac{d^{-2\epsilon}\mu}{(2\pi)^{-2\epsilon}}\int\frac{d^4\bar{l}_1}{(2\pi)^4}\prod_{i=1}^4\delta(\bar{l}_i^2-\mu^2)A_1A_2A_3A_4 \notag \\
&=(4\pi)^{2-\epsilon}\int\frac{d^{-2\epsilon}\mu}{(2\pi)^{-2\epsilon}}\sum_{\sigma}A_1A_2A_3A_4(\bar{l}_1^{\sigma}),
\label{eq:4partCut}
\end{align}
where the sum over $\sigma$ denotes the sum over the two solutions to the on-shell conditions.  Only pentagon and box integrals contain quadruple cuts, so
\begin{align}
(4\pi)^{2-\epsilon}&\int\frac{d^{-2\epsilon}\mu}{(2\pi)^{-2\epsilon}}\frac{1}{2}\sum_{\sigma}A_1A_2A_3A_4(\bar{l}_1^{\sigma}) \notag \\
&=\sum_{K_5}C_{5;K_5}^{[2]}I_{5;K_5}^{\mathrm{cut}}[\mu^2]+\sum_{K_4}C_{4;K_4}^{[0]}I_{4;K_4}^{\mathrm{cut}}+\sum_{K_4}C_{4;K_4}^{[2]}I_{4;K_4}^{\mathrm{cut}}[\mu^2]+\sum_{K_4}C_{4;K_4}^{[4]}I_{4;K_4}^{\mathrm{cut}}[\mu^4],
\end{align}
where $I_n^{\mathrm{cut}}[f(\mu^2)]$ is the quadruple-cut version of the $(4-2\epsilon)$-dimensional integral in eq.~\eqref{Ints}, and a factor of $\frac{1}{2}$ has been included to account for the averaging over solutions.  In simple cases, the integrand is a polynomial in $\mu^2$.  This is the case for four-point amplitudes with massless particles, which also do not contain a pentagon contribution.  After integrating the cut integrals over the delta functions, we can easily identify our coefficients by matching orders of $\mu^2$ in
\begin{align}
C_4^{[0]}+\mu^2C_4^{[2]}+\mu^4C_4^{[4]}&=\frac{i}{2}\sum_{\sigma}A_1A_2A_3A_4(\bar{l}_1^{\sigma}).
\label{eq:4ptBoxCoeffApp}
\end{align}
Even in cases where we have a nonzero $C_5^{[2]}$, it would not have to be disentangled from $C_4^{[2]}$ since the integrals associated with them are of order $\epsilon$.  In more general cases, the quadruple cut does not fall naturally into a polynomial in $\mu^2$, and we must identify each coefficient.  We can find $C_4^{[0]}$, our cut-constructible piece, by taking $\mu^2\rightarrow 0$,
\begin{align}
C_4^{[0]}&=\frac{i}{2}\sum_{\sigma}\left.A_1A_2A_3A_4(\bar{l}_1^{\sigma})\right|_{\mu^2\rightarrow 0},
\label{eq:C40App}
\end{align}
which effectively mimics a four-dimensional generalized unitarity procedure.  To find $C_4^{[4]}$, following ref.~\cite{Badger}, we express our integral with a term containing the behavior at infinity and residue terms contained in a sum over poles in $\mu^2$,
\begin{align}
(4\pi)&{}^{2-\epsilon}\int\frac{d^{-2\epsilon}\mu}{(2\pi)^{-2\epsilon}}\sum_{\sigma}A_1A_2A_3A_4(\bar{l}_1^{\sigma}) \notag \\
&=(4\pi)^{2-\epsilon}\int\frac{d^{-2\epsilon}\mu}{(2\pi)^{-2\epsilon}}\sum_{\sigma}\left([\mathrm{Inf}_{\mu^2}A_1A_2A_3A_4(\bar{l}_1^{\sigma})](\mu^2)+\sum_{\mathrm{poles}\{i\}}\frac{\mathrm{Res}_{\mu^2=\mu_i^2}A_1A_2A_3A_4(\bar{l}_1^{\sigma})}{\mu^2-\mu_i^2}\right)\!.
\end{align}
As mentioned in section \ref{IntExtract}, the Inf term can be expressed as a polynomial in $\mu^2$ with maximum order $\mu^4$,
\begin{align}
[\mathrm{Inf}_{\mu^2}A_1A_2A_3A_4](\mu^2)=\sum_{i=0}^2c_i\mu^{2i}.
\end{align}
To express the residue terms as a polynomial in $\mu^2$, we partial fraction and find at most a term of order $\mu^2$,
\begin{align}
\frac{\mathrm{Res}_{\mu^2=\mu_i^2}(A_1A_2A_3A_4(\bar{l}_1^{\sigma}))}{\mu^2-\mu_i^2}=\frac{\mu^2\mathrm{Res}_{\mu^2=\mu_i^2}(A_1A_2A_3A_4(\bar{l}_1^{\sigma}))}{\mu_i^2(\mu^2-\mu_i^2)}-\frac{\mathrm{Res}_{\mu^2=\mu_i^2}(A_1A_2A_3A_4(\bar{l}_1^{\sigma}))}{\mu_i^2}.
\label{eq:PartialApp}
\end{align}
Evidently the first term in eq.~\eqref{eq:PartialApp} combines with the order $\mu^2$ term in the boundary polynomial to give $C_4^{[2]}$ entangled with the cut pentagon, and the second term combines with the order $\mu^0$ term to give $C_4^{[0]}$.  $C_4^{[0]}$ could be found by combining these terms, but eq.~\eqref{eq:C40App} is much simpler.  The crucial point is that only the boundary term has a term of order $\mu^4$, so we find
\begin{align}
C_4^{[4]}&=\frac{i}{2}\sum_{\sigma}\left.[\mathrm{Inf}_{\mu^2}A_1A_2A_3A_4](\mu^2)\right|_{\mu^4},
\label{eq:BoxExtractApp}
\end{align}
where the Inf operator has been restricted to the coefficient of the $\mu^4$ term.  Along with eq.~\eqref{eq:C40App}, we have the box coefficient extraction formulas in eq.~\eqref{eq:BoxExtract}.

\subsection{Triangle integral coefficients} \label{TrianglesApp}

Recall that the triple-cut momentum solution has a free complex parameter $t$,
\begin{align}
\bar{l}_1^{\mu}&=aK_3^{\flat\mu}+bK_1^{\flat\mu}+\frac{t}{2}\langle K_3^{\flat -}|\gamma^{\mu}|K_1^{\flat -}\rangle+\frac{\gamma_{13}ab-\mu^2}{2t\gamma_{13}}\langle K_1^{\flat -}|\gamma^{\mu}|K_3^{\flat -}\rangle.
\label{eq:TriCutApp}
\end{align}
We can then express a generic triple cut as
\begin{align}
(4\pi)^{2-\epsilon}&(-2\pi i)^3\int\frac{d^{4-2\epsilon}l_1}{(2\pi)^{4-2\epsilon}}\prod_{i=1}^3\delta(l_i^2)A_1A_2A_3 \notag \\
=&(4\pi)^{2-\epsilon}(-2\pi i)^3\int \frac{d^{-2\epsilon}\mu}{(2\pi)^{-2\epsilon}}\int \frac{d^4\bar{l}_1}{(2\pi)^4}\prod_{i=1}^3\delta(\bar{l}_i^2-\mu^2)A_1A_2A_3 \notag \\
=&i(4\pi)^{2-\epsilon}\int \frac{d^{-2\epsilon}\mu}{(2\pi)^{-2\epsilon}}\int dt J_t\sum_{\sigma}\left([\mathrm{Inf}_tA_1A_2A_3(\bar{l}_1^{\sigma})](t)+\sum_{\mathrm{poles}\,\{j\}}\frac{\mathrm{Res}_{t=t_j}A_1A_2A_3(\bar{l}_1^{\sigma})}{t-t_j}\right).
\end{align}
In going from the second line to the last line, we have performed the integral transformation from $l^{\mu}$ to $t$ and integrated over the three delta functions (thereby applying the on-shell constraints which give rise to the sum over solutions, including solutions for $\gamma_{13}$ and the conjugate momentum solution).  The Jacobian of the transformation is $J_t$ (which also absorbs a factor of $1/2\pi$).  As before, the Inf term contains the information from the boundary of the $t$ contour integral,
\begin{align}
\lim_{t\rightarrow\infty}([\mathrm{Inf}_tA_1A_2A_3](t)-A_1(t)A_2(t)A_3(t))=0,
\end{align}
while the second term is a sum over poles in $t$.

Poles in $t$ in the cut integrand come from propagator terms of the form $1/(l-K)^2$.  The only contributions from the residue terms occur where $t=t_j$, which corresponds to putting another propagator on shell $(l-K)^2=(\bar{l}-K)^2-\mu^2=0$.  Therefore the second term contains only box and pentagon coefficients, and we can ignore it in our search for triangle coefficients,
\begin{align}
(4\pi)^{2-\epsilon}&(-2\pi i)^3\int\frac{d^{4-2\epsilon}l}{(2\pi)^{4-2\epsilon}}\prod_{i=1}^3\delta(l_i^2)A_1A_2A_3 \notag \\
&=i(4\pi)^{2-\epsilon}\int \frac{d^{-2\epsilon}\mu}{(2\pi)^{-2\epsilon}}\int dt J_t\sum_{\sigma}[\mathrm{Inf}_tA_1A_2A_3(\bar{l}_1^{\sigma})](t) +\mathrm{box/pent.~terms}.
\end{align}

For the first term, much like before we can expand around infinity and write it as a polynomial in $t$,
\begin{align}
[\mathrm{Inf}_tA_1A_2A_3](t)=\sum_{i=0}^mf_it^i.
\end{align}
Because of our particular parametrization of the loop momentum \eqref{eq:TriCutApp}, all integrals over $t^n$ for $n\neq 0$ vanish \cite{Forde},
\begin{align}
\int dt\,J_tt^n=0,\,n\neq 0,
\end{align}
allowing us to drop all terms in the expansion not of order $t^0$.  Our triple cut is then
\begin{align}
(4\pi)^{2-\epsilon}&(-2\pi i)^3\int\frac{d^{4-2\epsilon}l}{(2\pi)^{4-2\epsilon}}\prod_{i=1}^3\delta(l_i^2)A_1A_2A_3 \notag \\
&=i(4\pi)^{2-\epsilon}\int \frac{d^{-2\epsilon}\mu}{(2\pi)^{-2\epsilon}}\sum_{\sigma}f_0(\bar{l}_1^{\sigma})\int dt J_t + \mathrm{box/pent.~terms} \notag \\
&=i(4\pi)^{2-\epsilon}\int \frac{d^{-2\epsilon}\mu}{(2\pi)^{-2\epsilon}}\sum_{\sigma}\left[[\mathrm{Inf}_tA_1A_2A_3(\bar{l}_1^{\sigma})](t)|_{t\rightarrow 0}\right]\int dt J_t + \mathrm{box/pent.~terms}.
\end{align}
The triple-cut triangle integral is precisely
\begin{align}
-(4\pi)^{2-\epsilon}\int\frac{d^{-2\epsilon}\mu}{(2\pi)^{-2\epsilon}}f(\mu^2)\int dt\,J_t=i(4\pi)^{2-\epsilon}(-2\pi i)^3\int\frac{d^{4-2\epsilon}l_1}{(2\pi)^{4-2\epsilon}}f(\mu^2)\prod_{i=1}^3\delta(l_i^2),
\end{align}
which allows us to identify the triangle coefficients and the appropriate sign.  For simple cases where the Inf term falls into a polynomial in $\mu^2$, we can match orders of $\mu$.  Averaging over solutions (both of $\gamma_{13}$ and the conjugate momentum solution for fixed $\gamma_{13}$), we have
\begin{align}
C_3^{[0]}+\mu^2C_3^{[2]}&=-\frac{1}{2n_{\gamma}}\sum_{\sigma}\left.[\mathrm{Inf}_tA_1A_2A_3(\bar{l}_1^{\sigma})](t)\right|_{t\rightarrow 0},
\end{align}
where $n_{\gamma}$ denotes the number of solutions for $\gamma_{13}$, which is either $1$ or $2$.  In more general cases where $\left.[\mathrm{Inf}_tA_1A_2A_3(\bar{l}_1^{\sigma})](t)\right|_{t\rightarrow 0}$ is not a polynomial in $\mu^2$, we use a procedure similar to that of section \ref{BoxesApp}.  We take $\mu^2\rightarrow 0$ for $C_3^{[0]}$, and for $C_3^{[2]}$ we use a polynomial expansion in $\mu^2$,
\begin{align}
\left.[\mathrm{Inf}_{\mu^2}[\mathrm{Inf}_tA_1A_2A_3(\bar{l}_1^{\sigma})](t)](\mu^2)\right|_{t\rightarrow 0}=\sum_{i=0}^1 c_i\mu^{2i},
\end{align}
and restrict to the coefficient of the $\mu^2$ term, recovering eq.~\eqref{eq:TriCoeffs},
\begin{align}
C_3^{[0]}&=-\frac{1}{2n_{\gamma}}\sum_{\sigma}\left.[\mathrm{Inf}_tA_1A_2A_3(\bar{l}_1^{\sigma})](t)\right|_{\mu^2\rightarrow 0,t\rightarrow 0}, \notag \\
C_3^{[2]}&=-\frac{1}{2n_{\gamma}}\sum_{\sigma}\left.[\mathrm{Inf}_{\mu^2}[\mathrm{Inf}_tA_1A_2A_3(\bar{l}_1^{\sigma})](t)](\mu^2)\right|_{\mu^2,t\rightarrow 0}.
\end{align}

\subsection{Bubble integral coefficients} \label{BubblesApp}

The double-cut momentum solution \eqref{eq:DoubCut} has two free parameters $t$ and $y$, so the double cut is given by
\begin{align}
(4\pi)^{2-\epsilon}&(-2\pi i)^2\int\frac{d^{4-2\epsilon}l}{(2\pi)^{4-2\epsilon}}\prod_{i=1}^2\delta(l_i^2)A_1A_2 \notag \\
=&-(4\pi)^{2-\epsilon}\int\frac{d^{-2\epsilon}\mu}{(2\pi)^{-2\epsilon}}\int dt dy J_{t,y}\sum_{\sigma}\left([\mathrm{Inf}_yA_1A_2(\bar{l}_1^{\sigma})](y)+\sum_{\mathrm{poles}\{j\}}\frac{\mathrm{Res}_{y=y_j}A_1A_2(\bar{l}_1^{\sigma})}{y-y_j}\right) \notag \\
=&-(4\pi)^{2-\epsilon}\int\frac{d^{-2\epsilon}\mu}{(2\pi)^{-2\epsilon}}\int dt dy J_{t,y} \notag \\
&\times\sum_{\sigma}\left([\mathrm{Inf}_t[\mathrm{Inf}_yA_1A_2(\bar{l}_1^{\sigma})](y)](t)+\left[\mathrm{Inf}_t\left(\sum_{\mathrm{poles}\{j\}}\frac{\mathrm{Res}_{y=y_j}A_1A_2(\bar{l}_1^{\sigma})}{y-y_j}\right)\right](t)\vphantom{\frac{\mathrm{Res}_{t=t_l}\left[\frac{\mathrm{Res}_{y=y_j}A_1A_2(\bar{l}_1^{\sigma}))}{y-y_j}\right]}{t-t_l}}\right. \notag \\
&\left.+\sum_{\mathrm{poles}\{l\}}\frac{\mathrm{Res}_{t=t_l}[\mathrm{Inf}_yA_1A_2(\bar{l}_1^{\sigma})](y)}{t-t_l}+\sum_{\mathrm{poles}\{j\},\{l\}}\frac{\mathrm{Res}_{t=t_l}\left[\frac{\mathrm{Res}_{y=y_j}A_1A_2(\bar{l}_1^{\sigma}))}{y-y_j}\right]}{t-t_l}\right),
\label{eq:twoPartCutApp}
\end{align}
where we have integrated over the delta functions and performed an integral transformation from $l^{\mu}$ to $t$ and $y$, similar to our treatment of the triple cut.  The last term of the final expression has two additional propagators on shell, and its numerator has no dependence on $t$ or $y$.  It therefore corresponds to box and pentagon terms only.  We might then believe that the second and third terms correspond to triangle coefficients only and grab the first term for the bubbles, but the fact that there is dependence on the free parameters $t$ and $y$ in the numerators is significant.

These single residue terms have a third propagator on shell, which indeed allows us to correspond them to a triple cut.  We can, without loss of generality, fix $y$ to satisfy the constraint and use a triple cut to find the residue terms rather than computing them directly.  However, this parametrization of the loop momentum may differ from that of section \ref{TrianglesApp} depending on our choice of $\chi$.  With the new parametrization, integrals over positive powers of $t$ that conveniently vanished in section \ref{TrianglesApp} will not necessarily vanish here,
\begin{align}
\int dt J_t' t^n\neq 0,
\end{align}
where $J_t'$ is the Jacobian corresponding our new loop momentum parametrization.  This effectively means that there are powers of the loop momentum (containing more than just the extra-dimensional pieces $\mu$) in the numerators, and our triple cut produces tensor triangle integrals.  These tensor integrals are not in our integral basis and must be reduced.  Passarino-Veltman reduction shows us that there are both scalar triangle and scalar bubble integrals within these tensor integrals,
\begin{align}
\int dt J_t' t^n=C_3I_3^{4;\mathrm{cut}}+C_2I_2^{4;\mathrm{cut}},
\end{align}
so we must extract the bubble integral coefficients.

Imposing a third on-shell condition in our momentum parametrization, eq.~\eqref{eq:DoubCut} gives for $y$,
\begin{align}
y_{\pm}=\frac{B_1\pm\sqrt{B_1^2+4B_0B_2}}{2B_2},
\label{eq:ySolnsApp}
\end{align}
where
\begin{align}
B_2&=S_1\langle \chi^-|\s{K}_3|K_1^{\flat -}\rangle, \notag \\
B_1&=\bar\gamma t\langle K_1^{\flat -}|\s{K}_3|K_1^{\flat -}\rangle-S_1t\langle\chi^-|\s{K}_3|\chi^-\rangle+S_1\langle\chi^-|\s{K}_3|K_1^{\flat -}\rangle, \notag \\
B_0&=\bar\gamma t^2\langle K_1^{\flat -}|\s{K}_3|\chi^-\rangle-\mu^2\langle\chi^-|\s{K}_3|K_1^{\flat -}\rangle+\bar\gamma t S_3+t S_1\langle\chi^-|\s{K}_3|\chi^-\rangle.
\end{align}
After dropping box and pentagon terms, the triple cut is given by
\begin{align}
i(4\pi)^{2-\epsilon}\int\frac{d^{-2\epsilon}\mu}{(2\pi)^{-2\epsilon}}\sum_{\sigma_y}\int dt J_t'[\mathrm{Inf}_tA_1A_2A_3(\bar{l}_1^{\sigma_y})](t),
\end{align}
where the sum is over the solutions in eq.~\eqref{eq:ySolnsApp}.  Once again we can expand around $t\rightarrow\infty$,
\begin{align}
[\mathrm{Inf}_tA_1A_2A_3](t)=\sum_{i=0}^mf_it^i,
\end{align}
to perform our integrals.  The $t^0$ term gives us the double-cut scalar triangle, while positive powers of $t$ return, among other pieces, cut scalar bubble coefficients. We evaluate the integrals over positive powers of $t$ and retain only the contributing bubble integral to our particular double cut,
\begin{align}
(4\pi)^2\int dt J_t't^j&=T_jI_2^{4;\mathrm{cut}} \notag \\
&=-\left(\frac{S_1}{\bar{\gamma}}\right)^j\frac{\langle\chi^-|\s{K}_3|K_1^{\flat,-}\rangle^j(K_1\cdot K_3)^{j-1}}{\Delta^j}\left(\sum_{l=1}^j\mathrm{\calligra{C}}_{jl}\frac{S_3^{l-1}}{(K_1\cdot K_3)^{l-1}}\right)I_2^{4;\mathrm{cut}},
\end{align}
where
\begin{align}
\mathrm{\calligra{C}}_{11}&=\frac{1}{2}, \notag \\
\mathrm{\calligra{C}}_{21}&=\frac{3}{8},\hspace{8mm} \mathrm{\calligra{C}}_{22}=\frac{3}{8}, \notag \\
\mathrm{\calligra{C}}_{31}&=-\frac{1}{12}\frac{\Delta}{(K_1\cdot K_3)^2}\left(1-4\frac{\mu^2}{S_1}\right)+\frac{5}{16},\hspace{8mm} \mathrm{\calligra{C}}_{32}=\frac{5}{8},\hspace{8mm} \mathrm{\calligra{C}}_{33}=\frac{5}{16}, \notag
\end{align}
\begin{align}
&\Delta=(K_1\cdot K_3)^2-S_1S_3, \notag \\
I_2^{4;\mathrm{cut}}=(&-i)(4\pi)^2(-2\pi i)^2\int\frac{d^4\bar{l}_1}{(2\pi)^4}\prod_{i=1}^2\delta(\bar{l}_i).
\end{align}
For QCD we can have terms up to order $t^3$, so our relevant terms are
\begin{align}
T_1&=-\frac{S_1\langle\chi^-|\s{K}_3|K_1^{\flat -}\rangle}{2\bar\gamma\Delta}, \notag \\
T_2&=-\frac{3S_1\langle\chi^-|\s{K}_3|K_1^{\flat -}\rangle^2}{8\bar{\gamma}^2\Delta^2}(S_1S_3+K_1\cdot K_3S_1), \notag \\
T_3&=-\frac{\langle\chi^-|\s{K}_3|K_1^{\flat -}\rangle^3}{48\bar{\gamma}^3\Delta^3}\left(15 S_1^3S_3^2+30 K_1\cdot K_3 S_1^3S_3+11(K_1\cdot K_3)^2S_1^3+4 S_1^4 S_3+16\mu^2 S_1^2\Delta\right),
\end{align}
recovering eq.~\eqref{eq:T}.  We then have the contribution of the single residue terms to the bubble coefficient.  Defining $T_0=0$, we have
\begin{align}
-\frac{1}{2}\sum_{\sigma_y}[\mathrm{Inf}_tA_1A_2A_3](t)|_{t^j\rightarrow T_j},
\end{align}
after averaging over solutions.  Single residue terms arise from every possible third leg that can go on-shell, i.e. every triple cut that shares two of its cuts with our two-particle cut, so we must find a term for every such possible configuration.

Lastly we need the bubble coefficients due to the first term in eq.~\eqref{eq:twoPartCutApp},
\begin{align}
-(4\pi)^{2-\epsilon}\int\frac{d^{-2\epsilon}\mu}{(2\pi)^{-2\epsilon}}\int dt dy J_{t,y}\sum_{\sigma}[\mathrm{Inf}_t[\mathrm{Inf}_yA_1A_2(\bar{l}_1^{\sigma})](y)](t).
\end{align}
Once again we expand around infinity to evaluate these integrals.  Because of our choice of momentum parametrization, integrals of positive order in $t$ vanish.  This is not the case with $y$, so we must keep terms of the form $t^0y^m$.  These evaluate to
\begin{align}
Y_m&=(2\pi)^2\int dt dy J_{t,y} y^m \notag \\
&=\frac{1}{m+1}\sum_{i=0}^{\lfloor m/2\rfloor}\left(\begin{array}{cc} m-i \\ i \end{array}\right)\left(\frac{-\mu^2}{S_1}\right)^i.
\end{align}
QCD can give terms up to order $y^4$, so our relevant integrals are
\begin{align}
Y_0=1, \hspace{4mm} Y_1=\frac{1}{2}, \hspace{4mm} Y_2=\frac{1}{3}\left(1-\frac{\mu^2}{S_1}\right), \hspace{4mm} Y_3=\frac{1}{4}\left(1-2\frac{\mu^2}{S_1}\right), \hspace{4mm} Y_4=\frac{1}{5}\left(1-3\frac{\mu^2}{S_1}+\frac{\mu^4}{S_1^2}\right).
\end{align}
The term is then
\begin{align}
-(4\pi)^{2-\epsilon}\int&\frac{d^{-2\epsilon}\mu}{(2\pi)^{-2\epsilon}}\int dt dy J_{t,y}\sum_{\sigma}[\mathrm{Inf}_t[\mathrm{Inf}_yA_1A_2(\bar{l}_1^{\sigma})](y)](t) \notag \\
&=-(4\pi)^{2-\epsilon}\int\frac{d^{-2\epsilon}\mu}{(2\pi)^{-2\epsilon}}\sum_{\sigma}\left[[\mathrm{Inf}_t[\mathrm{Inf}_yA_1A_2(\bar{l}_1^{\sigma})](y)](t)|_{t\rightarrow 0, y^m\rightarrow Y_m}\right]\int dt dy J_{t,y},
\end{align}
in which we identify the double-cut scalar bubble integral,
\begin{align}
i(4\pi)^{2-\epsilon}\int\frac{d^{-2\epsilon}\mu}{(2\pi)^{-2\epsilon}}\int dt dy J_{t,y}=(-i)(4\pi)^{2-\epsilon}(-2\pi i)^2\int\frac{d^{4-2\epsilon}l_1}{(2\pi)^{4-2\epsilon}}\prod_{i=1}^2\delta(l_i^2).
\end{align}
Adjusting a factor of $i$, we can easily identify the bubble coefficient from the first term of eq.~\eqref{eq:twoPartCutApp} as
\begin{align}
-i[\mathrm{Inf}_t[\mathrm{Inf}_yA_1A_2](y)](t)|_{t\rightarrow 0, y^m\rightarrow Y_m}.
\end{align}
We then have for our total bubble coefficient, in the case where our calculation falls into a polynomial in $\mu^2$,
\begin{align}
C_2^{[0]}+\mu^2C_2^{[2]}=-i[\mathrm{Inf}_t[\mathrm{Inf}_yA_1A_2](y)](t)|_{t\rightarrow 0, y^m\rightarrow Y_m}-\frac{1}{2}\sum_{C_{\mathrm{tri}}}\sum_{\sigma_y}[\mathrm{Inf}_tA_1A_2A_3](t)|_{t^j\rightarrow T_j},
\label{eq:BubbCoeffApp}
\end{align}
where $C_{\mathrm{tri}}$ denotes a sum over all possible triangles attainable from cutting one more leg of our two-particle cut.  Most generally, we have
\begin{align}
C_2^{[0]}=&-i[\mathrm{Inf}_t[\mathrm{Inf}_yA_1A_2](y)](t)|_{\mu^2\rightarrow 0,t\rightarrow 0, y^m\rightarrow Y_m}-\frac{1}{2}\sum_{C_{\mathrm{tri}}}\sum_{\sigma_y}[\mathrm{Inf}_tA_1A_2A_3](t)|_{\mu^2\rightarrow 0,t^j\rightarrow T_j}, \notag \\
C_2^{[2]}=&-i[\mathrm{Inf}_{\mu^2}[\mathrm{Inf}_t[\mathrm{Inf}_yA_1A_2](y)](t)](\mu^2)|_{\mu^2,t\rightarrow 0, y^m\rightarrow Y_m} \notag \\
&-\frac{1}{2}\sum_{C_{\mathrm{tri}}}\sum_{\sigma_y}[\mathrm{Inf}_{\mu^2}[\mathrm{Inf}_tA_1A_2A_3](t)](\mu^2)|_{\mu^2,t^j\rightarrow T_j},
\end{align}
in agreement with eq.~\eqref{eq:BubbCoeff}.

%%%%%%%%%%%%%%%%%%%%%%%%%%%%%%%%%%%%%%%%%%%%%%%%%%%%%%%%%%%%


\begin{thebibliography}{99}

%+% 2 refs
\bibitem{NLOMultileg}
  Z.~Bern {\it et al.} [NLO Multileg Working Group Collaboration],
  %``The NLO multileg working group: Summary report,''
  [arXiv:0803.0494 [hep-ph]].
  %%CITATION = ARXIV:0803.0494;%%

%+% 1 ref
\bibitem{BDDK1}
  Z.~Bern, L.~J.~Dixon, D.~C.~Dunbar and D.~A.~Kosower,
  %``One loop n point gauge theory amplitudes, unitarity and collinear limits,''
  Nucl.\ Phys.\ B\ {\bf 425}, 217  (1994)
  [hep-ph/9403226].
  %%CITATION = NUPHA,B425,217;%%

%+% 3 refs
\bibitem{BDDK2}
  Z.~Bern, L.~J.~Dixon, D.~C.~Dunbar and D.~A.~Kosower,
  %``Fusing gauge theory tree amplitudes into loop amplitudes,''
  Nucl.\ Phys.\ B\ {\bf 435}, 59  (1995)
  [hep-ph/9409265].
  %%CITATION = NUPHA,B435,59;%%

%+% 2 refs
\bibitem{Zqqgg}
  Z.~Bern, L.~J.~Dixon and D.~A.~Kosower,
  %``One-loop amplitudes for e+ e- to four partons,''
  Nucl.\ Phys.\  B {\bf 513}, 3 (1998)
  [hep-ph/9708239].
%%CITATION = NUPHA,B513,3;%%

%+% 7 refs
\bibitem{BernMorgan}
  Z.~Bern and A.~G.~Morgan,
  %``Massive loop amplitudes from unitarity,''
  Nucl.\ Phys.\  B {\bf 467}, 479 (1996)
  [hep-ph/9511336].
  %%CITATION = NUPHA,B467,479;%%

%+% 1 ref
\bibitem{genUnit}
  R.~Britto, F.~Cachazo and B.~Feng,
  %``Generalized unitarity and one-loop amplitudes in N=4 super-Yang-Mills,''
  Nucl.\ Phys.\  B {\bf 725}, 275 (2005)
  [hep-th/0412103].
  %%CITATION = NUPHA,B725,275;%%

%+% 3 refs
\bibitem{OPP1}
  G.~Ossola, C.~G.~Papadopoulos and R.~Pittau,
  %``Reducing full one-loop amplitudes to scalar integrals at the integrand
  %level,''
  Nucl.\ Phys.\  B {\bf 763}, 147 (2007)
  [hep-ph/0609007].
  %%CITATION = NUPHA,B763,147;%%

%+% 4 refs
\bibitem{Forde}
  D.~Forde,
  %``Direct extraction of one-loop integral coefficients,''
  Phys.\ Rev.\ D\ {\bf 75}, 125019  (2007)
  [arXiv:0704.1835 [hep-ph]].
  %%CITATION = PHRVA,D75,125019;%%

%+% 2 refs
\bibitem{AguilaPittau}
  F.~del Aguila and R.~Pittau,
  %``Recursive numerical calculus of one-loop tensor integrals,''
  JHEP {\bf 0407}, 017 (2004)
  [hep-ph/0404120].
  %%CITATION = JHEPA,0407,017;%%

%+% 2 refs
\bibitem{BlackHatI}
C.~F.~Berger {\it et al.},
% Z.~Bern, L.~J.~Dixon, F.~Febres~Cordero, D.~Forde, H.~Ita,
% D.~A.~Kosower, D.~Ma\^{\i}tre
%``An Automated Implementation of On-Shell Methods for One-Loop
%Amplitudes,''
Phys.\ Rev.\ D {\bf 78}, 036003 (2008)  [arXiv:0803.4180 [hep-ph]].
%%CITATION = PHRVA,D78,036003;%%

%+% 1 ref
\bibitem{RecentFeynman}
A.~Bredenstein, A.~Denner, S.~Dittmaier and S.~Pozzorini,
%``NLO QCD corrections to top anti-top bottom anti-bottom production at the
% LHC: 1. quark-antiquark annihilation,''
JHEP {\bf 0808}, 108 (2008) [arXiv:0807.1248 [hep-ph]],
%%CITATION = JHEPA,0808,108;%%
%
%A.~Bredenstein, A.~Denner, S.~Dittmaier and S.~Pozzorini,
%``NLO QCD corrections to pp -> t anti-t b anti-b + X at the LHC,''
Phys.\ Rev.\ Lett.\  {\bf 103}, 012002 (2009) [arXiv:0905.0110 [hep-ph]],
%%CITATION = PRLTA,103,012002;%%
%
%A.~Bredenstein, A.~Denner, S.~Dittmaier and S.~Pozzorini,
%``NLO QCD corrections to top anti-top bottom anti-bottom production at the
%LHC: 2. full hadronic results,''
JHEP {\bf 1003}, 021 (2010) [arXiv:1001.4006 [hep-ph]].
%%CITATION = JHEPA,1003,021;%%

%+% 1 ref
\bibitem{SpinorHelicity}
F.~A.~Berends, R.~Kleiss, P.~De Causmaecker, R.~Gastmans and T.~T.~Wu,
%``Single Bremsstrahlung Processes In Gauge Theories,''
Phys.\ Lett.\ B {\bf 103}, 124 (1981);\\
%%CITATION = PHLTA,B103,124;%%
%
P.~De Causmaecker, R.~Gastmans, W.~Troost and T.~T.~Wu,
%``Multiple Bremsstrahlung In Gauge Theories At High-Energies. 1. General
%Formalism For Quantum Electrodynamics,''
Nucl.\ Phys.\ B {\bf 206}, 53 (1982);\\
%%CITATION = NUPHA,B206,53;%%
%
Z.~Xu, D.~H.~Zhang and L.~Chang,
%``Helicity Amplitudes For Multiple Bremsstrahlung In Massless Nonabelian
% Gauge Theory. 1. New Definition Of Polarization Vector And Formulation Of
% Amplitudes In Grassmann Algebra,''
TUTP-84/3-TSINGHUA;\\
%\href{http://www.slac.stanford.edu/spires/find/hep/www?r=tutp-84\
%2F3-tsinghua}{SPIRES entry}
%
R.~Kleiss and W.~J.~Stirling,
%``Spinor Techniques For Calculating P Anti-P $\to$ W+- / Z0 + Jets,''
Nucl.\ Phys.\ B {\bf 262}, 235 (1985);\\
%%CITATION = NUPHA,B262,235;%%
%
J.~F.~Gunion and Z.~Kunszt,
% ``Improved Analytic Techniques For Tree Graph Calculations And The G G Q
% Anti-Q Lepton Anti-Lepton Subprocess,''
Phys.\ Lett.\ B {\bf 161}, 333 (1985);\\
%%CITATION = PHLTA,B161,333;%%
%
Z.~Xu, D.~H.~Zhang and L.~Chang,
%``Helicity Amplitudes For Multiple Bremsstrahlung In Massless Nonabelian
%Gauge Theories,''
Nucl.\ Phys.\ B {\bf 291}, 392 (1987).
%%CITATION = NUPHA,B291,392;%%

%+% 1 ref
\bibitem{ParkeTaylor}
  S.~J.~Parke, T.~R.~Taylor,
  %``An Amplitude for $n$ Gluon Scattering,''
  Phys.\ Rev.\ Lett.\  {\bf 56}, 2459 (1986);\\
  %%CITATION = PRLTA,56,2459;%%
  F.~A.~Berends, W.~T.~Giele,
  %``Recursive Calculations for Processes with n Gluons,''
  Nucl.\ Phys.\  {\bf B306}, 759 (1988).
%%CITATION = NUPHA,B306,759;%%

%+% 3 refs
\bibitem{Wp4Jets}
  C.~F.~Berger {\it et al.},
  %``Precise Predictions for W + 4 Jet Production at the Large Hadron
  %Collider,''
  Phys.\ Rev.\ Lett.\  {\bf 106}, 092001 (2011)
  [arXiv:1009.2338 [hep-ph]].
  %%CITATION = PRLTA,106,092001;%%

%+% 1 ref
\bibitem{recursionRational}
  Z.~Bern, L.~J.~Dixon and D.~A.~Kosower,
  %``On-shell recurrence relations for one-loop QCD amplitudes,''
  Phys.\ Rev.\  D {\bf 71}, 105013 (2005)
  [hep-th/0501240];\\
  %%CITATION = PHRVA,D71,105013;%%
  Z.~Bern, L.~J.~Dixon and D.~A.~Kosower,
  %``Bootstrapping multi-parton loop amplitudes in QCD,''
  Phys.\ Rev.\  D {\bf 73}, 065013 (2006)
  [hep-ph/0507005];\\
  %%CITATION = PHRVA,D73,065013;%%
  D.~Forde and D.~A.~Kosower,
  %``All-multiplicity amplitudes with massive scalars,''
  Phys.\ Rev.\  D {\bf 73}, 065007 (2006)
  [hep-th/0507292];\\
  %%CITATION = PHRVA,D73,065007;%%
  D.~Forde and D.~A.~Kosower,
  %``All-multiplicity one-loop corrections to MHV amplitudes in QCD,''
  Phys.\ Rev.\  D {\bf 73}, 061701 (2006)
  [hep-ph/0509358];\\
  %%CITATION = PHRVA,D73,061701;%%
  C.~F.~Berger, Z.~Bern, L.~J.~Dixon, D.~Forde and D.~A.~Kosower,
  %``All One-loop Maximally Helicity Violating Gluonic Amplitudes in QCD,''
  Phys.\ Rev.\  D {\bf 75}, 016006 (2007)
  [hep-ph/0607014].
  %%CITATION = PHRVA,D75,016006;%%

%+% 2 refs
\bibitem{BernLastOfTheFinite}
  Z.~Bern, L.~J.~Dixon, D.~A.~Kosower,
  %``The last of the finite loop amplitudes in QCD,''
  Phys.\ Rev.\  D {\bf 72}, 125003 (2005).
  [hep-ph/0505055].
  %%CITATION = PHRVA,D72,125003;%%

%+% 2 refs
\bibitem{BDDK3}
  Z.~Bern, L.~J.~Dixon, D.~C.~Dunbar and D.~A.~Kosower,
  %``One loop selfdual and N=4 superYang-Mills,''
  Phys.\ Lett.\  B {\bf 394}, 105 (1997)
  [hep-th/9611127].
  %%CITATION = PHLTA,B394,105;%%

%+% 1 ref
\bibitem{OtherDDimUnitarity}
R.~Britto, B.~Feng and P.~Mastrolia,
%``The cut-constructible part of QCD amplitudes,''
Phys.\ Rev.\  D {\bf 73}, 105004 (2006)
[hep-ph/0602178];\\
%%CITATION = PHRVA,D73,105004;%%
%
C.~Anastasiou, R.~Britto, B.~Feng, Z.~Kunszt and P.~Mastrolia,
%``D-dimensional unitarity cut method,''
Phys.\ Lett.\  B {\bf 645}, 213 (2007)
[hep-ph/0609191], JHEP {\bf 0703}, 111 (2007)
  [hep-ph/0612277];\\
%
%%CITATION = PHLTA,B645,213;%%
%C.~Anastasiou, R.~Britto, B.~Feng, Z.~Kunszt and P.~Mastrolia,
%``Unitarity cuts and reduction to master integrals in d dimensions for
%one-loop amplitudes,''
%%CITATION = HEP-PH/0612277;%%
%
P.~Mastrolia,
%``On triple-cut of scattering amplitudes,''
Phys.\ Lett.\  B {\bf 644}, 272 (2007)
[hep-th/0611091];\\
%%CITATION = PHLTA,B644,272;%%
%
R.~Britto, B.~Feng and G.~Yang,
%``Polynomial Structures in One-Loop Amplitudes,''
JHEP {\bf 0809}, 089 (2008)
[arXiv:0803.3147 [hep-ph]].
 %%CITATION = JHEPA,0809,089;%%

%+% 1 ref
\bibitem{RationFeynmanRules}
 P.~Draggiotis, M.~V.~Garzelli, C.~G.~Papadopoulos and R.~Pittau,
  %``Feynman Rules for the Rational Part of the QCD 1-loop amplitudes,''
  JHEP {\bf 0904}, 072 (2009)
  [arXiv:0903.0356 [hep-ph]].
  %%CITATION = JHEPA,0904,072;%%

%+% 1 ref
\bibitem{KunsztNumerical}
  R.~K.~Ellis, W.~T.~Giele and Z.~Kunszt,
  %``A Numerical Unitarity Formalism for Evaluating One-Loop Amplitudes,''
  JHEP {\bf 0803}, 003 (2008)
  [arXiv:0708.2398 [hep-ph]].
  %%CITATION = JHEPA,0803,003;%%

%+% 9 refs
\bibitem{FullOneLoopGiele}
  W.~T.~Giele, Z.~Kunszt and K.~Melnikov,
  %``Full one-loop amplitudes from tree amplitudes,''
  JHEP {\bf 0804}, 049 (2008)
  [arXiv:0801.2237 [hep-ph]].
  %%CITATION = JHEPA,0804,049;%%

%+% 1 ref
\bibitem{KunsztFermionsDUnitarity}
  R.~K.~Ellis, W.~T.~Giele, Z.~Kunszt and K.~Melnikov,
  %``Masses, fermions and generalized $D$-dimensional unitarity,''
  Nucl.\ Phys.\  B {\bf 822}, 270 (2009)
  [arXiv:0806.3467 [hep-ph]].
  %%CITATION = NUPHA,B822,270;%%

%+% 12 refs
\bibitem{Badger}
  S.~D.~Badger,
  %``Direct Extraction Of One Loop Rational Terms,''
  JHEP {\bf 0901}, 049 (2009)
  [arXiv:0806.4600 [hep-ph]].
  %%CITATION = JHEPA,0901,049;%%

%+% 1 ref
\bibitem{Wp3Jets}
C.~F.~Berger {\it et al.},
%``Next-to-Leading Order QCD Predictions for W+3-Jet Distributions at Hadron
%Colliders,''
Phys.\ Rev.\  D {\bf 80}, 074036 (2009) [arXiv:0907.1984 [hep-ph]].
%%CITATION = PHRVA,D80,074036;%%

%+% 10 refs
\bibitem{COC6D}
  C.~Cheung and D.~O'Connell,
  %``Amplitudes and Spinor-Helicity in Six Dimensions,''
  JHEP {\bf 0907}, 075 (2009)
  [arXiv:0902.0981 [hep-th]].
  %%CITATION = JHEPA,0907,075;%%

%+% 1 ref
\bibitem{BernFDH}
Z.~Bern and D.~A.~Kosower,
%``The Computation of loop amplitudes in gauge theories,''
Nucl.\ Phys.\  B {\bf 379}, 451 (1992);\\
%%CITATION = NUPHA,B379,451;%%
%
Z.~Bern, A.~De Freitas, L.~J.~Dixon and H.~L.~Wong,
%``Supersymmetric regularization, two-loop QCD amplitudes and coupling
%shifts,''
Phys.\ Rev.\  D {\bf 66}, 085002 (2002)
[hep-ph/0202271].
%%CITATION = PHRVA,D66,085002;%%

%+% 4 refs
\bibitem{Bern6D}
  Z.~Bern, J.~J.~Carrasco, T.~Dennen, Y.~t.~Huang and H.~Ita,
  %``Generalized Unitarity and Six-Dimensional Helicity,''
  Phys.\ Rev.\  D {\bf 83}, 085022 (2011)
  [arXiv:1010.0494 [hep-th]].
  %%CITATION = PHRVA,D83,085022;%%

%+% 2 refs
\bibitem{Tristan}
  T.~Dennen, Y.~-t.~Huang and W.~Siegel,
  %``Supertwistor space for 6D maximal super Yang-Mills,''
  JHEP\ {\bf 1004}, 127  (2010)
  [arXiv:0910.2688 [hep-th]].
  %%CITATION = JHEPA,1004,127;%%

%+% 1 ref
\bibitem{Koschade}
  A.~Brandhuber, D.~Korres, D.~Koschade and G.~Travaglini,
  %``One-loop Amplitudes in Six-Dimensional (1,1) Theories from Generalised Unitarity,''
  JHEP\ {\bf 1102}, 077  (2011)
  [arXiv:1010.1515 [hep-th]].
  %%CITATION = JHEPA,1102,077;%%

%+% 2 refs
\bibitem{GeorgiGlashow}
  H.~M.~Georgi, S.~L.~Glashow, M.~E.~Machacek, D.~V.~Nanopoulos,
  %``Higgs Bosons from Two Gluon Annihilation in Proton Proton Collisions,''
  Phys.\ Rev.\ Lett.\  {\bf 40}, 692 (1978).
  %%CITATION = PRLTA,40,692;%%

%+% 2 refs
\bibitem{Djouadi}
  A.~Djouadi, M.~Spira, P.~M.~Zerwas,
  %``Production of Higgs bosons in proton colliders: QCD corrections,''
  Phys.\ Lett.\  {\bf B264}, 440-446 (1991).
  %%CITATION = PHLTA,B264,440;%%

%+% 2 refs
\bibitem{Dawson}
  S.~Dawson,
  %``Radiative corrections to Higgs boson production,''
  Nucl.\ Phys.\  {\bf B359}, 283-300 (1991).
  %%CITATION = NUPHA,B359,283;%%

%+% 3 refs
\bibitem{Wilczek}
  F.~Wilczek,
  %``Decays of Heavy Vector Mesons Into Higgs Particles,''
  Phys.\ Rev.\ Lett.\  {\bf 39}, 1304 (1977).
  %%CITATION = PRLTA,39,1304;%%

%+% 1 ref
\bibitem{Shifman}
  M.~A.~Shifman, A.~I.~Vainshtein, V.~I.~Zakharov,
  %``Remarks on Higgs Boson Interactions with Nucleons,''
  Phys.\ Lett.\  {\bf B78}, 443 (1978).
  %%CITATION = PHLTA,B78,443;%%

%+% 1 ref
\bibitem{Higgs4Parton}
  S.~D.~Badger and E.~W.~N.~Glover,
  %``One-loop helicity amplitudes for H ---> gluons: The All-minus
  %configuration,''
  Nucl.\ Phys.\ Proc.\ Suppl.\  {\bf 160}, 71 (2006)
  [hep-ph/0607139];\\
  %%CITATION = NUPHZ,160,71;%%
  S.~D.~Badger, E.~W.~N.~Glover and K.~Risager,
  %``One-loop phi-MHV amplitudes using the unitarity bootstrap,''
  JHEP {\bf 0707}, 066 (2007)
  [arXiv:0704.3914 [hep-ph]];\\
  %%CITATION = JHEPA,0707,066;%%
  E.~W.~N.~Glover, P.~Mastrolia and C.~Williams,
  %``One-loop phi-MHV amplitudes using the unitarity bootstrap: The General
  %helicity case,''
  JHEP {\bf 0808}, 017 (2008)
  [arXiv:0804.4149 [hep-ph]];\\
  %%CITATION = JHEPA,0808,017;%%
  C.~F.~Berger, V.~Del Duca and L.~J.~Dixon,
  %``Recursive Construction of Higgs-Plus-Multiparton Loop Amplitudes: The Last
  %of the Phi-nite Loop Amplitudes,''
  Phys.\ Rev.\  D {\bf 74}, 094021 (2006)
  [Erratum-ibid.\  D {\bf 76}, 099901 (2007)]
  [hep-ph/0608180];\\
  %%CITATION = PHRVA,D74,094021;%%
  S.~Badger, E.~W.~Nigel Glover, P.~Mastrolia and C.~Williams,
  %``One-loop Higgs plus four gluon amplitudes: Full analytic results,''
  JHEP {\bf 1001}, 036 (2010)
  [arXiv:0909.4475 [hep-ph]];\\
  %%CITATION = JHEPA,1001,036;%%
  L.~J.~Dixon and Y.~Sofianatos,
  %``Analytic one-loop amplitudes for a Higgs boson plus four partons,''
  JHEP {\bf 0908}, 058 (2009)
  [arXiv:0906.0008 [hep-ph]];\\
  %%CITATION = JHEPA,0908,058;%%
  S.~Badger, J.~M.~Campbell, R.~K.~Ellis and C.~Williams,
  %``Analytic results for the one-loop NMHV Hqqgg amplitude,''
  JHEP {\bf 0912}, 035 (2009)
  [arXiv:0910.4481 [hep-ph]].
  %%CITATION = JHEPA,0912,035;%%

%+% 1 ref
\bibitem{Dixon}
  L.~J.~Dixon,
  %``Calculating scattering amplitudes efficiently,''
  [hep-ph/9601359].
  %%CITATION = HEP-PH/9601359;%%

%+% 1 ref
\bibitem{Boels}
  R.~Boels,
  %``Covariant representation theory of the Poincare algebra and some of its extensions,''
  JHEP\ {\bf 1001}, 010  (2010)
  [arXiv:0908.0738 [hep-th]];
  %%CITATION = JHEPA,1001,010;%%

\bibitem{BCFW}
  R.~Britto, F.~Cachazo and B.~Feng,
  %``New recursion relations for tree amplitudes of gluons,''
  Nucl.\ Phys.\ B\ {\bf 715}, 499  (2005)
  [hep-th/0412308]; \\
  %%CITATION = NUPHA,B715,499;%%
  R.~Britto, F.~Cachazo, B.~Feng and E.~Witten,
  %``Direct proof of tree-level recursion relation in Yang-Mills theory,''
  Phys.\ Rev.\ Lett.\ \ {\bf 94}, 181602  (2005)
  [hep-th/0501052].
  %%CITATION = PRLTA,94,181602;%%

%+% 3 refs
\bibitem{PentInts}
  Z.~Bern, L.~J.~Dixon and D.~A.~Kosower,
  %``Dimensionally regulated pentagon integrals,''
  Nucl.\ Phys.\  B {\bf 412}, 751 (1994)
  [hep-ph/9306240].
  %%CITATION = NUPHA,B412,751;%%

%+% 2 refs
\bibitem{Kilgore}
  W.~B.~Kilgore,
  %``One-loop Integral Coefficients from Generalized Unitarity,''
  [arXiv:0711.5015 [hep-ph]].
  %%CITATION = ARXIV:0711.5015;%%

%+% 1 ref
\bibitem{OPP2}
  G.~Ossola, C.~G.~Papadopoulos and R.~Pittau,
  %``CutTools: A Program implementing the OPP reduction method to compute
  %one-loop amplitudes,''
  JHEP {\bf 0803}, 042 (2008)
  [arXiv:0711.3596 [hep-ph]].
  %%CITATION = JHEPA,0803,042;%%

%+% 1 ref
\bibitem{BernInf}
  C.~F.~Berger, Z.~Bern, L.~J.~Dixon, D.~Forde and D.~A.~Kosower,
  %``Bootstrapping One-Loop QCD Amplitudes with General Helicities,''
  Phys.\ Rev.\  D {\bf 74}, 036009 (2006)
  [hep-ph/0604195].
  %%CITATION = PHRVA,D74,036009;%%

%+% 1 ref
\bibitem{Bern5Point}
  Z.~Bern, L.~J.~Dixon and D.~A.~Kosower,
  %``One loop corrections to five gluon amplitudes,''
  Phys.\ Rev.\ Lett.\  {\bf 70}, 2677 (1993)
  [hep-ph/9302280].
  %%CITATION = PRLTA,70,2677;%%

%+% 1 ref
\bibitem{Bern2g3f}
  Z.~Bern, L.~J.~Dixon and D.~A.~Kosower,
  %``One loop corrections to two quark three gluon amplitudes,''
  Nucl.\ Phys.\  B {\bf 437}, 259 (1995)
  [hep-ph/9409393].
  %%CITATION = NUPHA,B437,259;%%

%+% 2 refs
\bibitem{KunsztResults}
  Z.~Kunszt, A.~Signer and Z.~Trocsanyi,
  %``One loop helicity amplitudes for all 2 ---> 2 processes in QCD and N=1
  %supersymmetric Yang-Mills theory,''
  Nucl.\ Phys.\  B {\bf 411}, 397 (1994)
  [hep-ph/9305239].
  %%CITATION = NUPHA,B411,397;%%

%+% 1 ref
\bibitem{Kramer}
  M.~Kramer, E.~Laenen and M.~Spira,
  %``Soft gluon radiation in Higgs boson production at the LHC,''
  Nucl.\ Phys.\  B {\bf 511}, 523 (1998)
  [hep-ph/9611272].
  %%CITATION = NUPHA,B511,523;%%

%+% 1 ref
\bibitem{DelDucaKilgore}
  V.~Del Duca, W.~Kilgore, C.~Oleari, C.~Schmidt and D.~Zeppenfeld,
  %``Higgs + 2 jets via gluon fusion,''
  Phys.\ Rev.\ Lett.\  {\bf 87}, 122001 (2001)
  [hep-ph/0105129].
  %%CITATION = PRLTA,87,122001;%%

%+% 2 refs
\bibitem{Schmidt}
  C.~R.~Schmidt,
  %``H ---> g g g (g q anti-q) at two loops in the large M(t) limit,''
  Phys.\ Lett.\  B {\bf 413}, 391 (1997)
  [hep-ph/9707448].
  %%CITATION = PHLTA,B413,391;%%

%+% 2 refs
\bibitem{PassarinoVeltman}
  G.~Passarino, M.~J.~G.~Veltman,
  %``One Loop Corrections for e+ e- Annihilation Into mu+ mu- in the Weinberg Model,''
  Nucl.\ Phys.\  {\bf B160}, 151 (1979).
  %%CITATION = NUPHA,B160,151;%%

%+% 1 ref
\bibitem{BrittoReview}
  R.~Britto,
  %``Loop amplitudes in gauge theories: modern analytic approaches,''
  [arXiv:1012.4493 [hep-th]].
  %%CITATION = ARXIV:1012.4493;%%

%+% 1 ref
\bibitem{DimRegInts}
  Z.~Bern, L.~J.~Dixon and D.~A.~Kosower,
  %``Dimensionally regulated one loop integrals,''
  Phys.\ Lett.\  B {\bf 302}, 299 (1993)
  [Erratum-ibid.\  B {\bf 318}, 649 (1993)]
  [hep-ph/9212308].
  %%CITATION = PHLTA,B302,299;%%

%+% 1 ref
\bibitem{ScalarBasis1}
  R.~Britto, B.~Feng and P.~Mastrolia,
  %``Closed-Form Decomposition of One-Loop Massive Amplitudes,''
  Phys.\ Rev.\  D {\bf 78}, 025031 (2008)
  [arXiv:0803.1989 [hep-ph]].
  %%CITATION = PHRVA,D78,025031;%%

%+% 1 ref
\bibitem{ScalarBasis2}
  G.~Ossola, C.~G.~Papadopoulos and R.~Pittau,
  %``On the Rational Terms of the one-loop amplitudes,''
  JHEP {\bf 0805}, 004 (2008)
  [arXiv:0802.1876 [hep-ph]].
  %%CITATION = JHEPA,0805,004;%%

\end{thebibliography}
\end{document}